\documentclass[prl,twocolumn,float,aps]{revtex4-1}
\usepackage{graphicx,amsfonts,amssymb,amsmath,hyperref,hypcap,enumerate}
\usepackage{color}
\usepackage{ulem}

\newcommand\T{\rule{0pt}{2.6ex}}          
\newcommand\B{\rule[-1.2ex]{0pt}{0pt}}  

\makeatletter
\renewcommand\section{\@startsection {section}{1}{\z@}%
	{-2.5ex \@plus -1ex \@minus -.2ex}%
	{0.4ex \@plus.2ex}%
	{\normalfont\bfseries}}
\makeatother

\makeatletter
\renewcommand\subsection{%
	\@startsection{subsubsection}{3}{\z@}{3.25ex \@plus1ex \@minus.2ex}%
	{-1em}{\normalfont\normalsize\bfseries}}
\makeatother

\begin{document}
\title{Topological Superconducting Vortex From Trivial Electronic Bands}
\author{Lun-Hui Hu$^{1,2}$}
\author{Rui-Xing Zhang$^{1,3,2}$} 
\email{ruixing@utk.edu}

\affiliation{$^1$Department of Physics and Astronomy, The University of Tennessee, Knoxville, Tennessee 37996, USA}
\affiliation{$^2$Institute for Advanced Materials and Manufacturing, The University of Tennessee, Knoxville, Tennessee 37920, USA}
\affiliation{$^3$Department of Materials Science and Engineering, The University of Tennessee, Knoxville, Tennessee 37996, USA}


\begin{abstract}
Superconducting vortices are promising traps to confine non-Abelian Majorana quasi-particles. It has been widely believed that bulk-state topology, of either normal-state or superconducting ground-state wavefunctions, is crucial for enabling Majorana zero modes in solid-state systems. This common belief has shaped two major search directions for Majorana modes, in either intrinsic topological superconductors or trivially superconducting topological materials. Here we show that Majorana-carrying superconducting vortex is not exclusive to bulk-state topology, but can arise from topologically trivial quantum materials as well. We predict that the trivial bands in superconducting HgTe-class materials are responsible for inducing anomalous vortex topological physics that goes beyond any existing theoretical paradigms. A feasible scheme of strain-controlled Majorana engineering and experimental signatures for vortex Majorana modes
are also discussed. Our work provides new guidelines for vortex-based Majorana search in general superconductors.
\end{abstract}

\maketitle

\section{Introduction}

In condensed matter systems, the marriage of topology and electron correlations allows for fractionalizing electronic degrees of freedom into exotic non-Abelian quasiparticles such as Majorana zero modes (MZMs)~\cite{Kitaev_aop_2003,nayak_rmp_2008}. Research efforts in the past two decades have together established superconductors (SCs) with certain topological properties as the best venue for trapping and manipulating MZMs, with which quantum information can be processed in a topologically protected manner. For example, a topological SC (TSC) can host zero-dimensional (0D) MZMs bound to either its geometric boundary~\cite{Kitaev_pu_2001} or the superconducting vortex~\cite{read2000paired}, a manifestation of the bulk-boundary correspondence principle. This scenario has motivated enormous research efforts in unconventional SCs and ferromagnet-SC heterostructures~\cite{lutchyn_prl_2010,sau_prl_2010,Mourik_science_2012,kezilebieke2020topological}, where natural and artificial TSCs are believed to exist, respectively. Remarkably, such a topological requirement can be further relaxed for vortex-trapped MZMs if the bulk electronic band structure, instead of the superconductivity itself, carries a non-trivial topological index~\cite{fu_prl_2008,hosur_prl_2011}. This spirit also inspires another intensive search of topological band materials with intrinsic yet non-topological SC~\cite{pacholski_prl_2018,	konig_prl_2019,qin_prl_2019,yan_prl_2020,ghazaryan_prb_2020,kobayashi2020double,giwa_prl_2021,hu_arxiv_2021}, with many promising candidates discovered~\cite{sun_prl_2016,Wang_science_2018,Kong_np_2019,Liu_nc_2020}. However, as far as we know, the possibility of trapping MZMs in trivial $s$-wave SCs with trivial electronic band structures has been rarely explored in the literature.

In this work, we show that a three-dimensional (3D) $s$-wave spin-singlet SC, with certain non-topological normal states, is capable of harboring Majorana-carrying topological vortices. This conclusion is explicitly demonstrated in the superconducting phase of 3D Luttinger semimetal (LSM)~\cite{luttinger_pr_1956} as a proof of concept, whose normal-state semimetallicity is of trivial topology. Topological superconducting vortex-line states with either 0D end-localized MZMs or a 1D Dirac-nodal dispersion are found to be ubiquitous in the vortex phase diagram of LSMs, shedding new light on this 60-year-old classical band system. The vortex line topology here manifests a distinct origin from known vortex Majorana theories~\cite{fu_prl_2008,hosur_prl_2011,chiu_prl_2012,xu_prl_2016,yan_prl_2017,chan_prl_2017,chan20172ndChern,pacholski_prl_2018,konig_prl_2019,qin_prl_2019,yan_prl_2020,kobayashi2020double,ghazaryan_prb_2020,giwa_prl_2021}, most of which would require topological band inversion in the normal states.
Furthermore, a tensile-strained LSM is found to be a bulk-trivial yet vortex-exotic band insulator, which harbors distinct topological vortex phases in the presence of electron and hole dopings, respectively.

LSMs generally show up as the $\Gamma_8$ quartet in HgTe-class materials, where the inversion between $\Gamma_8$ and $\Gamma_6$ bands usually creates a zero-gap topological insulator (TI). The composition of TI and LSM bands offers a minimal exemplar to visualize the competition between topological and trivial bulk bands for deciding the vortex topology. While a topological-band-only analysis anticipates a Majorana-carrying Kitaev vortex, our new vortex paradigm predicts a Majorana-free topological nodal vortex instead, further confirmed by our numerical simulations. We propose lattice strain effect as a promising control knob to detect and engineer vortex MZMs in superconducting HgTe-class materials. Experimental signatures of the proposed vortex topological physics are discussed in the details.
We conclude by highlighting the potentially crucial role of low-energy trivial bands in deciding the vortex topology in general SCs and further providing suggestions on the ongoing Majorana search.

\section{Results}

\subsection{$C_n$-symmetric vortex topology.} 


\begin{table}[tb]
	\begin{tabular}{c  c  c  c  c  c}  
		\hline \hline
		Symmetry  & $C_1$ & $C_2$ & $C_3$ & $C_4$ & $C_6$ \T\B  \\    \hline      
		Classification   & $\mathbb{Z}_2$  & $\mathbb{Z}_2\times \mathbb{Z}_2$  & $\mathbb{Z}_2\times \mathbb{Z}$ & $(\mathbb{Z}_2 )^2\times \mathbb{Z}$ & $(\mathbb{Z}_2 )^2\times (\mathbb{Z})^2$  \T \B \\      
		Invariant & $\nu_0$ & $\nu_{0,1}$ & $(\nu_0, {\cal Q}_1)$  & $(\nu_{0,2}, {\cal Q}_1)$ & $(\nu_{0,3}, {\cal Q}_{1,2})$  \T \B \\  \hline \hline
	\end{tabular}
	\caption{Vortex topological classification of $C_n$-invariant 
		$s$-wave spin-singlet superconductors. $\nu_{J_z}\in\mathbb{Z}_2$ is a symmetry-indexed topological invariant signaling the presence ($\nu_{J_z}$=1) or absence ($\nu_{J_z}=0$) of a $J_z$-labeled vortex Majorana zero mode (MZM).  The $C_n$ topological charge ${\cal Q}_{J_z}\in \mathbb{Z}$ characterizes the symmetry-protected vortex band crossings (i.e., a nodal vortex) near the zero energy. In principle, a vortex line is capable of carrying multiple 0D vortex MZMs and nodal bands that do not interfere with each other, as long as they are supported by distinct topological indices.
	}
	\label{Table1}
\end{table}


We start with a general topological discussion on the superconducting vortex-line states. A superconducting vortex in a 3D Bogoliubov-de Gennes (BdG) system is a 1D line defect that traps low-energy Caroli-de Gennes-Matricon (CdGM) bound states. Generated by an external magnetic field ${\bf B}$, the CdGM states disperse along ${\bf k_B} \parallel {\bf B}$ to form an effective 1D system in symmetry class D, as described by a vortex-line Hamiltonian $h_\text{vort}(k_{\bf B})$. Throughout this work, we will denote $\hat{z}$ as the magnetic field direction for simplicity. Besides the built-in particle-hole symmetry (PHS), $h_{\text{vort}}$ can additionally respect ${\cal G}_{\bf B}$, a subgroup of the 3D crystalline group ${\cal G}$ in the zero-field limit. The band topology of $h_{\text{vort}}$ is protected by both PHS $\Xi$ and ${\cal G}_{\bf B}$. 

We focus in this work on general $s$-wave spin-singlet superconductors,
where ${\cal G}_{\bf B}$ is a $n$-fold rotation group $C_n$ and every CdGM state carries a $C_n$ index $J_z\in\{0,1,2,...,n-1\}$, i.e., the $\hat{z}$-directional angular momentum modulo $n$. CdGM states with different $J_z$ labels are decoupled from each other along $k_z$ and each $J_z$ sector can be characterized by its own 1D topological index. With an $s$-wave pairing,
$J_z\in \{0, \frac{n}{2}\}$ sectors are PHS invariant themselves and carry a $\mathbb{Z}_2$ Pfaffian index $\nu_{J_z}\in\{0,1\}$~\cite{Kitaev_pu_2001}. 
Note that for systems with a non-$s$-wave pairing, the PHS-invariant $J_z$ sectors might be different from the above.
When $\nu_{J_z}=1$, all $J_z$-indexed CdGM states constitute a 1D TSC phase that is equivalent to a Kitaev Majorana chain, contributing to a $J_z$-labeled vortex MZM on the sample surface. We dub this gapped vortex phase a Kitaev vortex. On the other hand, $J_z$ and $n-J_z$ form particle-hole conjugate sectors if $J_z\notin \{0,\frac{n}{2}\}$ and together carry a $\mathbb{Z}$-type topological index, 
\begin{equation}
	{\cal Q}_{J_z} = n_{J_z}^{(v)}(0) - n_{J_z}^{(v)} (\pi),
\end{equation}
where $n_{J_z}^{(v)}(k_z)$ counts the number of $J_z$-carrying CdGM states with a negative energy at $k_z$. A derivation of ${\cal Q}_{J_z}$ is provided in the Supplemetary Note 1. Physically, ${\cal Q}_{J_z}$ indicates the number of pairs of $C_n$-protected BdG nodal points along $k_z$, signaling a band-inverted gapless vortex state dubbed a nodal vortex. Kitaev and nodal vortices are elementary building blocks to construct general $C_n$-protected vortex topological phenomena. 

We now demonstrate our classification scheme. For instance, $C_2$ group possesses two PHS-invariant $J_z$ sectors $J_z=0$ and $J_z=1$, and a general $C_2$-invariant vortex can only harbor Kitaev vortices but not the nodal ones. The vortex topology is then characterized by $\nu_{0,1}$, thus being $\mathbb{Z}_2 \times \mathbb{Z}_2$ classified. When $\nu_0=\nu_1=1$, a Majorana doublet emerges in the surface vortex core and the two MZMs will not mix for carrying distinct $J_z$ labels. Take $C_6$ as another example, the $(\mathbb{Z}_2)^2 = \mathbb{Z}_2 \times \mathbb{Z}_2$ part is contributed by the PHS-invariant sectors $J_z=0$ and $J_z=3$, similar to that in the $C_2$ case. In addition, $(J_z=1, J_z=5)$ and $(J_z=2, J_z=4)$ form two pairs of particle-hole conjugate sectors indicated by ${\cal Q}_1$ and ${\cal Q}_2$, so that only nodal vortices can occur in these sectors. This leads to another $\mathbb{Z}\times \mathbb{Z}$ contribution, promoting the classification of $C_6$-symmetric vortices to $(\mathbb{Z}_2)^2\times (\mathbb{Z} )^2$. We summarize the vortex topological classification and characterization for all $C_n$ groups in Table.~\ref{Table1}.     	

Notably, the protection of vortex-line topology is decided by both the bulk crystalline symmetry group and the magnetic field orientation. Thus, it is possible to realize distinct vortex topological states in a single superconducting material by simply rotating the applied magnetic field. This clearly implies the absence of an exact one-to-one mapping between bulk-state and vortex-line topologies. This observation motivates us to explore the possibility of topological vortices inside a \textit{completely trivial} SC, whose topological triviality manifests in both its Cooper-pair and normal-state wavefunctions.  	

\subsection{Vortex topology from trivial bulk bands.}

\begin{figure*}[t]
	\centering
	\includegraphics[width=\linewidth]{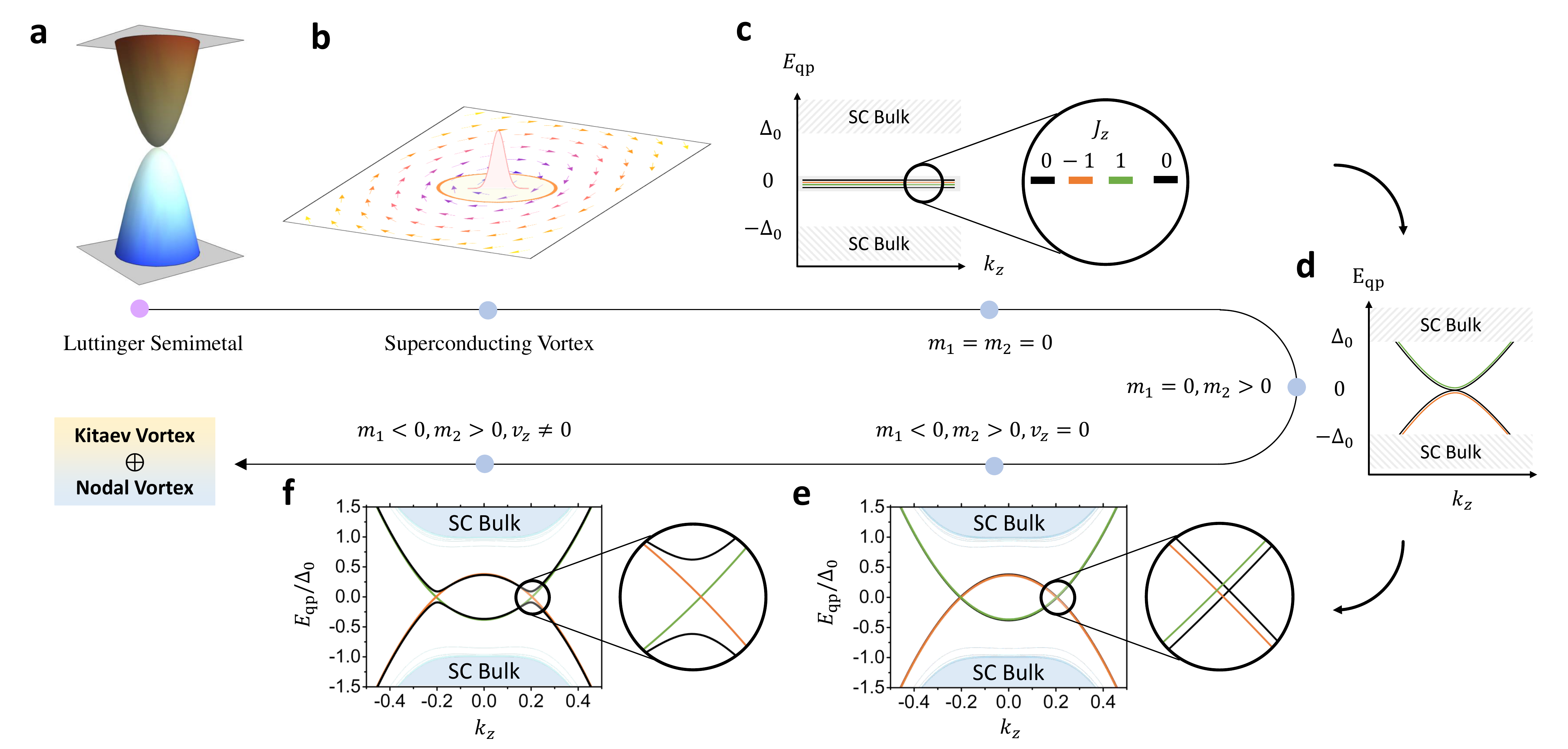}
	\caption{Topological superconducting vortex in a Luttinger semimetal (LSM). (a) shows the quadratic band touching around $\Gamma$ point of a LSM. In (b), the superconducting (SC) pairing function $\Delta(x,y)$ is illustrated for the $k_z=0$ plane, where the vortex phase winding is denoted by in-plane arrows surrounding the vortex core. Four vortex zero modes are expected to occur for LSM at $k_z=0$ due to an emergent chiral winding number. The vortex-line low-energy spectra $E_{\text{qp}}/\Delta_0$ are illustrated in (c) for  $m_1=m_2=v_z=0$ with four zero-energy flat bands labeled by angular momenta $J_z$; and in (d) for $m_1=0,m_2>0,v_z=0$. Two pairs of vortex nodal bands show up in (e) for $v_z=0$, while only the ones formed by $J_z=\pm 1$ are symmetry protected. Turning on $v_z\neq0$ will gap out the unprotected crossings within $J_z=0$ sector, as shown in (f), leading to a Kitaev vortex.  The final vortex state of a LSM consists of a nodal vortex coexisting with a Kitaev vortex. (e) and (f) are numerically simulated in a disk geometry with band parameters $m_1=-1, m_2=2, v_\parallel=\sqrt{3}, v_z=2\sqrt{3}, \Delta_0=0.2$.
	}
	\label{fig1}
\end{figure*}

Our target trivial-band system is a 3D Luttinger semimetal (LSM), which is defined by a single four-fold degenerate quadratic band touching at $\Gamma$~\cite{luttinger_pr_1956,murakami_prb_2004}, i.e., the origin of the Brillouin zone (BZ).
This band degeneracy arises from a 4D double-valued irreducible representation (irrep) $\Gamma_8$ of point groups such as $O$, $O_h$ and $T_d$. Unlike traditional topological semimetals~\cite{bansil_rmp_2016,armitage_rmp_2018,lv_rmp_2021}, the point node of a LSM does not serve as a topological quantum critical point between two distinct lower-dimensional gapped topological phases, and is thus trivial in the topological sense. 
Remarkably, such a trivial band set, together with isotropic $s$-wave superconductivity, will give rise to nontrivial vortex topologies, which we will show below. 

The $\Gamma_8$-bands are captured by the atomic basis $\vert \Psi_{\Gamma_8} \rangle = (\vert p_+,\uparrow\rangle, \vert p_+,\downarrow\rangle, \vert p_-,\uparrow\rangle, \vert p_-,\downarrow\rangle)^{T}$ with $\uparrow,\downarrow$ denoting the electron spin and $p_\pm = p_x \pm i p_y$ orbitals. Under this basis, we consider a ${\bf k \cdot p}$ model Hamiltonian around $\Gamma$ that respects inversion, time-reversal, and around-$\hat{z}$-axis full rotation symmetries. In particular, $\mathcal{H}_{\text{LSM}} = \lambda_1 k^2\gamma_0 + M({\bf k})\gamma_5 + v_z k_z (k_x \gamma_{45}+ k_y \gamma_{35}) - \sqrt{3}\lambda_2 ((k_x^2-k_y^2)\gamma_{25} + 2k_xk_y\gamma_{15})$. 
Here, $M({\bf k})=m_1(k_x^2+k_y^2)+m_2k_z^2$ and the $4\times4$ $\gamma$-matrices are defined as $\gamma_1 = \sigma_x \otimes s_z,\ \gamma_2 = \sigma_y \otimes s_z,\ \gamma_3 = \sigma_0 \otimes s_x,\ \gamma_4 = \sigma_0 \otimes s_y,\ \gamma_5 = \sigma_z \otimes s_z $ with $\gamma_{mn}=-i\gamma_m\gamma_n$ and $\gamma_0=\sigma_0\otimes s_0$ the identity matrix.
$\sigma$ and $s$ are Pauli matrices denoting the orbital and spin degrees of freedom, respectively.
Without loss of generality, we set $\lambda_1=0$ in the following discussion, and the four bulk band dispersions are $E_{\pm}({\bf k}) =\pm \sqrt{(m_1^2 + 3\lambda_2^2)k_\parallel^4 + (2m_1m_2 + v_z^2)k_z^2k_\parallel^2 + m_2^2k_z^4}$ with $k_\parallel^2 = k_x^2 + k_y^2$. Therefore, ${\cal H}_{LSM}$ describes a quadratic semimetal with different in-plane and out-of-plane dispersions, serving as an anisotropic generalization of the conventional isotropic LSM model~\cite{luttinger_pr_1956, murakami_prb_2004}. The isotropic limit can be achieved with $m_1=-\tfrac{1}{2}m_2=\lambda_2$ and $v_z=-2\sqrt{3}\lambda_2$, leading to $E_{\pm}({\bf k}) = \pm 2\vert \lambda_2\vert k^2$ with $k^2=k_\parallel^2 + k_z^2$.
A dispersion plot for the isotropic LSM phase is shown in Fig.~\ref{fig1} (a).  Superconductivity of LSMs is described by generalizing $\mathcal{H}_{\text{LSM}}$ into a BdG form,
\begin{equation}\label{eq-bdg-ham-lsm}
	\mathcal{H}_\text{BdG} = \begin{pmatrix}
		\mathcal{H}_{\text{LSM}}(\mathbf{k})-\mu  & \mathcal{H}_\Delta \\
		\mathcal{H}_\Delta^\dagger & \mu-\mathcal{H}_{\text{LSM}}^\ast(-\mathbf{k})
	\end{pmatrix}, 
\end{equation}
where $\mu$ is the chemical potential. $\mathcal{H}_\Delta= i\Delta(\mathbf{r})\gamma_{13}$ describes an isotropic $s$-wave spin-singlet pairing, making ${\cal H}_\text{BdG}$ carry a trivial bulk topology. A superconducting vortex line centering at $r=0$ can be generated by $\Delta(\mathbf{r})=\Delta_0\tanh(r/\xi_0) e^{i\theta}$, with $(r,\theta)$ being the in-plane polar coordinates and $\xi_0$ the SC coherence length. 

Origin of topological vortex-line modes in LSMs can be understood in a perturbative manner, which is schematically depicted in Fig.~\ref{fig1}.  This is motivated by a key observation that the normal state ${\cal H}_\text{LSM}({\bf k}) = h^{(0)} ({\bf k}_{\parallel}) + h^{(1)} ({\bf k}_{\parallel}, k_z)$ with
\begin{equation}
	h^{(0)}({\bf k}_\parallel) =\begin{pmatrix}
		0 &  -\sqrt{3}\lambda_2 k_-^2 \\
		-\sqrt{3}\lambda_2 k_+^2 & 0
	\end{pmatrix} \otimes  s_0. 
\end{equation}
Here $k_{\pm}=k_x \pm i k_y$. The unperturbed part $h^{(0)}$ describes two identical copies of 2D massless quadratic Dirac fermions, each of which carries a $2\pi$ Berry phase and is similar to those live in bilayer graphene~\cite{McCann_RPP_2013} 
and on the surfaces of topological crystalline insulators~\cite{fu_prl_2011,zhang_prb_2015}. While a 2D linear Dirac fermion carries a single vortex MZM~\cite{fu_prl_2008}, we naturally expect $h^{(0)}$ to support four vortex MZMs if going superconducting, with each quadratic Dirac fermion contributing a pair of MZMs in Fig.~\ref{fig1} (c). 

This conjecture is confirmed by exactly mapping the 2D vortex problem of $h^{(0)} ({\bf k}_{\parallel})$ to a 3D chiral topological insulator~\cite{teo2010topodefect}, thanks to an emergent chiral symmetry ${\cal S}$ of the system. This allows us to exploit the 3D chiral winding number ${\cal N_S}$~\cite{schnyder2008class} to topologically quantify the zero modes, with the spatial polar angle $\theta$ acting as an extra dimension in addition to $k_x$ and $k_y$. As discussed in Methods, we analytically calculate ${\cal N_S} =4$, confirming these four vortex zero modes. We further simulate the superconducting vortex of $h^{(0)} ({\bf k}_{\parallel})$ on a large disc geometry to numerically confirm the zero modes, and find that they are $J_z$-labeled. In particular, two zero modes form a PHS-related pair and carry $J_z=\pm 1$, while the other two are both labeled by $J_z=0$.  

\begin{figure*}[t]
	\centering
	\includegraphics[width=0.7\linewidth]{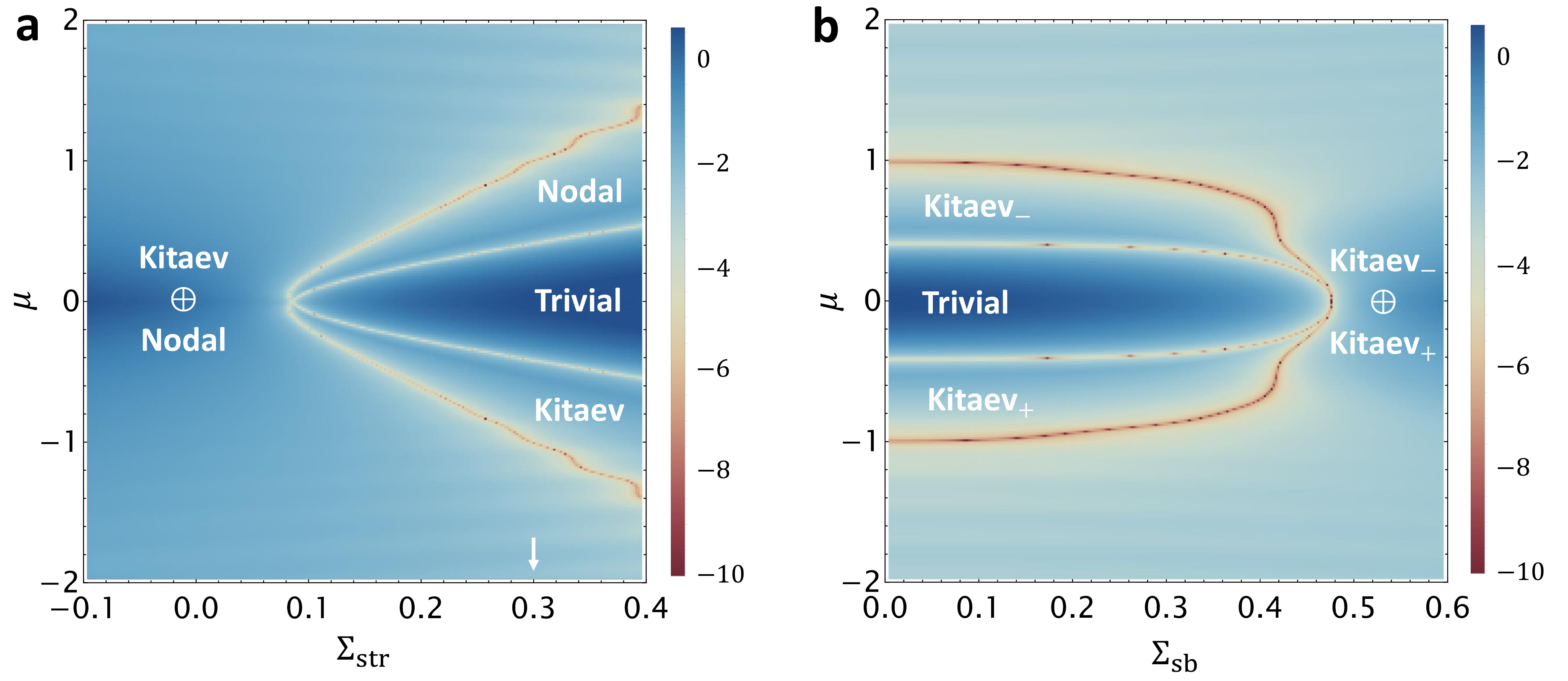}
	\caption{Vortex topological phase diagrams (VTPD) of a strained LSM.  
	Both VTPDs are mapped out by calculating the vortex-state energy gap at $k_z=0$, whose logarithmic value is shown by the colors in (a) and (b). Vortex topology changes whenever the vortex-state gap closes. (a) shows the VTPD as a function of $\Sigma_{str}$ and $\mu$. 
	Specifically, the normal state is a topologically trivial insulator for $\Sigma_{\text{str}}>0$ and a Dirac semimetal for $\Sigma_{\text{str}}<0$.
		(b) shows the VTPD as a function of $\Sigma_{\text{sb}}$ and $\mu$, with a fixed $\Sigma_{\text{str}}=0.3$ [white arrow in (a)]. The rotational symmetry breaking induced by $\Sigma_{\text{sb}}$ updates the nodal vortex in (a) to the Kitaev$_-$ vortex in (b). Here $\pm$ is used to represent the eigenvalue of the two-fold rotational symmetry. The model parameters for both calculations are the same as those in Fig.~\ref{fig1} (f). 
	}
	\label{newfig2}
\end{figure*}

Taking into account $h^{(1)} ({\bf k}_{\parallel}, k_z)$, the four zero modes start to hybridize, split, and disperse along $k_z$. Crucially, we note that in $h^{(1)}$, $M({\bf k}) = m_1(k_x^2+k_y^2)+m_2k_z^2$ features $m_1m_2=-2\lambda_2^2<0$ for an isotropic LSM. As we rigorously prove in the Supplementary Note 3, a negative $m_1$ will send two zero modes with $J_z=0,1$ [i.e. colored in black and green in Fig.~\ref{fig1} (c)] to a negative energy. 
Meanwhile, a  positive $m_2$ will make sure the same zero modes to quadractically disperse along $k_z$, but with a positive mass. The PHS requires the other two zero modes with $J_z=0, -1$ to behave oppositely. As a result, the original quartet of zero modes evolves into two pairs of 1D inverted CdGM bands, as numerically shown in Fig.~\ref{fig1} (e). The inverted bands with $J_z=\pm1$ feature a pair of rotation-protected band crossings, forming a nodal vortex state. The $J_z=0$ bands, however, will open up a topological gap as the $v_z$ term of $h^{(1)}$ is included~[see Fig.~\ref{fig1} (f)], which forms a Majorana-carrying Kitaev vortex. Moreover, this exotic vortex-line physics holds in the isotropic limit as well, which we confirm numerically by mapping out the vortex topological phase diagram in the Fig.~\ref{newfig2}.
Therefore, we have managed to prove that a superconducting anisotropic or isotropic LSM will simultaneously carry topological Kitaev and nodal vortices, i.e., $\nu_0 = {\cal Q} _1= 1$, 
despite the trivial nature of its normal-state electron bands. 


As a 4D irrep of the crystalline group, the quadratic band touching of LSM is unstable against lattice strain effects. It is natural to ask about the stability of the LSM-origined vortex topological phases under strain-induced perturbations. Motivated by this, we consider to perturb the original LSM Hamiltonian with two different strain effects described by $\mathcal{H}_{\text{LSM}}' = - \Sigma_{\text{str}} \gamma_5 + \Sigma_{\text{sb}} \gamma_{15}$. In particular, a positive (negative) $\Sigma_{\text{str}}$ describes a uniaxial tensile (compressive) strain that reduce the original $O(3)$ symmetry to an around-$\hat{z}$ continuous rotation symmetry $C_\infty$. Meanwhile, $\Sigma_{\text{sb}}$ further breaks $C_\infty$ down to a two-fold rotation $C_2$. Both terms preserve inversion symmetry ${\cal P}=\gamma_0$ of the normal-state Hamiltonian. 
In Fig.~\ref{newfig2}, we numerically map out the vortex topological phase diagrams (VTPDs) as a function of $\mu$, $\Sigma_{\text{str}}$, and $\Sigma_{\text{sb}}$. This is achieved by regularizing the vortex-inserted LSM Hamiltonian ($\mathcal{H}_{\text{LSM}}+\mathcal{H}_{\text{LSM}}'$) on a $80\times 80$ square latttice and calculating its CdGM energy spectrum along $k_z$. As elaborated in the Supplementary Note 4, the VTPDs for lattice-regularized models generally agree well with those of the continuum models in a quantitative manner. Whenever the CdGM gap closes at $k_z=0$, vortex-line topology will simultaneously change.

Let us start with the $\Sigma_{\text{str}}$-$\mu$ VTPD in Fig.~\ref{newfig2}~(a) with $\Sigma_{\text{sb}}=0$. At the bulk-band level, $\Sigma_{\text{str}}<0$ creates a new band inversion around $\Gamma$, leading to a Dirac semimetal phase with a pair of linearly dispersing 3D Dirac nodes on the $k_z$ axis~\cite{xu_prl_2017}. Unlike Na$_3$Bi or Cd$_3$As$_2$, this Dirac semimetal phase does not feature any topological surface state, because of ${\cal P}=\gamma_0$. Remarkably, the VTPD is governed by the coexistence of Kitaev and nodal vortex phases (denoted as Kitaev $\oplus$ Nodal) for $\Sigma_{\text{str}}\le 0$, as shown in Fig.~\ref{newfig2} (a). 
This agrees with our analytical perturbation theory derived in the Supplementary Note 3, where a negative $\Sigma_{\text{str}}$ enhances the band inversions of CdGM bands and thus stabilizes the Kitaev $\oplus$ Nodal phase. 
Conversely, a positive $\Sigma_{\text{str}}$ would destabilize this phase at small $\mu$. Because $\Sigma_{\text{str}}>0$ energetically shifts the electron bands in the opposite way, driving the system into a trivial band insulator. When $\mu$ lies inside the band gap ($|\mu|<\Sigma_{\text{str}}$), the vortex-line topology is guaranteed to be trivial for having neither bulk nor surface states at the Fermi level, further forming a fan-shaped trivial vortex regime as confirmed in Fig.~\ref{newfig2}~(a). Strikingly, hole (electron) doping of this trivial insulator will enable a topological Kitaev (nodal) vortex phase. 

Switching on $\Sigma_{\text{sb}}$ generally spoils symmetry protection of the nodal vortex phase by introducing a topological gap for the CdGM states. Due to the PHS and the remaining $C_2$, this new gapped vortex state necessarily carries a nontrivial Kitaev $Z_2$ index $\nu_1=1$ in the $C_2=-1$ sector. Therefore, this $\Sigma_{\text{sb}}$-induced Kitaev phase is topologically distinct from the preexisting Kitaev vortex phase that carries $\nu_0=1$, a manifestation of the $C_2$-stabilized $\mathbb{Z}_2\times \mathbb{Z}_2$ vortex topological classification shown in Table.~\ref{Table1}. We thus dub a Kitaev vortex phase living in the $C_2=\pm 1$ sector a Kitaev$_\pm$ vortex phase, to highlight its symmetry-eigenvalue label. For a fixed $\Sigma_{\text{str}}=0.3$ (i.e., the normal state is the trivial insulator phase), we numerically map out the $\Sigma_{\text{sb}}$-$\mu$ VTPD, as shown in Fig.~\ref{newfig2}~(b). Interestingly, the VTPD contains all four gapped vortex phases dictated by the set of $\mathbb{Z}_2\times \mathbb{Z}_2$ topological indices $(\nu_0, \nu_1)$: trivial phase with $(0,0)$, Kitaev$_+$ phase with $(1,0)$, Kitaev$_-$ phase with $(0,1)$, and Kitaev$_- \oplus$ Kitaev$_+$ phase with $(1,1)$. In the Supplementary Note 2.3, we numerically calculate the surface local density of states for both Kitaev$_\pm$ vortex phases using the recursive Green's function method~\cite{sancho_JPF_1985}. The existence of vortex Majorana zero mode for each phase is confirmed by the presence of a zero-bias-peak at the vortex core center. This unambiguously demonstrates how a variety of vortex-line topologies, as well as their accompanied Majorana modes, can arise from a doped trivial band insulator with $s$-wave superconductivity. 

\subsection{Material realization.} 

\begin{figure*}[t]
	\centering
	\includegraphics[width=0.88\linewidth]{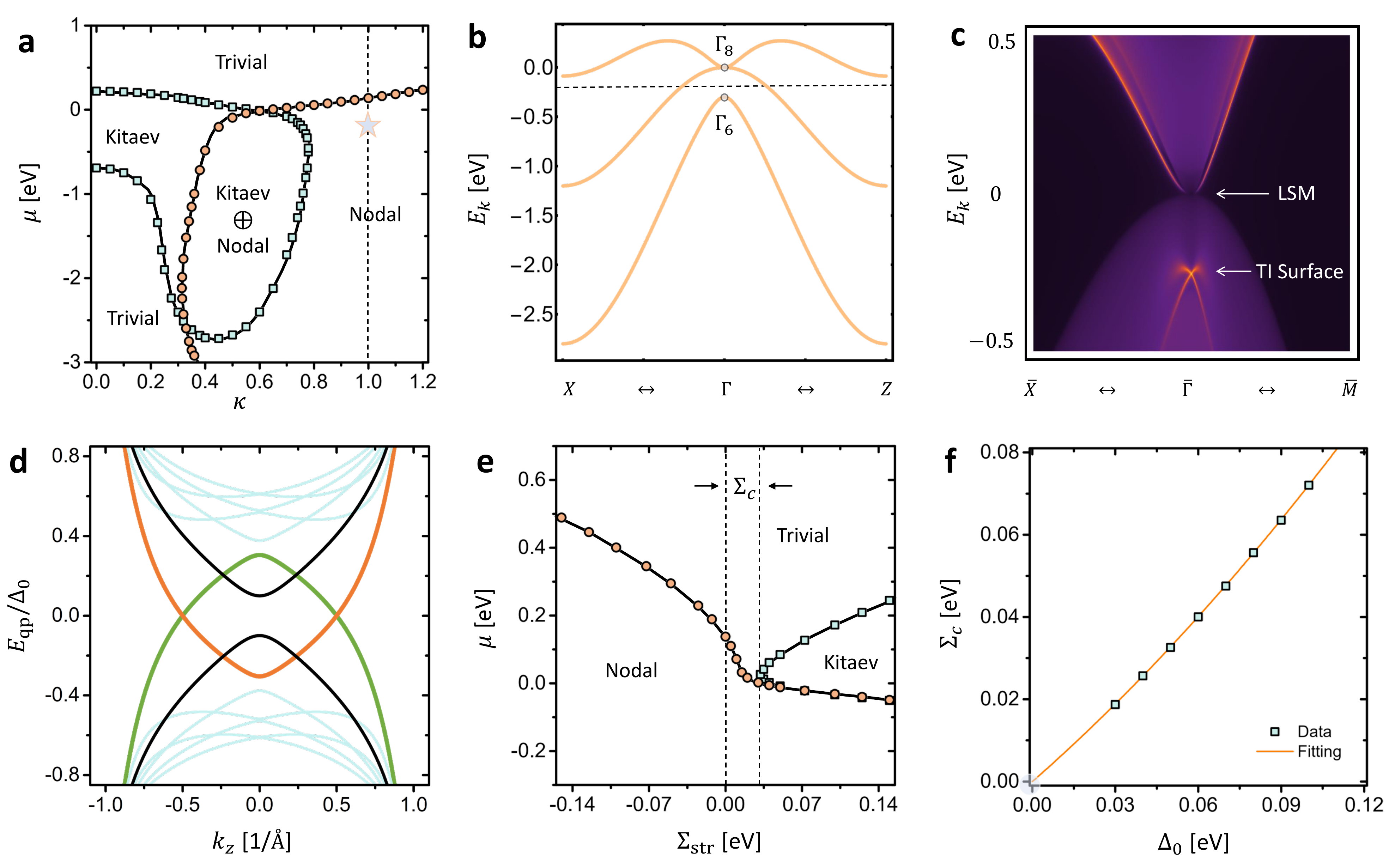}
	\caption{Vortex phase diagram of HgTe. In (a), we show the phase diagram as a function of inter-band coupling $\kappa$ and the chemical potential $\mu$, which includes Kitaev vortex (small $\kappa$), Kitaev $\oplus$ nodal vortex (intermediate $\kappa$), nodal vortex (large $\kappa$) and trivial vortex. $\kappa=1$ is the Luttinger semimetal (LSM) limit, which recovers the realistic model parameters for HgTe (dashed black line). (b) and (c) show the bulk and (001) surface dispersions of HgTe based on a realistic 6-band Kane model, which clearly reveals the coexisting topological insulator (TI) and LSM physics. In (d), the nodal vortex spectrum $E_{\text{qp}}/\Delta_0$ is calculated for the star location in (a), with two bands carrying $J_z=-1$	(orange line) and $J_z=+1$ (green line) crossing at zero-energy. The strain-controlled topological phase diagram is shown in (e) as a function of the strain strength $\Sigma_{\text{str}}$ and $\mu$, where the crictial strain strength $\Sigma_c$ is defined. (f) shows the scaling behavior of $\Sigma_c$ as a function of $\Delta_0$. The fitting function in orange dashed line is exactly extrapolated to the origin.
	}
	\label{fig2}
\end{figure*}


The LSM-band physics has been experimentally established in HgTe-class materials, including HgTe~\cite{novik_prb_2005}, $\alpha$-Sn~\cite{groves_1963_prl,xu_prl_2017}, pyrochlore iridates such as Pr$_2$Ir$_2$O$_7$~\cite{kondo2015quadratic}, half-Heusler alloys such as LaPtBi~\cite{yan2014half}, etc. As shown in Fig.~\ref{fig2} (b), the typical bulk band structure of HgTe-class materials is well captured by a six-band Kane model, which consists of a pair of $s$-type $\Gamma_6$ electron bands with $J_z=\pm 1/2$ and a quartet of $p$-type $\Gamma_8$ hole bands with $J_z=\pm 1/2$ [light holes (LHs)] and $J_z=\pm 3/2$ [heavy holes (HHs)]. To achieve LSM bands, the band order between $\Gamma_6$ and LH-bands needs to be inverted when comparing to that in semiconductors such as CdTe. This band inversion makes $\Gamma_6$ and LHs a typical TI band set, sitting right below the $\Gamma_8$ band touching (i.e., LSM). As a result, LSM and TI bands always coexist near the Fermi level in HgTe-class materials, as shown in the surface spectrum of HgTe in Fig.~\ref{fig2} (c). 

Given the Dirac surface state in Fig.~\ref{fig2} (c), a direct application of the Fu-Kane theory would immediately predict the existence of gapped Kitaev vortex topology in the vortex phase diagram. Such a prediction, however, is oversimplified for dropping both the HH band and the relevant LSM physics. In addition to the TI-induced Kitaev vortex, we expect the $\Gamma_8$ quartet itself will contribute to one additional nodal vortex state, as well as another Kitaev vortex state, following the analysis in Fig.~\ref{fig1}. As a result, we predict that HgTe-class material will only host a single nodal vortex instead of a Kitaev one, since 
\begin{eqnarray}
	\underbrace{\text{Kitaev vortex}\times 2}_{\text{TI} \oplus \text{LSM}} \oplus \underbrace{\text{nodal vortex}}_\text{LSM}  \equiv \underbrace{\text{nodal vortex}}_\text{HgTe}.
	\label{eq-HgTe-vortex}
\end{eqnarray}
Here, two Kitaev vortices annihilate with each other topologically due to their $\mathbb{Z}_2$ topological classification.

To verify Eq.~\eqref{eq-HgTe-vortex}, our strategy is to start with a TI-based vortex system with well-defined Fu-Kane physics, and then gradually turn on the LSM physics to explore the evolution of vortex topology. This motivates us to define a generalized six-band Kane model with a new coupling parameter $\kappa$, which serves as an effective measure of the overall coupling strength between HH bands and the remaining TI bands. In particular, we have
\begin{equation}\label{eq-kane-normal-ham}
	\mathcal{H}_\text{Kane}(\kappa, {\bf k})= \begin{pmatrix} 
		h_\text{TI}({\bf k}) & \kappa T({\bf k}) \\
		\kappa T^\dagger ({\bf k}) &h_\text{HH} ({\bf k})
	\end{pmatrix}.
\end{equation}
The TI bands are described by ${\cal H}_\text{TI} = E_+ \gamma_0 + E_- \gamma_{12} + v/\sqrt{6} (k_y \gamma_{24} - k_x \gamma_{23} + 2 k_z \gamma_{25})$. We also denote  $h_\text{HH} = E_{8}s_0$ and $E_\pm  = (E_6\pm E_8)/2$, with $E_6 = E_c + \lambda_3 k^2$ and $E_8= \lambda_1 k^2 - \lambda_2 (k_x^2+k_y^2-2k_z^2)$. Controlled by $\kappa$, the inter-band-coupling term is given by
\begin{equation} \label{eq-T-matrix-in-Kane}
	\frac{T({\bf k} ) ^{\dagger}}{\sqrt{3} \lambda_2}=  \begin{pmatrix}
		0 & -\frac{v}{\sqrt{6}\lambda_2} k_+ & -k_+^2 & 2 k_z k_+ \\
		-\frac{v}{\sqrt{6}\lambda_2} k_- & 0 & -2 k_z k_- & -k_-^2 \\
	\end{pmatrix}.
\end{equation}
Notably, the limit with $\kappa=0$ turns off all the couplings between HH bands and TI bands, which is dubbed a decoupling limit. As $\kappa$ increases, LSM physics is gradually turned on among the $\Gamma_8$ bands until it eventually reaches the isotropic limit of LSM at $\kappa=1$, which is dubbed the LSM limit. Without loss of generality, we choose the realistic parameter set of bulk HgTe~\cite{novik_prb_2005} in all our numerical simulations below. Other members in the HgTe class will have slightly different model parameters, which will only quantitatively, but not qualitatively, modify our phase diagram of the topological vortices.

The vortex topological phase diagram (VTPD) of HgTe with an isotropic $s$-wave spin-singlet pairing is mapped out as a function of $\kappa$ and the chemical potential $\mu$ in Fig.~\ref{fig2} (a). The vortex physics of ${\cal H}_\text{Kane}(\kappa, {\bf k})$ is numerically simulated in a disk geometry with the Bessel function expansion technique (see Methods). In the decoupling limit $\kappa=0$, only Kitaev vortex phase is found in the VTPD for $\mu\in[-0.69 \text{ eV},0.22 \text{ eV}]$, which exists around the energy window of the topological gap between $\Gamma_6$ and LH bands. Since the TI physics dominates at $\kappa=0$, the appearance of a Kitaev vortex agrees well with both the Fu-Kane theory and the $\pi$-Berry-phase criterion in Ref.~\cite{hosur_prl_2011}. As we increase $\kappa$ from zero, the Kitaev vortex region expands rapidly~\cite{chiu_prl_2012} and suddenly vanishes at $\kappa= 0.779$. This observation of Kitaev-vortex cancellation matches our expectation in Eq.~\eqref{eq-HgTe-vortex}.

Meanwhile, a new topological region with the nodal vortex start to emerge at $\kappa=0.314$ and continues to expand as $\kappa$ grows. Finally, in the isotropic LSM limit with $\kappa=1$ [i.e., the dashed line in Fig.~\ref{fig2} (a)], only a nodal vortex phase is found in the $\kappa$-$\mu$ VTPD for a large range of $\mu$, in an excellent agreement with our prediction in Eq.~\eqref{eq-HgTe-vortex}. Nodal vortex dispersion with $\kappa=1$ and $\mu=-0.15$ eV is shown in Fig.~\ref{fig2} (d), which clearly illustrates a pair of 1D Dirac points formed by the $J_z=\pm 1$ CdGM states. We further find this nodal vortex state indicated by  ${\cal Q}_1=-1$, confirming its topological stability. Note that ${\cal Q}_1=1$ in Fig.~\ref{fig1} is due to a different parameter choice in the LSM model, which we elaborate in the Supplementary Note 2.1.
Therefore, despite the fact that HgTe is a zero-gap TI, our calculation predicts a topological nodal phase to show up in its superconducting vortices. This deviation from existing TI-based Majorana vortex paradigms is a direct consequence of trivial-band-induced vortex topology.

\subsection{Strain-controlled Majorana engineering.} 

Given the richness of topological physics in the strain-controlled VTPDs for LSM, we are motivated to explore the physical consequence of perturbing the six-band Kane-model system in Eq.~\eqref{eq-kane-normal-ham} with similar lattice strains. An experimentally relevant in-plane strain effect is described by $\mathcal{H}_{\text{\text{str}}} = \text{diag}[0,0,\Sigma_{\text{str}},\Sigma_{\text{str}},-\Sigma_{\text{str}},-\Sigma_{\text{str}}]$~\cite{xu_prl_2017}. This coincides with the $\Sigma_{\text{str}}$ perturbation considered earlier for LSM, and we thus adopt the same notation here.

In Fig.~\ref{fig2} (e), we numerically map out the VTPD as a function of the strain parameter $\Sigma_{\text{str}}$ and $\mu$. The LSM limit $\kappa=1$ is imposed to match the realistic parameters of HgTe. Similar to the scenario of LSM,
a compressive strain with $\Sigma_{\text{str}}<0$ creates a new band inversion between LH and HH bands. This drives the $\Gamma_8$ bands into a 3D Dirac semimetal state with a pair of linear Dirac nodes, coexisting with the $\Gamma_6$-LH TI state~\cite{xu_prl_2017}. Interestingly, as shown in Fig.~\ref{fig2} (e), such a compressive strain will lead to a rapid expansion of the nodal vortex region, while no Kitaev vortex phase shows up for any value of $\mu$, similar to the zero-strain limit. Thus, a compressive strain appears to further stabilize the LSM-induced vortex topological physics, instead of spoiling it, which agrees with our LSM-based VTPD in Fig.~\ref{newfig2}.

A tensile strain with $\Sigma_{\text{str}}>0$ allows LH and HH bands to detach from each other. In this case, the HH bands behave as a set of trivial bands floating inside the topological gap formed by $\Gamma_6$ and LH bands, without touching any of them. Notably, TI surface state is now the only electron state inside the strain-induced energy gap $E_g\sim 2\Sigma_{\text{str}}$ between LHs and HHs. Inside this energy window $E_g$, we expect an emergence of Kitaev vortex as required by the Fu-Kane paradigm. Indeed, Fig.~\ref{fig2} (e) shows a fan-shaped Kitaev-vortex dome for $\Sigma_{\text{str}}>0$, exactly around $E_g$. Right below the Kitaev-vortex dome, LSM-induced nodal vortex state remains to be the dominating vortex phase. Together with the $\Sigma_{\text{str}}$-$\mu$ VTPD in the compressive region, we conclude that  the LSM-induced vortex topological physics is robust against lattice strain effect, even though the bulk LSM bands are not. 

Remarkably, the Kitaev-vortex dome shows up only after a finite positive critical strain $\Sigma_c$ [i.e., the distance between two black dashed lines in Fig.~\ref{fig2} (e)]. While Fu-Kane theory predicts a Kitaev vortex region for an arbitrarily small $\Sigma_{\text{str}}>0$, violation of the Fu-Kane theory occurs when $0<\Sigma_{\text{str}}<\Sigma_c$. We remark that this interesting discrepancy arises from the break-down of weak-pairing limit in our numerical simulation, which, however, appears as a basic assumption in the Fu-Kane theory. Specifically, the region where the Fu-Kane picture gets violated in the $\Sigma_{\text{str}}$-$\mu$ VTPD is also where both $\Sigma_{\text{str}}$ and $\mu$ are smaller than the numerical value of SC order parameter $\Delta_0=0.05$ eV in our calculation. Practically, the strong finite-size effect makes it challenging to scale the value of $\Delta_0$ down to a realistic experimental value (e.g., 1 meV) in our simulation. Therefore, it is exactly this finite-pairing effect that allows us to deviate from the Fu-Kane theory. When $\Sigma_{\text{str}}>\Delta_0$, we start to approach the weak-pairing limit and this is why the Kitaev-vortex physics begins to show up, signaling a recovery of the Fu-Kane physics.

To eliminate this finite-pairing effect and further test the limit of the Fu-Kane theory, we carry out a careful scaling analysis of $\Sigma_c$ as a function of $\Delta_0$. As shown in Fig.~\ref{fig2} (f), the scaling relation fits nicely to a simple quadratic relation that is well extrapolated to the origin with $\Sigma_c=\Delta_0=0$,
\begin{equation}
	\Sigma_c = \chi_1 \Delta_0 + \chi_2 \Delta_0^2,
	\label{eq-scaling}
\end{equation}
where $\chi_1=0.59$ and $\chi_2=1.31$ meV$^{-1}$. Physically, the scaling relation implies a monotonic shrink of the  Fu-Kane-violation region as the pairing amplitude $\Delta_0$ decreases. When the weak-pairing limit is reached at $\Delta_0 \rightarrow 0^+$, the Fu-Kane limit is fully restored with $\Sigma_c\rightarrow 0^+$. Crucially, we note that $\Delta_0$ is always small but finite in realistic superconducting systems. For example, an experimentally relevant $\Delta_0\sim 1$ meV will lead to $\Sigma_c\sim 0.6$ meV following Eq.~\eqref{eq-scaling}. This immediately leads to two important experimental consequences: 
\begin{enumerate}
	\item[(i)] The absence of Kitaev vortex in a unstrained HgTe generally holds for any small but finite $\Delta_0$;
	\item[(ii)] Vortex MZMs can be recovered via a strain control, and the critical strain trigger $\Sigma_c\sim 0.6$ meV is experimentally accessible~\cite{xu_prl_2017}.
\end{enumerate}

\subsection{Experimental signatures.}


\begin{figure*}[t]
	\centering
	\includegraphics[width=\linewidth]{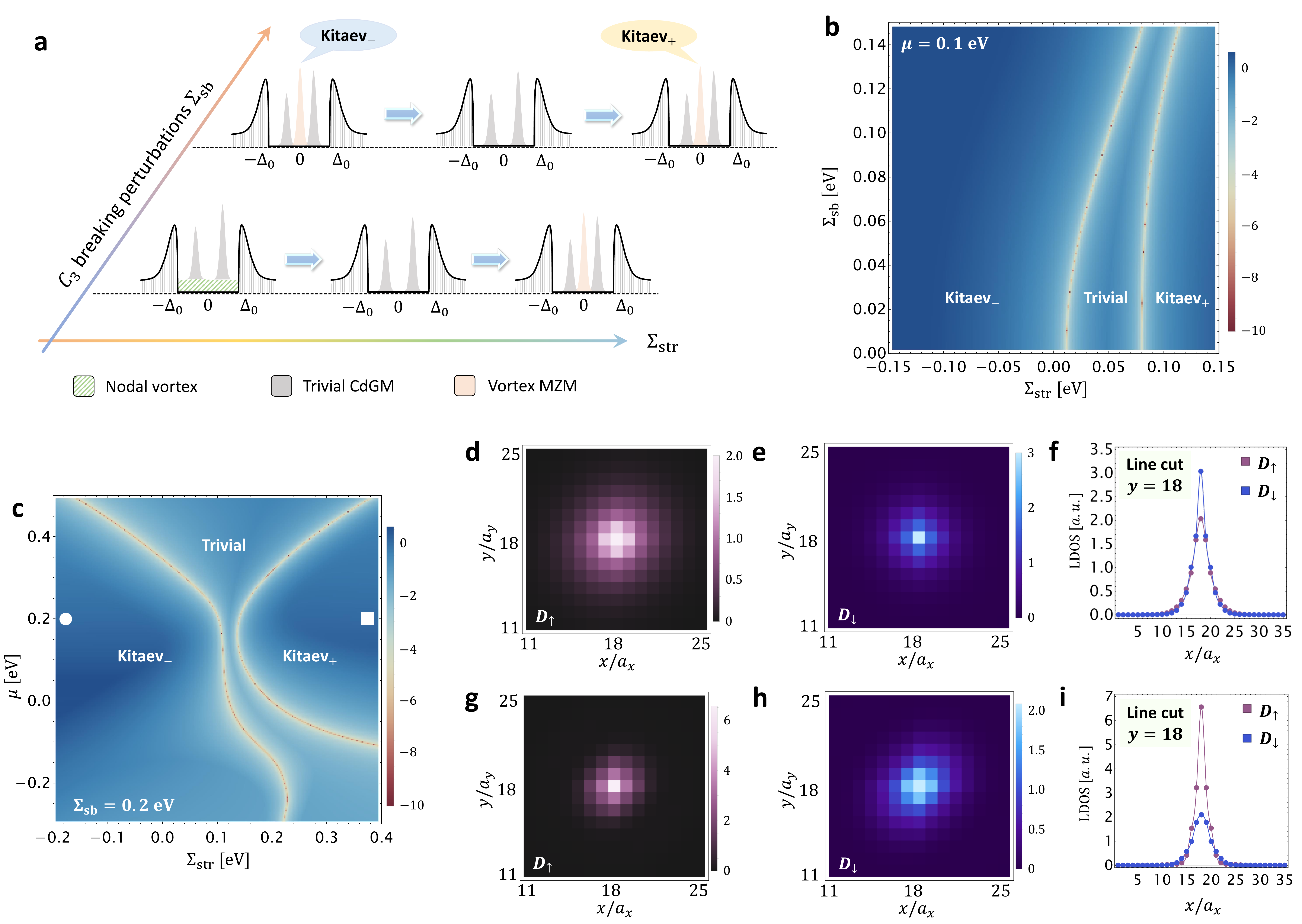}
	\caption{
		Strain-controlled Majorana engineering of HgTe-class materials. 
		(a) schematically shows the evolution of local density of state (LDOS) at the vortex core center as a function of bias voltage by tuning both the in-plane lattice strain strength $\Sigma_{\text{str}}$ and the $C_3$-symmetry breaking perturbation $\Sigma_{\text{sb}}$. The Kitaev-trivial-Kitaev transition with vortex Majorana zero mode (MZM) of a Kitaev vortex in (a) is numerically verified by both mapping the $\Sigma_{\text{str}}$-$\Sigma_{\text{sb}}$ VTPD in (b) at a fixed $\mu=0.1$ eV, and the $\Sigma_{\text{str}}$-$\mu$ VTPD in (c) at a fixed $\Sigma_{\text{sb}}=0.2$ eV. 
		The colors in (b) and (c) represent the logarithmic value of the vortex energy gap at $k_z=0$. The color map plots of the spatial spin-resolved surface LDOS (a.u.=arbitrary units) at a zero-bias voltage are numerically calculated for the Kitaev$_-$ vortex in (d) - (f) and the Kitaev$_+$ vortex in (g) - (i), respectively. These two topologically equivalent Kitaev vortex phases can be clearly distinguished by their distinct zero-bias spin textures as shown in (f) with $D_\uparrow < D_\downarrow$ and (i) with $D_\uparrow > D_\downarrow$ at the vortex core center ${\bf r}_c=(18,18)$ in unit of in-plane lattice constants $a_x$ and $a_y$. 
	}
	\label{fig3}
\end{figure*}


The $\Sigma_{\text{str}}$-$\mu$ VTPD in Fig.~\ref{fig2} (e) sheds light on the detection and manipulation of vortex MZMs. By continuously tuning the strain from a compressive type to a tensile type, the vortex of an electron-doped HgTe (e.g., $\mu\sim 0.1$ eV) will undergo a series of vortex topological phase transitions, from Majorana-free nodal and trivial vortices to a Majorana-carrying Kitaev vortex. 
Consequently, probing the local density of state (LDOS) at the surface vortex core with a scanning tunneling microscope (STM) will reveal a single transition at $\Sigma_c$, after which a zero-bias peak (ZBP) emerges in the tunneling spectrum, as schematically shown in the bottom panel of Fig.~\ref{fig3} (a).

While a nodal vortex does not carry MZMs, breaking the around-axis rotation symmetry spoils the vortex nodal structure and further leads to a Kitaev vortex~\cite{hu_arxiv_2021}. 
Such a symmetry breaking effect can be feasibly generated by tilting the applied magnetic field ${\bf B}$, or applying an in-plane lattice strain $\Sigma_{\text{sb}}$ following ${\cal H}_{\text{LSM}}'$ defined for LSM [i.e.,~replacing $\sqrt{3}\lambda_2 k_xk_y$ with $\sqrt{3}\lambda_2 k_xk_y + \Sigma_{\text{sb}}$ in Eq.~\eqref{eq-T-matrix-in-Kane}].  We note that most HgTe-class materials respect either a space group $F\bar{4}3m$ (No.~216) or $Fd3m$ (No.~227), the highest-fold rotation symmetry of which is $C_3$ along (111) direction. Perturbing HgTe-class systems with $\Sigma_{\text{sb}}$ will directly break $C_3$ down to $C_1$, which admits a single $\mathbb{Z}_2$ index $\nu_0$. This is crucially different from the fully rotational symmetric LSM considered in the previous sections where $\Sigma_{\text{sb}}: C_\infty \mapsto C_2$. Following our notation in Fig.~\ref{newfig2}, we still denote the nodal-origined Kitaev vortex as Kitaev$_-$ and the preexisting Kitaev vortex as Kitaev$_+$ for convenience. However, one should keep in mind that the Kitaev$_{\pm}$ vortex phases here are topologically indistinguishable due to the lack of $C_2$ symmetry.

By tuning $\Sigma_{\text{str}}$, we expect a Kitaev-trivial-Kitaev transition for a finite $\Sigma_{\text{sb}}$. As schematically shown in the top panel of Fig.~\ref{fig3} (a), a MZM-induced ZBP from the Kitaev$_{-}$ vortex will first vanish in the LDOS after entering the trivial phase, and will eventually reappear when the Kitaev$_{+}$ vortex is turned on. This transition for a fixed $\mu=0.1$ eV is explicitly verified by numerically mapping out the VTPD as a function of $\Sigma_{\text{sb}}$ and $\Sigma_{\text{str}}$, which we summarize in Fig.~\ref{fig3}~(b). Here, we have regularized the strained HgTe model on a $50\times 50$ 2D square lattice, while keeping $k_z$ a good quantum number. $\Delta_0=0.1$ eV is applied to eliminate any possible finite size effect. We further numerically explore the VTPD for a fixed $\Sigma_{\text{sb}}=0.2$ eV by varying both $\mu$ and $\Sigma_{\text{str}}$ and have observed the same Kitaev-trivial-Kitaev transition, as shown in Fig.~\ref{fig3}~(c). 

Finally, we wonder if the Kitaev$_\pm$ phases in HgTe, despite their topological equivalence, could be locallly distinguished from each other through surface LDOS measurements. Using the recursive Green's function method, we numerically calculate the spatial spin-resolved surface LDOS $D_{\uparrow}({\bf r}_\parallel)$ and $D_{\downarrow}({\bf r}_\parallel)$ at a zero-bias voltage for the strained HgTe model in a semi-infinite geometry along the $\hat{z}$ direction. Open boundary conditions are imposed for both in-plane directions with $N_x=N_y=35$ and we have chosen $\Sigma_{\text{sb}}=0.2$ eV, $\mu=0.2$ eV and $\Delta_0=0.2$ eV for all calculations to eliminate the in-plane finite size effect. Here, ${\bf r}_\parallel=(x,y)$ and the vortex core center locates at ${\bf r}_c=(18,18)$ in unit of in-plane lattice constant $a_x=a_y=6.46$ \AA. The spin-resolved LDOS plots for a representative Kitaev$_-$ vortex phase [the white dot in Fig.~\ref{fig3} (c)] are shown in Fig.~\ref{fig3} (d) - (f). In particular, $D_{\downarrow}({\bf r}_\parallel)$ shows a greater ZBP than that of $D_{\uparrow}({\bf r}_\parallel)$ at ${\bf r}_c$. 
In contrast, the zero-bias spin texture for the Kitaev$_+$ vortex phase [the white square in Fig.~\ref{fig3}~(c)] is exactly opposite, where the ZBP of $D_{\uparrow}({\bf r}_\parallel)$ is significantly higher than $D_{\downarrow}({\bf r}_\parallel)$ at ${\bf r}_c$. Therefore, a state-of-the-art spin-polarized STM should be capable of extracting the distinct spin patterns for the Kitaev$_{\pm}$ phases in HgTe-class materials. We furthe note that the spin pattern for Kitaev$_-$ phase here is consistent with that of the Kitaev$_-$ vortex phase of LSM [see Fig.~3 of the Supplementary Note 2.3], agreeing with the fact that the Kitaev$_{-}$ phase of the Kane model arises from the overall trivial LSM-dominant bands. Observing the above wavefunction information, together with strain-induced ZBP transitions, will provde a rather compelling experimental evidence for the Majorana nature of these topological vortices.

\section{Discussion} 

We have demonstrated the possibility of topological nontrivial superconducting vortices based on a set of topology-free electronic bands. On the material side, we have established HgTe-class materials as an unprecedented playground to study trivial-band-induced vortex topology. We notice that intrinsic or proximity-induced superconductivity has already been observed in several members of this material family, including HgTe/Nb heterostructure~\cite{maier2012induced}, $\alpha$-Sn/PbTe heterostructure~\cite{Liao_np_2018,falson2020type}, and half-Heusler alloys such as LaPtBi~\cite{Goll_pbcm_2008}, YPtBi~\cite{butch_prb_2011}, and RPdBi with $R=$ Lu, Tm, Er, Ho~\cite{Nakajima_sa_2015}. Our theory will serve as an important guidance to detect, control, and engineer Majorana modes in these candidate superconducting systems.

Our results further suggest several new guidelines for the ongoing vortex-based Majorana search. First of all, we note that most topological-band-based SC candidates have coexisting trivial bands near the Fermi level, while most literatures choose to drop the trivial bands to simplify the vortex topology analysis. Our finding, however, suggests that trivial bands in a topological-band SC should have also been in the spotlight, without which the Majorana interpretation of the material could be fallacious. Second, we should not limit the Majorana-oriented material search to intrinsic TSCs or topological-band SCs, since Majorana vortices can exist in certain types of bulk-topology irrelevant SCs as well. We hope that our work will motivate more theoretical and experimental research efforts under the spirit of Majorana from trivial bands and further initiates a new journey of the Majorana research in this large uncharted territory, the trivial superconductors. 

\section{Methods}

\subsection{Bessel Function Expansion.}
The Bessel function expansion technique enables the calculation of vortex energy spectrum for continuum models, which we will describe below. In a rotation-symmetric disk or cylinder geometry, a BdG Hamiltonian $\mathcal{H}_{\text{BdG}}$ is characterized by two good quantum numbers, $z$-directional crystal momentum $k_z$ and $z$-component total angular momentum $J_z$. In particular, the angular momentum operator is 
\begin{align}
	\hat{J}_z = (-i \partial_\theta ) {I}_{2N_h\times 2N_h} + J_{\text{basis}} + J_{\text{vortex}},
\end{align}
where $I_{2N_h\times2N_h}$ is the $2N_h$-by-$2N_h$ identity matrix with $N_h$ the dimension of the normal-state Hamiltonian and $(r,\theta)$ denote the in-plane polar coordinates. For the 4-band LSM ($N_h=4$), we have
\begin{align}
	J_{\text{basis}} = \text{diag}[\tfrac{3}{2}, -\tfrac{1}{2}, \tfrac{3}{2}, -\tfrac{1}{2}, \tfrac{1}{2}, -\tfrac{3}{2}, \tfrac{1}{2}, -\tfrac{3}{2}].
\end{align} 
Here, $J_\text{vortex}$ arises from the vortex phase winding,
\begin{align}
	J_{\text{vortex}}=\text{diag}[-\tfrac{1}{2}, \tfrac{1}{2}, \tfrac{1}{2}, -\tfrac{1}{2}, -\tfrac{1}{2}, \tfrac{1}{2}, \tfrac{1}{2}, -\tfrac{1}{2}].
\end{align} 
Clearly, $[\hat{J}_z, \mathcal{H}_{\text{BdG}} ]=0$, and the BdG Hamiltonian matrix can be decomposed into $J_z$-labeled matrix blocks,
\begin{align}\label{sm-eq-bdg-reduce-2d}
	\mathcal{H}_{\text{BdG}} &= \sum_{J_z}\oplus H_{J_z}(r,\theta).
\end{align}
As a result, we only need to solve $H_{J_z}(r,\theta) \vert \Phi(J_z,r,\theta)\rangle = E \vert \Phi(J_z,r,\theta)\rangle$, where
a general energy eigenstate is $J_z$ labeled and further takes the following form,
\begin{align}\label{em-eq-ansatz}
\begin{split}
	&\vert \Phi(J_z,r,\theta)\rangle =  e^{i(J_z-1)\theta} [ u_1(J_z-1,r), u_2(J_z,r)e^{i\theta},   \\
	&u_3(J_z-2,r)e^{-i\theta}, u_4(J_z+1,r)e^{2i\theta}, v_1(J_z,r) e^{i\theta},  \\
	& v_2(J_z+1,r)e^{2i\theta},  v_3(J_z-1,r) , v_4(J_z+2,r)e^{3i\theta} ]^T, 
\end{split}
\end{align}
where both $u_{i}(n,r)$ and $v_{i}(n,r)$ with $i=1,2,3,4$ yield the following expansions,
\begin{subequations}
\begin{align}\label{sm-eq-besselj-expansion-uv}
		u(J_z,r) = \sum_{j=1}^{N} c_{j,J_z} \phi(J_z,r,\alpha_j), \\
		v(J_z,r) = \sum_{j=1}^{N} c_{j,J_z}^\prime \phi(J_z,r,\alpha_j). 
\end{align}
\end{subequations}
Here, the normalized Bessel function is defined as 
\begin{equation}
	\phi(J_z,r,\alpha_i)=\frac{\sqrt{2}}{R} \mathcal{J}_{J_z}(\alpha_i r/R)/\mathcal{J}_{J_z+1}(\alpha_i),
\end{equation}
where $\mathcal{J}_n$ is the Bessel function of the first kind. \(\alpha_i\) and $R$ denote the \(i^{\text{th}}\) zero of \(\mathcal{J}_{J_z}(r)\) and the radius of the disk, respectively. $c$ and $c^\prime$ are expansion coefficients that are yet to be numerically calculated. We further note that in the polar coordinate system, the crystal momenta
$k_\pm = k_x \pm i k_y$ become
\begin{subequations}
\begin{align}
 k_{+} &=e^{i\theta}\left\lbrack -i\frac{\partial}{\partial r}+\frac{1}{r}\frac{\partial}{\partial \theta} \right\rbrack,  \\
 k_{-} &=e^{-i\theta}\left\lbrack -i\frac{\partial}{\partial r}-\frac{1}{r}\frac{\partial}{\partial\theta} \right\rbrack,
\end{align}
\end{subequations}
which satisfy
\begin{subequations}
\begin{align}
 k_{+}\left(e^{in\theta}\mathcal{J}_n(\alpha r)\right)  &= i\alpha e^{i(n+1)\theta}\mathcal{J}_{n+1}(\alpha r), \\
 k_{-}\left(e^{in\theta}\mathcal{J}_n(\alpha r)\right)  &= -i\alpha e^{i(n-1)\theta}\mathcal{J}_{n-1}(\alpha r).
\end{align}
\end{subequations}
It is also easy to show that
\begin{equation}
	(k_x^2+k_y^2)[e^{in\theta}\mathcal{J}_n(\alpha r)]  = \alpha^2 [e^{in\theta}\mathcal{J}_n(\alpha r)].
\end{equation}

The energy eigen-equation is now essentially a set of 1D radial equations for fixed $k_z$ and $J_z$. In addition, the disk geometry with hard-wall boundary conditions requires $\vert \Phi(J_z,r,\theta)\rangle$ to satisfy $u_i(r=R)=v_i(r=R)=0$.
Notably, a Bessel functions with a large $\alpha_i$ will oscillate rapidly and we expect it to contribute little to the low-energy vortex bound states.
Therefore, for a reasonaly large $N\in\mathbb{Z}_{>0}$, we can truncate the zeros of the Bessel functions at $\alpha_N$, making the dimension of each decoupled Hilbert subspace to be $8N$. Physically, this truncation can be interpreted as a Debye frequency cutoff around the Fermi energy. Solving these radial equations leads us to the vortex-bound states and their energy relations for a general vortex problem.

The vortex simulation of LSM model in the continuum limit is performed using the above Bessel function expansion technique with $R_{\text{disk}}=250$. We further truncate the zeros of Bessel function at $N=250$ and numerically confirm the validity of this truncation. As discussed in the Supplementary Note 4, the continuum model approach agrees quantitatively with the discrete tight-binding model approach.

As for the 6-band Kane model ($N_h=6$), a general vortex wavefunction that respects the rotation symmetry is given by
\begin{align}
\begin{split}
		& \vert \Phi_{\text{Kane}}(J_z,r,\theta)\rangle =  e^{iJ_z\theta} [u_1(J_z,r), u_2(J_z+1,r)e^{i\theta},   \\
		& u_3(J_z,r), u_4(J_z+1,r)e^{i\theta},  u_5(J_z+2,r)e^{2i\theta},  \\
		& u_6(J_z-1,r)e^{-i\theta},  v_1(J_z,r),  v_2(J_z-1,r)e^{-i\theta},  \\
		& v_3(J_z,r) , v_4(J_z-1,r)e^{-i\theta}, v_5(J_z-2,r)e^{-2i\theta} ,\\
		& v_6(J_z+1,r)e^{i\theta} ]^T,  
\end{split}
\end{align}
where the components $u_{i}(J_z,r)$ and $v_{i}(J_z,r)$ with $i=1,2,...,6$ can be both expanded by the normalized Bessel functions, as we discussed earlier. To eliminate the finite size effect that is induced by a small $\Delta_0$, we consider a large disk radius of $R_{\text{disk}}=2100$ in unit of the in-plane lattice constant. The truncation of the zeros of the Bessel function is $N=385$ and the dimension of Hilbert space in our simulation is $12N = 4620$.

We finally remark on the particle-hole symmetry $\Xi$ of $\vert \Phi(J_z,r,\theta)\rangle$. Starting from an eigenstate at $k_z=0$ with $H_{J_z} \vert \Phi(J_z,r,\theta)\rangle= E_{J_z} \vert \Phi(J_z,r,\theta)\rangle$, we have 
\begin{subequations}
\begin{align}
 \vert \Phi'(-J_z,r,\theta)\rangle &= \Xi \vert \Phi(J_z,r,\theta) \rangle,  \\
 H_{J_z} \vert \Phi'(-J_z,r,\theta) \rangle &= -E_{J_z} \vert \Phi'(-J_z,r,\theta) \rangle.
\end{align}
\end{subequations}
Since our continuum models with isotropic $s$-wave spin-singlet pairings feature a full rotation symmetry, the $J_z=0$ subspace $H_{J_z=0}$ is the only sector that respects particle-hole symmetry, while a $J_z\neq 0$ subspace is related to the $-J_z$ one via particle-hole symmetry.

\subsection{Chiral Winding Number and Vortex Zero Modes.}

We discuss the winding number argument to understand the existence of vortex zero modes of LSM in Fig.~\ref{fig1}~(c). As shown in the Supplementary Note 2.2, it is suggestive to separate Eq.~\eqref{eq-bdg-ham-lsm} into a direct sum of two matrix blocks $H_0=h_\Delta ({\bf k}_\parallel,\theta)\oplus h_{-\Delta} ({\bf k}_\parallel,\theta)$ and a perturbation part $H_1({\bf k}_\parallel, k_z)$. In particular, 
\begin{align}\label{sm-eq-h0-4by4}
	h_\Delta({\bf k}_\parallel,\theta) &= \Delta(\cos\theta \tau_x\sigma_0 - \sin\theta \tau_y\sigma_z) \nonumber \\
	 &+ \tilde{v} \left\lbrack -(k_x^2-k_y^2)\tau_y\sigma_y  + 2k_xk_y \tau_y \sigma_x \right\rbrack.
\end{align}
It is easy to check that $h_\Delta({\bf k}_\parallel,\theta)$ respects an emergent chiral symmetry
\begin{align}
	\mathcal{S} = \tau_z \sigma_0,
\end{align}
which is independent of the sign of $\Delta$. A stable vortex zero mode is necessarily an eigenstate of ${\cal S}$ and carries a ${\cal S}=\pm 1$ label. Only zero modes that are differently ${\cal S}$-labeled can interact with each other and get hybridized, while those carrying the same label cannot get coupled. 

Now $h_\Delta({\bf k}_\parallel,\theta)$ manifests as an effective 3D Hamiltonian in the symmetry class AIII, whose topological behavior is characterized by a chiral winding number ${\cal N_S}\in\mathbb{Z}$~\cite{teo2010topodefect}. Physically, we have
\begin{equation}
	{\cal N_S} = {\cal N}_{+1} - {\cal N}_{-1}.
\end{equation}
Here ${\cal N}_{\pm 1}$ denotes the number of vortex zero modes that carry ${\cal S}=\pm 1$. Evaluation of ${\cal N_S}$ can be achieved by noting that $h_\Delta({\bf k}_\parallel,\theta)$ yields an off-block-diagonal form, as a result of the chiral symmetry,
\begin{equation}
	h_\Delta({\bf k}_\parallel,\theta) = \begin{pmatrix}
		0 & Q({\bf k}_{\parallel}, \theta) \\
		Q^{\dagger}({\bf k}_{\parallel}, \theta) & 0 \\
	\end{pmatrix},
\end{equation}
with 
\begin{equation}
	Q({\bf k}_{\parallel}, \theta) = \begin{pmatrix}
		\Delta e^{i\theta} & \tilde{v} k_-^2 \\
		-\tilde{v} k_+^2 & \Delta e^{-i\theta} \\
	\end{pmatrix}.
	\label{eq-chiral-Q}
\end{equation}
Then the chiral winding number can be written as 
\begin{equation}
	{\cal N_S} = -\frac{1}{24\pi^2}\int d^2 {\bf k} d\theta \epsilon^{\mu\nu\rho}\text{Tr}[(Q\partial_\mu Q^{\dagger})(Q\partial_\nu Q^{\dagger})(Q\partial_\rho Q^{\dagger})],
	\label{eq-chiral-winding}
\end{equation} 
where $\mu,\nu,\rho \in \{k_x, k_y, \theta\}$ and $\epsilon^{\mu\nu\rho}$ is the Levi-Civita tensor. Applying Eq.~\eqref{eq-chiral-winding} to Eq.~\eqref{eq-chiral-Q}, we arrive at
\begin{align}
	{\cal N_S} &= -\frac{1}{24\pi^2} \int_{0}^{2\pi} d\varphi \int_{0}^{2\pi} d\theta \int_0^\infty \frac{48\tilde{v}^2\Delta^2 k^2}{(\tilde{v}^2k^4-\Delta^2)^2}kdk \nonumber \\
	&= -\frac{1}{24\pi^2} (2\pi)^2 (-12) 
	= 2.
\end{align} 
Similarly, ${\cal N_S}=2$ also holds for the other $4\times 4$ block $h_{-\Delta}$ since the value of ${\cal N_S}$ is independent of the sign of $\Delta$. As a result, the net chiral winding number for $H_0$ is
\begin{equation}
	{\cal N_S}^{(\text{net})} = 4,
\end{equation}
indicating four robust zero-energy vortex bound states with ${\cal S}=+1$. Projecting $H_1({\bf k}_\parallel, k_z)$ onto the zero-mode basis will lead us to a perturbative understanding of the nontrivial vortex topology in superconducting LSM systems, as illustrated in Fig.~\ref{fig1}. The zero modes further serve as the basis for building an analytical perturbation theory for the vortex-line Hamiltonian of LSM, as shown in the Supplementary Note 3.

\vspace*{6mm}
\noindent{\bf Data availability} \\
The datasets generated during this study are available from the corresponding author on reasonable request.

\vspace*{6mm}
\noindent{\bf Code availability} \\
The custom codes generated during this study are available from the corresponding author on reasonable request.

\vspace*{6mm}
\noindent{\bf Acknowledgements} \\
We are grateful to L.-Y.~Kong, X.-Q.~Sun, J.~Yu, J.-S. Lee, B.~Seradjeh, P.~Ghaemi, T.~Hughes, and C.~Batista for stimulating discussions. We are particularly indebtful to J.-D.~Sau for his valuable comments and insight that motivate us to study the scaling relation of superconducting order parameter. This work is supported by a start-up fund at the University of Tennessee.

\vspace*{6mm}
\noindent{\bf Author contributions} \\
Both authors contributed essentially to the formulation and theoretical analysis of the problem and to writing the manuscript. L.-H. H performed numerical calculations with the help of R.-X. Z.

\vspace*{6mm}
\noindent{\bf Competing interests} \\
The authors declare no competing interests.


\begin{thebibliography}{49}%
	\makeatletter
	\providecommand \@ifxundefined [1]{%
		\@ifx{#1\undefined}
	}%
	\providecommand \@ifnum [1]{%
		\ifnum #1\expandafter \@firstoftwo
		\else \expandafter \@secondoftwo
		\fi
	}%
	\providecommand \@ifx [1]{%
		\ifx #1\expandafter \@firstoftwo
		\else \expandafter \@secondoftwo
		\fi
	}%
	\providecommand \natexlab [1]{#1}%
	\providecommand \enquote  [1]{``#1''}%
	\providecommand \bibnamefont  [1]{#1}%
	\providecommand \bibfnamefont [1]{#1}%
	\providecommand \citenamefont [1]{#1}%
	\providecommand \href@noop [0]{\@secondoftwo}%
	\providecommand \href [0]{\begingroup \@sanitize@url \@href}%
	\providecommand \@href[1]{\@@startlink{#1}\@@href}%
	\providecommand \@@href[1]{\endgroup#1\@@endlink}%
	\providecommand \@sanitize@url [0]{\catcode `\\12\catcode `\$12\catcode
		`\&12\catcode `\#12\catcode `\^12\catcode `\_12\catcode `\%12\relax}%
	\providecommand \@@startlink[1]{}%
	\providecommand \@@endlink[0]{}%
	\providecommand \url  [0]{\begingroup\@sanitize@url \@url }%
	\providecommand \@url [1]{\endgroup\@href {#1}{\urlprefix }}%
	\providecommand \urlprefix  [0]{URL }%
	\providecommand \Eprint [0]{\href }%
	\providecommand \doibase [0]{https://doi.org/}%
	\providecommand \selectlanguage [0]{\@gobble}%
	\providecommand \bibinfo  [0]{\@secondoftwo}%
	\providecommand \bibfield  [0]{\@secondoftwo}%
	\providecommand \translation [1]{[#1]}%
	\providecommand \BibitemOpen [0]{}%
	\providecommand \bibitemStop [0]{}%
	\providecommand \bibitemNoStop [0]{.\EOS\space}%
	\providecommand \EOS [0]{\spacefactor3000\relax}%
	\providecommand \BibitemShut  [1]{\csname bibitem#1\endcsname}%
	\let\auto@bib@innerbib\@empty
	\bibitem [{\citenamefont {Kitaev}(2003)}]{Kitaev_aop_2003}%
	\BibitemOpen
	\bibfield  {author} {\bibinfo {author} {\bibfnamefont {A.}~\bibnamefont
			{Kitaev}},\ }\bibfield  {title} {\bibinfo {title} {Fault-tolerant quantum
			computation by anyons},\ }\href
	{https://doi.org/10.1016/s0003-4916(02)00018-0} {\bibfield  {journal}
		{\bibinfo  {journal} {Annals of Physics}\ }\textbf {\bibinfo {volume}
			{303}},\ \bibinfo {pages} {2} (\bibinfo {year} {2003})}\BibitemShut {NoStop}%
	\bibitem [{\citenamefont {Nayak}\ \emph {et~al.}(2008)\citenamefont {Nayak},
		\citenamefont {Simon}, \citenamefont {Stern}, \citenamefont {Freedman},\ and\
		\citenamefont {Sarma}}]{nayak_rmp_2008}%
	\BibitemOpen
	\bibfield  {author} {\bibinfo {author} {\bibfnamefont {C.}~\bibnamefont
			{Nayak}}, \bibinfo {author} {\bibfnamefont {S.~H.}\ \bibnamefont {Simon}},
		\bibinfo {author} {\bibfnamefont {A.}~\bibnamefont {Stern}}, \bibinfo
		{author} {\bibfnamefont {M.}~\bibnamefont {Freedman}},\ and\ \bibinfo
		{author} {\bibfnamefont {S.~D.}\ \bibnamefont {Sarma}},\ }\bibfield  {title}
	{\bibinfo {title} {Non-abelian anyons and topological quantum computation},\
	}\href {https://doi.org/10.1103/revmodphys.80.1083} {\bibfield  {journal}
		{\bibinfo  {journal} {Reviews of Modern Physics}\ }\textbf {\bibinfo {volume}
			{80}},\ \bibinfo {pages} {1083} (\bibinfo {year} {2008})}\BibitemShut
	{NoStop}%
	\bibitem [{\citenamefont {Kitaev}(2001)}]{Kitaev_pu_2001}%
	\BibitemOpen
	\bibfield  {author} {\bibinfo {author} {\bibfnamefont {A.~Y.}\ \bibnamefont
			{Kitaev}},\ }\bibfield  {title} {\bibinfo {title} {Unpaired majorana fermions
			in quantum wires},\ }\href {https://doi.org/10.1070/1063-7869/44/10s/s29}
	{\bibfield  {journal} {\bibinfo  {journal} {Physics-Uspekhi}\ }\textbf
		{\bibinfo {volume} {44}},\ \bibinfo {pages} {131} (\bibinfo {year}
		{2001})}\BibitemShut {NoStop}%
	\bibitem [{\citenamefont {Read}\ and\ \citenamefont
		{Green}(2000)}]{read2000paired}%
	\BibitemOpen
	\bibfield  {author} {\bibinfo {author} {\bibfnamefont {N.}~\bibnamefont
			{Read}}\ and\ \bibinfo {author} {\bibfnamefont {D.}~\bibnamefont {Green}},\
	}\bibfield  {title} {\bibinfo {title} {Paired states of fermions in two
			dimensions with breaking of parity and time-reversal symmetries and the
			fractional quantum hall effect},\ }\href
	{https://link.aps.org/doi/10.1103/PhysRevB.61.10267} {\bibfield  {journal}
		{\bibinfo  {journal} {Phys. Rev. B}\ }\textbf {\bibinfo {volume} {61}},\
		\bibinfo {pages} {10267} (\bibinfo {year} {2000})}\BibitemShut {NoStop}%
	\bibitem [{\citenamefont {Lutchyn}\ \emph {et~al.}(2010)\citenamefont
		{Lutchyn}, \citenamefont {Sau},\ and\ \citenamefont
		{Das~Sarma}}]{lutchyn_prl_2010}%
	\BibitemOpen
	\bibfield  {author} {\bibinfo {author} {\bibfnamefont {R.~M.}\ \bibnamefont
			{Lutchyn}}, \bibinfo {author} {\bibfnamefont {J.~D.}\ \bibnamefont {Sau}},\
		and\ \bibinfo {author} {\bibfnamefont {S.}~\bibnamefont {Das~Sarma}},\
	}\bibfield  {title} {\bibinfo {title} {Majorana fermions and a topological
			phase transition in semiconductor-superconductor heterostructures},\ }\href
	{https://link.aps.org/doi/10.1103/PhysRevLett.105.077001} {\bibfield
		{journal} {\bibinfo  {journal} {Phys. Rev. Lett.}\ }\textbf {\bibinfo
			{volume} {105}},\ \bibinfo {pages} {077001} (\bibinfo {year}
		{2010})}\BibitemShut {NoStop}%
	\bibitem [{\citenamefont {Sau}\ \emph {et~al.}(2010)\citenamefont {Sau},
		\citenamefont {Lutchyn}, \citenamefont {Tewari},\ and\ \citenamefont
		{Das~Sarma}}]{sau_prl_2010}%
	\BibitemOpen
	\bibfield  {author} {\bibinfo {author} {\bibfnamefont {J.~D.}\ \bibnamefont
			{Sau}}, \bibinfo {author} {\bibfnamefont {R.~M.}\ \bibnamefont {Lutchyn}},
		\bibinfo {author} {\bibfnamefont {S.}~\bibnamefont {Tewari}},\ and\ \bibinfo
		{author} {\bibfnamefont {S.}~\bibnamefont {Das~Sarma}},\ }\bibfield  {title}
	{\bibinfo {title} {Generic new platform for topological quantum computation
			using semiconductor heterostructures},\ }\href
	{https://link.aps.org/doi/10.1103/PhysRevLett.104.040502} {\bibfield
		{journal} {\bibinfo  {journal} {Phys. Rev. Lett.}\ }\textbf {\bibinfo
			{volume} {104}},\ \bibinfo {pages} {040502} (\bibinfo {year}
		{2010})}\BibitemShut {NoStop}%
	\bibitem [{\citenamefont {Mourik}\ \emph {et~al.}(2012)\citenamefont {Mourik},
		\citenamefont {Zuo}, \citenamefont {Frolov}, \citenamefont {Plissard},
		\citenamefont {Bakkers},\ and\ \citenamefont
		{Kouwenhoven}}]{Mourik_science_2012}%
	\BibitemOpen
	\bibfield  {author} {\bibinfo {author} {\bibfnamefont {V.}~\bibnamefont
			{Mourik}}, \bibinfo {author} {\bibfnamefont {K.}~\bibnamefont {Zuo}},
		\bibinfo {author} {\bibfnamefont {S.~M.}\ \bibnamefont {Frolov}}, \bibinfo
		{author} {\bibfnamefont {S.~R.}\ \bibnamefont {Plissard}}, \bibinfo {author}
		{\bibfnamefont {E.~P. A.~M.}\ \bibnamefont {Bakkers}},\ and\ \bibinfo
		{author} {\bibfnamefont {L.~P.}\ \bibnamefont {Kouwenhoven}},\ }\bibfield
	{title} {\bibinfo {title} {Signatures of majorana fermions in hybrid
			superconductor-semiconductor nanowire devices},\ }\href
	{https://doi.org/10.1126/science.1222360} {\bibfield  {journal} {\bibinfo
			{journal} {Science}\ }\textbf {\bibinfo {volume} {336}},\ \bibinfo {pages}
		{1003} (\bibinfo {year} {2012})}\BibitemShut {NoStop}%
	\bibitem [{\citenamefont {Kezilebieke}\ \emph {et~al.}(2020)\citenamefont
		{Kezilebieke}, \citenamefont {Huda}, \citenamefont {Va{\v{n}}o},
		\citenamefont {Aapro}, \citenamefont {Ganguli}, \citenamefont {Silveira},
		\citenamefont {G{\l}odzik}, \citenamefont {Foster}, \citenamefont {Ojanen},\
		and\ \citenamefont {Liljeroth}}]{kezilebieke2020topological}%
	\BibitemOpen
	\bibfield  {author} {\bibinfo {author} {\bibfnamefont {S.}~\bibnamefont
			{Kezilebieke}}, \bibinfo {author} {\bibfnamefont {M.~N.}\ \bibnamefont
			{Huda}}, \bibinfo {author} {\bibfnamefont {V.}~\bibnamefont {Va{\v{n}}o}},
		\bibinfo {author} {\bibfnamefont {M.}~\bibnamefont {Aapro}}, \bibinfo
		{author} {\bibfnamefont {S.~C.}\ \bibnamefont {Ganguli}}, \bibinfo {author}
		{\bibfnamefont {O.~J.}\ \bibnamefont {Silveira}}, \bibinfo {author}
		{\bibfnamefont {S.}~\bibnamefont {G{\l}odzik}}, \bibinfo {author}
		{\bibfnamefont {A.~S.}\ \bibnamefont {Foster}}, \bibinfo {author}
		{\bibfnamefont {T.}~\bibnamefont {Ojanen}},\ and\ \bibinfo {author}
		{\bibfnamefont {P.}~\bibnamefont {Liljeroth}},\ }\bibfield  {title} {\bibinfo
		{title} {Topological superconductivity in a van der waals heterostructure},\
	}\href {https://www.nature.com/articles/s41586-020-2989-y} {\bibfield
		{journal} {\bibinfo  {journal} {Nature}\ }\textbf {\bibinfo {volume} {588}},\
		\bibinfo {pages} {424} (\bibinfo {year} {2020})}\BibitemShut {NoStop}%
	\bibitem [{\citenamefont {Fu}\ and\ \citenamefont {Kane}(2008)}]{fu_prl_2008}%
	\BibitemOpen
	\bibfield  {author} {\bibinfo {author} {\bibfnamefont {L.}~\bibnamefont
			{Fu}}\ and\ \bibinfo {author} {\bibfnamefont {C.~L.}\ \bibnamefont {Kane}},\
	}\bibfield  {title} {\bibinfo {title} {Superconducting proximity effect and
			majorana fermions at the surface of a topological insulator},\ }\href
	{https://link.aps.org/doi/10.1103/PhysRevLett.100.096407} {\bibfield
		{journal} {\bibinfo  {journal} {Phys. Rev. Lett.}\ }\textbf {\bibinfo
			{volume} {100}},\ \bibinfo {pages} {096407} (\bibinfo {year}
		{2008})}\BibitemShut {NoStop}%
	\bibitem [{\citenamefont {Hosur}\ \emph {et~al.}(2011)\citenamefont {Hosur},
		\citenamefont {Ghaemi}, \citenamefont {Mong},\ and\ \citenamefont
		{Vishwanath}}]{hosur_prl_2011}%
	\BibitemOpen
	\bibfield  {author} {\bibinfo {author} {\bibfnamefont {P.}~\bibnamefont
			{Hosur}}, \bibinfo {author} {\bibfnamefont {P.}~\bibnamefont {Ghaemi}},
		\bibinfo {author} {\bibfnamefont {R.~S.~K.}\ \bibnamefont {Mong}},\ and\
		\bibinfo {author} {\bibfnamefont {A.}~\bibnamefont {Vishwanath}},\ }\bibfield
	{title} {\bibinfo {title} {Majorana modes at the ends of superconductor
			vortices in doped topological insulators},\ }\href
	{https://link.aps.org/doi/10.1103/PhysRevLett.107.097001} {\bibfield
		{journal} {\bibinfo  {journal} {Phys. Rev. Lett.}\ }\textbf {\bibinfo
			{volume} {107}},\ \bibinfo {pages} {097001} (\bibinfo {year}
		{2011})}\BibitemShut {NoStop}%
	\bibitem [{\citenamefont {Pacholski}\ \emph {et~al.}(2018)\citenamefont
		{Pacholski}, \citenamefont {Beenakker},\ and\ \citenamefont
		{Adagideli}}]{pacholski_prl_2018}%
	\BibitemOpen
	\bibfield  {author} {\bibinfo {author} {\bibfnamefont {M.~J.}\ \bibnamefont
			{Pacholski}}, \bibinfo {author} {\bibfnamefont {C.~W.~J.}\ \bibnamefont
			{Beenakker}},\ and\ \bibinfo {author} {\bibfnamefont {i.~d.~I.}\ \bibnamefont
			{Adagideli}},\ }\bibfield  {title} {\bibinfo {title} {Topologically protected
			landau level in the vortex lattice of a weyl superconductor},\ }\href
	{https://link.aps.org/doi/10.1103/PhysRevLett.121.037701} {\bibfield
		{journal} {\bibinfo  {journal} {Phys. Rev. Lett.}\ }\textbf {\bibinfo
			{volume} {121}},\ \bibinfo {pages} {037701} (\bibinfo {year}
		{2018})}\BibitemShut {NoStop}%
	\bibitem [{\citenamefont {K\"onig}\ and\ \citenamefont
		{Coleman}(2019)}]{konig_prl_2019}%
	\BibitemOpen
	\bibfield  {author} {\bibinfo {author} {\bibfnamefont {E.~J.}\ \bibnamefont
			{K\"onig}}\ and\ \bibinfo {author} {\bibfnamefont {P.}~\bibnamefont
			{Coleman}},\ }\bibfield  {title} {\bibinfo {title}
		{Crystalline-symmetry-protected helical majorana modes in the iron
			pnictides},\ }\href {https://link.aps.org/doi/10.1103/PhysRevLett.122.207001}
	{\bibfield  {journal} {\bibinfo  {journal} {Phys. Rev. Lett.}\ }\textbf
		{\bibinfo {volume} {122}},\ \bibinfo {pages} {207001} (\bibinfo {year}
		{2019})}\BibitemShut {NoStop}%
	\bibitem [{\citenamefont {Qin}\ \emph {et~al.}(2019)\citenamefont {Qin},
		\citenamefont {Hu}, \citenamefont {Le}, \citenamefont {Zeng}, \citenamefont
		{Zhang}, \citenamefont {Fang},\ and\ \citenamefont {Hu}}]{qin_prl_2019}%
	\BibitemOpen
	\bibfield  {author} {\bibinfo {author} {\bibfnamefont {S.}~\bibnamefont
			{Qin}}, \bibinfo {author} {\bibfnamefont {L.}~\bibnamefont {Hu}}, \bibinfo
		{author} {\bibfnamefont {C.}~\bibnamefont {Le}}, \bibinfo {author}
		{\bibfnamefont {J.}~\bibnamefont {Zeng}}, \bibinfo {author} {\bibfnamefont
			{F.-c.}\ \bibnamefont {Zhang}}, \bibinfo {author} {\bibfnamefont
			{C.}~\bibnamefont {Fang}},\ and\ \bibinfo {author} {\bibfnamefont
			{J.}~\bibnamefont {Hu}},\ }\bibfield  {title} {\bibinfo {title} {Quasi-1d
			topological nodal vortex line phase in doped superconducting 3d dirac
			semimetals},\ }\href
	{https://link.aps.org/doi/10.1103/PhysRevLett.123.027003} {\bibfield
		{journal} {\bibinfo  {journal} {Phys. Rev. Lett.}\ }\textbf {\bibinfo
			{volume} {123}},\ \bibinfo {pages} {027003} (\bibinfo {year}
		{2019})}\BibitemShut {NoStop}%
	\bibitem [{\citenamefont {Yan}\ \emph {et~al.}(2020)\citenamefont {Yan},
		\citenamefont {Wu},\ and\ \citenamefont {Huang}}]{yan_prl_2020}%
	\BibitemOpen
	\bibfield  {author} {\bibinfo {author} {\bibfnamefont {Z.}~\bibnamefont
			{Yan}}, \bibinfo {author} {\bibfnamefont {Z.}~\bibnamefont {Wu}},\ and\
		\bibinfo {author} {\bibfnamefont {W.}~\bibnamefont {Huang}},\ }\bibfield
	{title} {\bibinfo {title} {Vortex end majorana zero modes in superconducting
			dirac and weyl semimetals},\ }\href
	{https://link.aps.org/doi/10.1103/PhysRevLett.124.257001} {\bibfield
		{journal} {\bibinfo  {journal} {Phys. Rev. Lett.}\ }\textbf {\bibinfo
			{volume} {124}},\ \bibinfo {pages} {257001} (\bibinfo {year}
		{2020})}\BibitemShut {NoStop}%
	\bibitem [{\citenamefont {Ghazaryan}\ \emph {et~al.}(2020)\citenamefont
		{Ghazaryan}, \citenamefont {Lopes}, \citenamefont {Hosur}, \citenamefont
		{Gilbert},\ and\ \citenamefont {Ghaemi}}]{ghazaryan_prb_2020}%
	\BibitemOpen
	\bibfield  {author} {\bibinfo {author} {\bibfnamefont {A.}~\bibnamefont
			{Ghazaryan}}, \bibinfo {author} {\bibfnamefont {P.~L.~S.}\ \bibnamefont
			{Lopes}}, \bibinfo {author} {\bibfnamefont {P.}~\bibnamefont {Hosur}},
		\bibinfo {author} {\bibfnamefont {M.~J.}\ \bibnamefont {Gilbert}},\ and\
		\bibinfo {author} {\bibfnamefont {P.}~\bibnamefont {Ghaemi}},\ }\bibfield
	{title} {\bibinfo {title} {Effect of zeeman coupling on the majorana vortex
			modes in iron-based topological superconductors},\ }\href
	{https://link.aps.org/doi/10.1103/PhysRevB.101.020504} {\bibfield  {journal}
		{\bibinfo  {journal} {Phys. Rev. B}\ }\textbf {\bibinfo {volume} {101}},\
		\bibinfo {pages} {020504} (\bibinfo {year} {2020})}\BibitemShut {NoStop}%
	\bibitem [{\citenamefont {Kobayashi}\ and\ \citenamefont
		{Furusaki}(2020)}]{kobayashi2020double}%
	\BibitemOpen
	\bibfield  {author} {\bibinfo {author} {\bibfnamefont {S.}~\bibnamefont
			{Kobayashi}}\ and\ \bibinfo {author} {\bibfnamefont {A.}~\bibnamefont
			{Furusaki}},\ }\bibfield  {title} {\bibinfo {title} {Double majorana vortex
			zero modes in superconducting topological crystalline insulators with surface
			rotation anomaly},\ }\href
	{https://link.aps.org/doi/10.1103/PhysRevB.102.180505} {\bibfield  {journal}
		{\bibinfo  {journal} {Phys. Rev. B}\ }\textbf {\bibinfo {volume} {102}},\
		\bibinfo {pages} {180505} (\bibinfo {year} {2020})}\BibitemShut {NoStop}%
	\bibitem [{\citenamefont {Giwa}\ and\ \citenamefont
		{Hosur}(2021)}]{giwa_prl_2021}%
	\BibitemOpen
	\bibfield  {author} {\bibinfo {author} {\bibfnamefont {R.}~\bibnamefont
			{Giwa}}\ and\ \bibinfo {author} {\bibfnamefont {P.}~\bibnamefont {Hosur}},\
	}\bibfield  {title} {\bibinfo {title} {Fermi arc criterion for surface
			majorana modes in superconducting time-reversal symmetric weyl semimetals},\
	}\href {https://link.aps.org/doi/10.1103/PhysRevLett.127.187002} {\bibfield
		{journal} {\bibinfo  {journal} {Phys. Rev. Lett.}\ }\textbf {\bibinfo
			{volume} {127}},\ \bibinfo {pages} {187002} (\bibinfo {year}
		{2021})}\BibitemShut {NoStop}%
	\bibitem [{\citenamefont {{Hu}}\ \emph {et~al.}(2021)\citenamefont {{Hu}},
		\citenamefont {{Wu}}, \citenamefont {{Liu}},\ and\ \citenamefont
		{{Zhang}}}]{hu_arxiv_2021}%
	\BibitemOpen
	\bibfield  {author} {\bibinfo {author} {\bibfnamefont {L.-H.}\ \bibnamefont
			{{Hu}}}, \bibinfo {author} {\bibfnamefont {X.}~\bibnamefont {{Wu}}}, \bibinfo
		{author} {\bibfnamefont {C.-X.}\ \bibnamefont {{Liu}}},\ and\ \bibinfo
		{author} {\bibfnamefont {R.-X.}\ \bibnamefont {{Zhang}}},\ }\bibfield
	{title} {\bibinfo {title} {{Competing Vortex Topologies in Iron-based
				Superconductors}},\ }\href {https://arxiv.org/abs/2110.11357} {\bibfield
		{journal} {\bibinfo  {journal} {arXiv:2110.11357}\ } (\bibinfo {year}
		{2021})}\BibitemShut {NoStop}%
	\bibitem [{\citenamefont {Sun}\ \emph {et~al.}(2016)\citenamefont {Sun},
		\citenamefont {Zhang}, \citenamefont {Hu}, \citenamefont {Li}, \citenamefont
		{Wang}, \citenamefont {Ma}, \citenamefont {Xu}, \citenamefont {Gao},
		\citenamefont {Guan}, \citenamefont {Li}, \citenamefont {Liu}, \citenamefont
		{Qian}, \citenamefont {Zhou}, \citenamefont {Fu}, \citenamefont {Li},
		\citenamefont {Zhang},\ and\ \citenamefont {Jia}}]{sun_prl_2016}%
	\BibitemOpen
	\bibfield  {author} {\bibinfo {author} {\bibfnamefont {H.-H.}\ \bibnamefont
			{Sun}}, \bibinfo {author} {\bibfnamefont {K.-W.}\ \bibnamefont {Zhang}},
		\bibinfo {author} {\bibfnamefont {L.-H.}\ \bibnamefont {Hu}}, \bibinfo
		{author} {\bibfnamefont {C.}~\bibnamefont {Li}}, \bibinfo {author}
		{\bibfnamefont {G.-Y.}\ \bibnamefont {Wang}}, \bibinfo {author}
		{\bibfnamefont {H.-Y.}\ \bibnamefont {Ma}}, \bibinfo {author} {\bibfnamefont
			{Z.-A.}\ \bibnamefont {Xu}}, \bibinfo {author} {\bibfnamefont {C.-L.}\
			\bibnamefont {Gao}}, \bibinfo {author} {\bibfnamefont {D.-D.}\ \bibnamefont
			{Guan}}, \bibinfo {author} {\bibfnamefont {Y.-Y.}\ \bibnamefont {Li}},
		\bibinfo {author} {\bibfnamefont {C.}~\bibnamefont {Liu}}, \bibinfo {author}
		{\bibfnamefont {D.}~\bibnamefont {Qian}}, \bibinfo {author} {\bibfnamefont
			{Y.}~\bibnamefont {Zhou}}, \bibinfo {author} {\bibfnamefont {L.}~\bibnamefont
			{Fu}}, \bibinfo {author} {\bibfnamefont {S.-C.}\ \bibnamefont {Li}}, \bibinfo
		{author} {\bibfnamefont {F.-C.}\ \bibnamefont {Zhang}},\ and\ \bibinfo
		{author} {\bibfnamefont {J.-F.}\ \bibnamefont {Jia}},\ }\bibfield  {title}
	{\bibinfo {title} {Majorana zero mode detected with spin selective andreev
			reflection in the vortex of a topological superconductor},\ }\href
	{https://link.aps.org/doi/10.1103/PhysRevLett.116.257003} {\bibfield
		{journal} {\bibinfo  {journal} {Phys. Rev. Lett.}\ }\textbf {\bibinfo
			{volume} {116}},\ \bibinfo {pages} {257003} (\bibinfo {year}
		{2016})}\BibitemShut {NoStop}%
	\bibitem [{\citenamefont {Wang}\ \emph {et~al.}(2018)\citenamefont {Wang},
		\citenamefont {Kong}, \citenamefont {Fan}, \citenamefont {Chen},
		\citenamefont {Zhu}, \citenamefont {Liu}, \citenamefont {Cao}, \citenamefont
		{Sun}, \citenamefont {Du}, \citenamefont {Schneeloch}, \citenamefont {Zhong},
		\citenamefont {Gu}, \citenamefont {Fu}, \citenamefont {Ding},\ and\
		\citenamefont {Gao}}]{Wang_science_2018}%
	\BibitemOpen
	\bibfield  {author} {\bibinfo {author} {\bibfnamefont {D.}~\bibnamefont
			{Wang}}, \bibinfo {author} {\bibfnamefont {L.}~\bibnamefont {Kong}}, \bibinfo
		{author} {\bibfnamefont {P.}~\bibnamefont {Fan}}, \bibinfo {author}
		{\bibfnamefont {H.}~\bibnamefont {Chen}}, \bibinfo {author} {\bibfnamefont
			{S.}~\bibnamefont {Zhu}}, \bibinfo {author} {\bibfnamefont {W.}~\bibnamefont
			{Liu}}, \bibinfo {author} {\bibfnamefont {L.}~\bibnamefont {Cao}}, \bibinfo
		{author} {\bibfnamefont {Y.}~\bibnamefont {Sun}}, \bibinfo {author}
		{\bibfnamefont {S.}~\bibnamefont {Du}}, \bibinfo {author} {\bibfnamefont
			{J.}~\bibnamefont {Schneeloch}}, \bibinfo {author} {\bibfnamefont
			{R.}~\bibnamefont {Zhong}}, \bibinfo {author} {\bibfnamefont
			{G.}~\bibnamefont {Gu}}, \bibinfo {author} {\bibfnamefont {L.}~\bibnamefont
			{Fu}}, \bibinfo {author} {\bibfnamefont {H.}~\bibnamefont {Ding}},\ and\
		\bibinfo {author} {\bibfnamefont {H.-J.}\ \bibnamefont {Gao}},\ }\bibfield
	{title} {\bibinfo {title} {Evidence for majorana bound states in an
			iron-based superconductor},\ }\href {https://doi.org/10.1126/science.aao1797}
	{\bibfield  {journal} {\bibinfo  {journal} {Science}\ }\textbf {\bibinfo
			{volume} {362}},\ \bibinfo {pages} {333} (\bibinfo {year}
		{2018})}\BibitemShut {NoStop}%
	\bibitem [{\citenamefont {Kong}\ \emph {et~al.}(2019)\citenamefont {Kong},
		\citenamefont {Zhu}, \citenamefont {Papaj}, \citenamefont {Chen},
		\citenamefont {Cao}, \citenamefont {Isobe}, \citenamefont {Xing},
		\citenamefont {Liu}, \citenamefont {Wang}, \citenamefont {Fan}, \citenamefont
		{Sun}, \citenamefont {Du}, \citenamefont {Schneeloch}, \citenamefont {Zhong},
		\citenamefont {Gu}, \citenamefont {Fu}, \citenamefont {Gao},\ and\
		\citenamefont {Ding}}]{Kong_np_2019}%
	\BibitemOpen
	\bibfield  {author} {\bibinfo {author} {\bibfnamefont {L.}~\bibnamefont
			{Kong}}, \bibinfo {author} {\bibfnamefont {S.}~\bibnamefont {Zhu}}, \bibinfo
		{author} {\bibfnamefont {M.}~\bibnamefont {Papaj}}, \bibinfo {author}
		{\bibfnamefont {H.}~\bibnamefont {Chen}}, \bibinfo {author} {\bibfnamefont
			{L.}~\bibnamefont {Cao}}, \bibinfo {author} {\bibfnamefont {H.}~\bibnamefont
			{Isobe}}, \bibinfo {author} {\bibfnamefont {Y.}~\bibnamefont {Xing}},
		\bibinfo {author} {\bibfnamefont {W.}~\bibnamefont {Liu}}, \bibinfo {author}
		{\bibfnamefont {D.}~\bibnamefont {Wang}}, \bibinfo {author} {\bibfnamefont
			{P.}~\bibnamefont {Fan}}, \bibinfo {author} {\bibfnamefont {Y.}~\bibnamefont
			{Sun}}, \bibinfo {author} {\bibfnamefont {S.}~\bibnamefont {Du}}, \bibinfo
		{author} {\bibfnamefont {J.}~\bibnamefont {Schneeloch}}, \bibinfo {author}
		{\bibfnamefont {R.}~\bibnamefont {Zhong}}, \bibinfo {author} {\bibfnamefont
			{G.}~\bibnamefont {Gu}}, \bibinfo {author} {\bibfnamefont {L.}~\bibnamefont
			{Fu}}, \bibinfo {author} {\bibfnamefont {H.-J.}\ \bibnamefont {Gao}},\ and\
		\bibinfo {author} {\bibfnamefont {H.}~\bibnamefont {Ding}},\ }\bibfield
	{title} {\bibinfo {title} {Half-integer level shift of vortex bound states in
			an iron-based superconductor},\ }\href
	{https://doi.org/10.1038/s41567-019-0630-5} {\bibfield  {journal} {\bibinfo
			{journal} {Nature Physics}\ }\textbf {\bibinfo {volume} {15}},\ \bibinfo
		{pages} {1181} (\bibinfo {year} {2019})}\BibitemShut {NoStop}%
	\bibitem [{\citenamefont {Liu}\ \emph {et~al.}(2020)\citenamefont {Liu},
		\citenamefont {Cao}, \citenamefont {Zhu}, \citenamefont {Kong}, \citenamefont
		{Wang}, \citenamefont {Papaj}, \citenamefont {Zhang}, \citenamefont {Liu},
		\citenamefont {Chen}, \citenamefont {Li}, \citenamefont {Yang}, \citenamefont
		{Kondo}, \citenamefont {Du}, \citenamefont {Cao}, \citenamefont {Shin},
		\citenamefont {Fu}, \citenamefont {Yin}, \citenamefont {Gao},\ and\
		\citenamefont {Ding}}]{Liu_nc_2020}%
	\BibitemOpen
	\bibfield  {author} {\bibinfo {author} {\bibfnamefont {W.}~\bibnamefont
			{Liu}}, \bibinfo {author} {\bibfnamefont {L.}~\bibnamefont {Cao}}, \bibinfo
		{author} {\bibfnamefont {S.}~\bibnamefont {Zhu}}, \bibinfo {author}
		{\bibfnamefont {L.}~\bibnamefont {Kong}}, \bibinfo {author} {\bibfnamefont
			{G.}~\bibnamefont {Wang}}, \bibinfo {author} {\bibfnamefont {M.}~\bibnamefont
			{Papaj}}, \bibinfo {author} {\bibfnamefont {P.}~\bibnamefont {Zhang}},
		\bibinfo {author} {\bibfnamefont {Y.-B.}\ \bibnamefont {Liu}}, \bibinfo
		{author} {\bibfnamefont {H.}~\bibnamefont {Chen}}, \bibinfo {author}
		{\bibfnamefont {G.}~\bibnamefont {Li}}, \bibinfo {author} {\bibfnamefont
			{F.}~\bibnamefont {Yang}}, \bibinfo {author} {\bibfnamefont {T.}~\bibnamefont
			{Kondo}}, \bibinfo {author} {\bibfnamefont {S.}~\bibnamefont {Du}}, \bibinfo
		{author} {\bibfnamefont {G.-H.}\ \bibnamefont {Cao}}, \bibinfo {author}
		{\bibfnamefont {S.}~\bibnamefont {Shin}}, \bibinfo {author} {\bibfnamefont
			{L.}~\bibnamefont {Fu}}, \bibinfo {author} {\bibfnamefont {Z.}~\bibnamefont
			{Yin}}, \bibinfo {author} {\bibfnamefont {H.-J.}\ \bibnamefont {Gao}},\ and\
		\bibinfo {author} {\bibfnamefont {H.}~\bibnamefont {Ding}},\ }\bibfield
	{title} {\bibinfo {title} {A new majorana platform in an fe-as bilayer
			superconductor},\ }\href {https://doi.org/10.1038/s41467-020-19487-1}
	{\bibfield  {journal} {\bibinfo  {journal} {Nature Communications}\ }\textbf
		{\bibinfo {volume} {11}} (\bibinfo {year} {2020})}\BibitemShut {NoStop}%
	\bibitem [{\citenamefont {Luttinger}(1956)}]{luttinger_pr_1956}%
	\BibitemOpen
	\bibfield  {author} {\bibinfo {author} {\bibfnamefont {J.~M.}\ \bibnamefont
			{Luttinger}},\ }\bibfield  {title} {\bibinfo {title} {Quantum theory of
			cyclotron resonance in semiconductors: General theory},\ }\href
	{https://link.aps.org/doi/10.1103/PhysRev.102.1030} {\bibfield  {journal}
		{\bibinfo  {journal} {Phys. Rev.}\ }\textbf {\bibinfo {volume} {102}},\
		\bibinfo {pages} {1030} (\bibinfo {year} {1956})}\BibitemShut {NoStop}%
	\bibitem [{\citenamefont {Chiu}\ \emph {et~al.}(2012)\citenamefont {Chiu},
		\citenamefont {Ghaemi},\ and\ \citenamefont {Hughes}}]{chiu_prl_2012}%
	\BibitemOpen
	\bibfield  {author} {\bibinfo {author} {\bibfnamefont {C.-K.}\ \bibnamefont
			{Chiu}}, \bibinfo {author} {\bibfnamefont {P.}~\bibnamefont {Ghaemi}},\ and\
		\bibinfo {author} {\bibfnamefont {T.~L.}\ \bibnamefont {Hughes}},\ }\bibfield
	{title} {\bibinfo {title} {Stabilization of majorana modes in magnetic
			vortices in the superconducting phase of topological insulators using
			topologically trivial bands},\ }\href
	{https://link.aps.org/doi/10.1103/PhysRevLett.109.237009} {\bibfield
		{journal} {\bibinfo  {journal} {Phys. Rev. Lett.}\ }\textbf {\bibinfo
			{volume} {109}},\ \bibinfo {pages} {237009} (\bibinfo {year}
		{2012})}\BibitemShut {NoStop}%
	\bibitem [{\citenamefont {Xu}\ \emph {et~al.}(2016)\citenamefont {Xu},
		\citenamefont {Lian}, \citenamefont {Tang}, \citenamefont {Qi},\ and\
		\citenamefont {Zhang}}]{xu_prl_2016}%
	\BibitemOpen
	\bibfield  {author} {\bibinfo {author} {\bibfnamefont {G.}~\bibnamefont
			{Xu}}, \bibinfo {author} {\bibfnamefont {B.}~\bibnamefont {Lian}}, \bibinfo
		{author} {\bibfnamefont {P.}~\bibnamefont {Tang}}, \bibinfo {author}
		{\bibfnamefont {X.-L.}\ \bibnamefont {Qi}},\ and\ \bibinfo {author}
		{\bibfnamefont {S.-C.}\ \bibnamefont {Zhang}},\ }\bibfield  {title} {\bibinfo
		{title} {Topological superconductivity on the surface of fe-based
			superconductors},\ }\href
	{https://link.aps.org/doi/10.1103/PhysRevLett.117.047001} {\bibfield
		{journal} {\bibinfo  {journal} {Phys. Rev. Lett.}\ }\textbf {\bibinfo
			{volume} {117}},\ \bibinfo {pages} {047001} (\bibinfo {year}
		{2016})}\BibitemShut {NoStop}%
	\bibitem [{\citenamefont {Yan}\ \emph {et~al.}(2017)\citenamefont {Yan},
		\citenamefont {Bi},\ and\ \citenamefont {Wang}}]{yan_prl_2017}%
	\BibitemOpen
	\bibfield  {author} {\bibinfo {author} {\bibfnamefont {Z.}~\bibnamefont
			{Yan}}, \bibinfo {author} {\bibfnamefont {R.}~\bibnamefont {Bi}},\ and\
		\bibinfo {author} {\bibfnamefont {Z.}~\bibnamefont {Wang}},\ }\bibfield
	{title} {\bibinfo {title} {Majorana zero modes protected by a hopf invariant
			in topologically trivial superconductors},\ }\href
	{https://link.aps.org/doi/10.1103/PhysRevLett.118.147003} {\bibfield
		{journal} {\bibinfo  {journal} {Phys. Rev. Lett.}\ }\textbf {\bibinfo
			{volume} {118}},\ \bibinfo {pages} {147003} (\bibinfo {year}
		{2017})}\BibitemShut {NoStop}%
	\bibitem [{\citenamefont {Chan}\ \emph {et~al.}(2017)\citenamefont {Chan},
		\citenamefont {Zhang}, \citenamefont {Poon}, \citenamefont {He},
		\citenamefont {Wang},\ and\ \citenamefont {Liu}}]{chan_prl_2017}%
	\BibitemOpen
	\bibfield  {author} {\bibinfo {author} {\bibfnamefont {C.}~\bibnamefont
			{Chan}}, \bibinfo {author} {\bibfnamefont {L.}~\bibnamefont {Zhang}},
		\bibinfo {author} {\bibfnamefont {T.~F.~J.}\ \bibnamefont {Poon}}, \bibinfo
		{author} {\bibfnamefont {Y.-P.}\ \bibnamefont {He}}, \bibinfo {author}
		{\bibfnamefont {Y.-Q.}\ \bibnamefont {Wang}},\ and\ \bibinfo {author}
		{\bibfnamefont {X.-J.}\ \bibnamefont {Liu}},\ }\bibfield  {title} {\bibinfo
		{title} {Generic theory for majorana zero modes in 2d superconductors},\
	}\href {https://link.aps.org/doi/10.1103/PhysRevLett.119.047001} {\bibfield
		{journal} {\bibinfo  {journal} {Phys. Rev. Lett.}\ }\textbf {\bibinfo
			{volume} {119}},\ \bibinfo {pages} {047001} (\bibinfo {year}
		{2017})}\BibitemShut {NoStop}%
	\bibitem [{\citenamefont {Chan}\ and\ \citenamefont
		{Liu}(2017)}]{chan20172ndChern}%
	\BibitemOpen
	\bibfield  {author} {\bibinfo {author} {\bibfnamefont {C.}~\bibnamefont
			{Chan}}\ and\ \bibinfo {author} {\bibfnamefont {X.-J.}\ \bibnamefont {Liu}},\
	}\bibfield  {title} {\bibinfo {title} {Non-abelian majorana modes protected
			by an emergent second chern number},\ }\href
	{https://link.aps.org/doi/10.1103/PhysRevLett.118.207002} {\bibfield
		{journal} {\bibinfo  {journal} {Phys. Rev. Lett.}\ }\textbf {\bibinfo
			{volume} {118}},\ \bibinfo {pages} {207002} (\bibinfo {year}
		{2017})}\BibitemShut {NoStop}%
	\bibitem [{\citenamefont {Murakami}\ \emph {et~al.}(2004)\citenamefont
		{Murakami}, \citenamefont {Nagosa},\ and\ \citenamefont
		{Zhang}}]{murakami_prb_2004}%
	\BibitemOpen
	\bibfield  {author} {\bibinfo {author} {\bibfnamefont {S.}~\bibnamefont
			{Murakami}}, \bibinfo {author} {\bibfnamefont {N.}~\bibnamefont {Nagosa}},\
		and\ \bibinfo {author} {\bibfnamefont {S.-C.}\ \bibnamefont {Zhang}},\
	}\bibfield  {title} {\bibinfo {title} {$\text{SU}(2)$ non-abelian holonomy
			and dissipationless spin current in semiconductors},\ }\href
	{https://link.aps.org/doi/10.1103/PhysRevB.69.235206} {\bibfield  {journal}
		{\bibinfo  {journal} {Phys. Rev. B}\ }\textbf {\bibinfo {volume} {69}},\
		\bibinfo {pages} {235206} (\bibinfo {year} {2004})}\BibitemShut {NoStop}%
	\bibitem [{\citenamefont {Bansil}\ \emph {et~al.}(2016)\citenamefont {Bansil},
		\citenamefont {Lin},\ and\ \citenamefont {Das}}]{bansil_rmp_2016}%
	\BibitemOpen
	\bibfield  {author} {\bibinfo {author} {\bibfnamefont {A.}~\bibnamefont
			{Bansil}}, \bibinfo {author} {\bibfnamefont {H.}~\bibnamefont {Lin}},\ and\
		\bibinfo {author} {\bibfnamefont {T.}~\bibnamefont {Das}},\ }\bibfield
	{title} {\bibinfo {title} {Colloquium: Topological band theory},\ }\href
	{https://link.aps.org/doi/10.1103/RevModPhys.88.021004} {\bibfield  {journal}
		{\bibinfo  {journal} {Rev. Mod. Phys.}\ }\textbf {\bibinfo {volume} {88}},\
		\bibinfo {pages} {021004} (\bibinfo {year} {2016})}\BibitemShut {NoStop}%
	\bibitem [{\citenamefont {Armitage}\ \emph {et~al.}(2018)\citenamefont
		{Armitage}, \citenamefont {Mele},\ and\ \citenamefont
		{Vishwanath}}]{armitage_rmp_2018}%
	\BibitemOpen
	\bibfield  {author} {\bibinfo {author} {\bibfnamefont {N.~P.}\ \bibnamefont
			{Armitage}}, \bibinfo {author} {\bibfnamefont {E.~J.}\ \bibnamefont {Mele}},\
		and\ \bibinfo {author} {\bibfnamefont {A.}~\bibnamefont {Vishwanath}},\
	}\bibfield  {title} {\bibinfo {title} {Weyl and dirac semimetals in
			three-dimensional solids},\ }\href
	{https://link.aps.org/doi/10.1103/RevModPhys.90.015001} {\bibfield  {journal}
		{\bibinfo  {journal} {Rev. Mod. Phys.}\ }\textbf {\bibinfo {volume} {90}},\
		\bibinfo {pages} {015001} (\bibinfo {year} {2018})}\BibitemShut {NoStop}%
	\bibitem [{\citenamefont {Lv}\ \emph {et~al.}(2021)\citenamefont {Lv},
		\citenamefont {Qian},\ and\ \citenamefont {Ding}}]{lv_rmp_2021}%
	\BibitemOpen
	\bibfield  {author} {\bibinfo {author} {\bibfnamefont {B.~Q.}\ \bibnamefont
			{Lv}}, \bibinfo {author} {\bibfnamefont {T.}~\bibnamefont {Qian}},\ and\
		\bibinfo {author} {\bibfnamefont {H.}~\bibnamefont {Ding}},\ }\bibfield
	{title} {\bibinfo {title} {Experimental perspective on three-dimensional
			topological semimetals},\ }\href
	{https://link.aps.org/doi/10.1103/RevModPhys.93.025002} {\bibfield  {journal}
		{\bibinfo  {journal} {Rev. Mod. Phys.}\ }\textbf {\bibinfo {volume} {93}},\
		\bibinfo {pages} {025002} (\bibinfo {year} {2021})}\BibitemShut {NoStop}%
	\bibitem [{\citenamefont {McCann}\ and\ \citenamefont
		{Koshino}(2013)}]{McCann_RPP_2013}%
	\BibitemOpen
	\bibfield  {author} {\bibinfo {author} {\bibfnamefont {E.}~\bibnamefont
			{McCann}}\ and\ \bibinfo {author} {\bibfnamefont {M.}~\bibnamefont
			{Koshino}},\ }\bibfield  {title} {\bibinfo {title} {The electronic properties
			of bilayer graphene},\ }\href {https://doi.org/10.1088/0034-4885/76/5/056503}
	{\bibfield  {journal} {\bibinfo  {journal} {Reports on Progress in Physics}\
		}\textbf {\bibinfo {volume} {76}},\ \bibinfo {pages} {056503} (\bibinfo
		{year} {2013})}\BibitemShut {NoStop}%
	\bibitem [{\citenamefont {Fu}(2011)}]{fu_prl_2011}%
	\BibitemOpen
	\bibfield  {author} {\bibinfo {author} {\bibfnamefont {L.}~\bibnamefont
			{Fu}},\ }\bibfield  {title} {\bibinfo {title} {Topological crystalline
			insulators},\ }\href
	{https://link.aps.org/doi/10.1103/PhysRevLett.106.106802} {\bibfield
		{journal} {\bibinfo  {journal} {Phys. Rev. Lett.}\ }\textbf {\bibinfo
			{volume} {106}},\ \bibinfo {pages} {106802} (\bibinfo {year}
		{2011})}\BibitemShut {NoStop}%
	\bibitem [{\citenamefont {Zhang}\ and\ \citenamefont
		{Liu}(2015)}]{zhang_prb_2015}%
	\BibitemOpen
	\bibfield  {author} {\bibinfo {author} {\bibfnamefont {R.-X.}\ \bibnamefont
			{Zhang}}\ and\ \bibinfo {author} {\bibfnamefont {C.-X.}\ \bibnamefont
			{Liu}},\ }\bibfield  {title} {\bibinfo {title} {Topological magnetic
			crystalline insulators and corepresentation theory},\ }\href
	{https://link.aps.org/doi/10.1103/PhysRevB.91.115317} {\bibfield  {journal}
		{\bibinfo  {journal} {Phys. Rev. B}\ }\textbf {\bibinfo {volume} {91}},\
		\bibinfo {pages} {115317} (\bibinfo {year} {2015})}\BibitemShut {NoStop}%
	\bibitem [{\citenamefont {Teo}\ and\ \citenamefont
		{Kane}(2010)}]{teo2010topodefect}%
	\BibitemOpen
	\bibfield  {author} {\bibinfo {author} {\bibfnamefont {J.~C.~Y.}\
			\bibnamefont {Teo}}\ and\ \bibinfo {author} {\bibfnamefont {C.~L.}\
			\bibnamefont {Kane}},\ }\bibfield  {title} {\bibinfo {title} {Topological
			defects and gapless modes in insulators and superconductors},\ }\href
	{https://doi.org/10.1103/PhysRevB.82.115120} {\bibfield  {journal} {\bibinfo
			{journal} {Phys. Rev. B}\ }\textbf {\bibinfo {volume} {82}},\ \bibinfo
		{pages} {115120} (\bibinfo {year} {2010})}\BibitemShut {NoStop}%
	\bibitem [{\citenamefont {Schnyder}\ \emph {et~al.}(2008)\citenamefont
		{Schnyder}, \citenamefont {Ryu}, \citenamefont {Furusaki},\ and\
		\citenamefont {Ludwig}}]{schnyder2008class}%
	\BibitemOpen
	\bibfield  {author} {\bibinfo {author} {\bibfnamefont {A.~P.}\ \bibnamefont
			{Schnyder}}, \bibinfo {author} {\bibfnamefont {S.}~\bibnamefont {Ryu}},
		\bibinfo {author} {\bibfnamefont {A.}~\bibnamefont {Furusaki}},\ and\
		\bibinfo {author} {\bibfnamefont {A.~W.~W.}\ \bibnamefont {Ludwig}},\
	}\bibfield  {title} {\bibinfo {title} {Classification of topological
			insulators and superconductors in three spatial dimensions},\ }\href
	{https://doi.org/10.1103/PhysRevB.78.195125} {\bibfield  {journal} {\bibinfo
			{journal} {Phys. Rev. B}\ }\textbf {\bibinfo {volume} {78}},\ \bibinfo
		{pages} {195125} (\bibinfo {year} {2008})}\BibitemShut {NoStop}%
	\bibitem [{\citenamefont {Xu}\ \emph {et~al.}(2017)\citenamefont {Xu},
		\citenamefont {Chan}, \citenamefont {Chen}, \citenamefont {Chen},
		\citenamefont {Wang}, \citenamefont {Dejoie}, \citenamefont {Wong},
		\citenamefont {Hlevyack}, \citenamefont {Ryu}, \citenamefont {Kee},
		\citenamefont {Tamura}, \citenamefont {Chou}, \citenamefont {Hussain},
		\citenamefont {Mo},\ and\ \citenamefont {Chiang}}]{xu_prl_2017}%
	\BibitemOpen
	\bibfield  {author} {\bibinfo {author} {\bibfnamefont {C.-Z.}\ \bibnamefont
			{Xu}}, \bibinfo {author} {\bibfnamefont {Y.-H.}\ \bibnamefont {Chan}},
		\bibinfo {author} {\bibfnamefont {Y.}~\bibnamefont {Chen}}, \bibinfo {author}
		{\bibfnamefont {P.}~\bibnamefont {Chen}}, \bibinfo {author} {\bibfnamefont
			{X.}~\bibnamefont {Wang}}, \bibinfo {author} {\bibfnamefont {C.}~\bibnamefont
			{Dejoie}}, \bibinfo {author} {\bibfnamefont {M.-H.}\ \bibnamefont {Wong}},
		\bibinfo {author} {\bibfnamefont {J.~A.}\ \bibnamefont {Hlevyack}}, \bibinfo
		{author} {\bibfnamefont {H.}~\bibnamefont {Ryu}}, \bibinfo {author}
		{\bibfnamefont {H.-Y.}\ \bibnamefont {Kee}}, \bibinfo {author} {\bibfnamefont
			{N.}~\bibnamefont {Tamura}}, \bibinfo {author} {\bibfnamefont {M.-Y.}\
			\bibnamefont {Chou}}, \bibinfo {author} {\bibfnamefont {Z.}~\bibnamefont
			{Hussain}}, \bibinfo {author} {\bibfnamefont {S.-K.}\ \bibnamefont {Mo}},\
		and\ \bibinfo {author} {\bibfnamefont {T.-C.}\ \bibnamefont {Chiang}},\
	}\bibfield  {title} {\bibinfo {title} {Elemental topological dirac semimetal:
			$\ensuremath{\alpha}$-sn on insb(111)},\ }\href
	{https://link.aps.org/doi/10.1103/PhysRevLett.118.146402} {\bibfield
		{journal} {\bibinfo  {journal} {Phys. Rev. Lett.}\ }\textbf {\bibinfo
			{volume} {118}},\ \bibinfo {pages} {146402} (\bibinfo {year}
		{2017})}\BibitemShut {NoStop}%
	\bibitem [{\citenamefont {Sancho}\ \emph {et~al.}(1985)\citenamefont {Sancho},
		\citenamefont {Sancho}, \citenamefont {Sancho},\ and\ \citenamefont
		{Rubio}}]{sancho_JPF_1985}%
	\BibitemOpen
	\bibfield  {author} {\bibinfo {author} {\bibfnamefont {M.~L.}\ \bibnamefont
			{Sancho}}, \bibinfo {author} {\bibfnamefont {J.~L.}\ \bibnamefont {Sancho}},
		\bibinfo {author} {\bibfnamefont {J.~L.}\ \bibnamefont {Sancho}},\ and\
		\bibinfo {author} {\bibfnamefont {J.}~\bibnamefont {Rubio}},\ }\bibfield
	{title} {\bibinfo {title} {Highly convergent schemes for the calculation of
			bulk and surface green functions},\ }\href
	{https://dx.doi.org/10.1088/0305-4608/15/4/009} {\bibfield  {journal}
		{\bibinfo  {journal} {Journal of Physics F: Metal Physics}\ }\textbf
		{\bibinfo {volume} {15}},\ \bibinfo {pages} {851} (\bibinfo {year}
		{1985})}\BibitemShut {NoStop}%
	\bibitem [{\citenamefont {Novik}\ \emph {et~al.}(2005)\citenamefont {Novik},
		\citenamefont {Pfeuffer-Jeschke}, \citenamefont {Jungwirth}, \citenamefont
		{Latussek}, \citenamefont {Becker}, \citenamefont {Landwehr}, \citenamefont
		{Buhmann},\ and\ \citenamefont {Molenkamp}}]{novik_prb_2005}%
	\BibitemOpen
	\bibfield  {author} {\bibinfo {author} {\bibfnamefont {E.~G.}\ \bibnamefont
			{Novik}}, \bibinfo {author} {\bibfnamefont {A.}~\bibnamefont
			{Pfeuffer-Jeschke}}, \bibinfo {author} {\bibfnamefont {T.}~\bibnamefont
			{Jungwirth}}, \bibinfo {author} {\bibfnamefont {V.}~\bibnamefont {Latussek}},
		\bibinfo {author} {\bibfnamefont {C.~R.}\ \bibnamefont {Becker}}, \bibinfo
		{author} {\bibfnamefont {G.}~\bibnamefont {Landwehr}}, \bibinfo {author}
		{\bibfnamefont {H.}~\bibnamefont {Buhmann}},\ and\ \bibinfo {author}
		{\bibfnamefont {L.~W.}\ \bibnamefont {Molenkamp}},\ }\bibfield  {title}
	{\bibinfo {title} {Band structure of semimagnetic
			${\mathrm{hg}}_{1\ensuremath{-}y}{\mathrm{mn}}_{y}\mathrm{Te}$ quantum
			wells},\ }\href {https://link.aps.org/doi/10.1103/PhysRevB.72.035321}
	{\bibfield  {journal} {\bibinfo  {journal} {Phys. Rev. B}\ }\textbf {\bibinfo
			{volume} {72}},\ \bibinfo {pages} {035321} (\bibinfo {year}
		{2005})}\BibitemShut {NoStop}%
	\bibitem [{\citenamefont {Groves}\ and\ \citenamefont
		{Paul}(1963)}]{groves_1963_prl}%
	\BibitemOpen
	\bibfield  {author} {\bibinfo {author} {\bibfnamefont {S.}~\bibnamefont
			{Groves}}\ and\ \bibinfo {author} {\bibfnamefont {W.}~\bibnamefont {Paul}},\
	}\bibfield  {title} {\bibinfo {title} {Band structure of gray tin},\ }\href
	{https://link.aps.org/doi/10.1103/PhysRevLett.11.194} {\bibfield  {journal}
		{\bibinfo  {journal} {Phys. Rev. Lett.}\ }\textbf {\bibinfo {volume} {11}},\
		\bibinfo {pages} {194} (\bibinfo {year} {1963})}\BibitemShut {NoStop}%
	\bibitem [{\citenamefont {Kondo}\ \emph {et~al.}(2015)\citenamefont {Kondo},
		\citenamefont {Nakayama}, \citenamefont {Chen}, \citenamefont {Ishikawa},
		\citenamefont {Moon}, \citenamefont {Yamamoto}, \citenamefont {Ota},
		\citenamefont {Malaeb}, \citenamefont {Kanai}, \citenamefont {Nakashima}
		\emph {et~al.}}]{kondo2015quadratic}%
	\BibitemOpen
	\bibfield  {author} {\bibinfo {author} {\bibfnamefont {T.}~\bibnamefont
			{Kondo}}, \bibinfo {author} {\bibfnamefont {M.}~\bibnamefont {Nakayama}},
		\bibinfo {author} {\bibfnamefont {R.}~\bibnamefont {Chen}}, \bibinfo {author}
		{\bibfnamefont {J.}~\bibnamefont {Ishikawa}}, \bibinfo {author}
		{\bibfnamefont {E.-G.}\ \bibnamefont {Moon}}, \bibinfo {author}
		{\bibfnamefont {T.}~\bibnamefont {Yamamoto}}, \bibinfo {author}
		{\bibfnamefont {Y.}~\bibnamefont {Ota}}, \bibinfo {author} {\bibfnamefont
			{W.}~\bibnamefont {Malaeb}}, \bibinfo {author} {\bibfnamefont
			{H.}~\bibnamefont {Kanai}}, \bibinfo {author} {\bibfnamefont
			{Y.}~\bibnamefont {Nakashima}}, \emph {et~al.},\ }\bibfield  {title}
	{\bibinfo {title} {Quadratic fermi node in a 3d strongly correlated
			semimetal},\ }\href {https://www.nature.com/articles/ncomms10042} {\bibfield
		{journal} {\bibinfo  {journal} {Nature communications}\ }\textbf {\bibinfo
			{volume} {6}},\ \bibinfo {pages} {1} (\bibinfo {year} {2015})}\BibitemShut
	{NoStop}%
	\bibitem [{\citenamefont {Yan}\ and\ \citenamefont
		{de~Visser}(2014)}]{yan2014half}%
	\BibitemOpen
	\bibfield  {author} {\bibinfo {author} {\bibfnamefont {B.}~\bibnamefont
			{Yan}}\ and\ \bibinfo {author} {\bibfnamefont {A.}~\bibnamefont
			{de~Visser}},\ }\bibfield  {title} {\bibinfo {title} {Half-heusler
			topological insulators},\ }\href {https://doi.org/10.1557/mrs.2014.198}
	{\bibfield  {journal} {\bibinfo  {journal} {MRS Bulletin}\ }\textbf {\bibinfo
			{volume} {39}},\ \bibinfo {pages} {859} (\bibinfo {year} {2014})}\BibitemShut
	{NoStop}%
	\bibitem [{\citenamefont {Maier}\ \emph {et~al.}(2012)\citenamefont {Maier},
		\citenamefont {Oostinga}, \citenamefont {Knott}, \citenamefont {Br\"une},
		\citenamefont {Virtanen}, \citenamefont {Tkachov}, \citenamefont
		{Hankiewicz}, \citenamefont {Gould}, \citenamefont {Buhmann},\ and\
		\citenamefont {Molenkamp}}]{maier2012induced}%
	\BibitemOpen
	\bibfield  {author} {\bibinfo {author} {\bibfnamefont {L.}~\bibnamefont
			{Maier}}, \bibinfo {author} {\bibfnamefont {J.~B.}\ \bibnamefont {Oostinga}},
		\bibinfo {author} {\bibfnamefont {D.}~\bibnamefont {Knott}}, \bibinfo
		{author} {\bibfnamefont {C.}~\bibnamefont {Br\"une}}, \bibinfo {author}
		{\bibfnamefont {P.}~\bibnamefont {Virtanen}}, \bibinfo {author}
		{\bibfnamefont {G.}~\bibnamefont {Tkachov}}, \bibinfo {author} {\bibfnamefont
			{E.~M.}\ \bibnamefont {Hankiewicz}}, \bibinfo {author} {\bibfnamefont
			{C.}~\bibnamefont {Gould}}, \bibinfo {author} {\bibfnamefont
			{H.}~\bibnamefont {Buhmann}},\ and\ \bibinfo {author} {\bibfnamefont {L.~W.}\
			\bibnamefont {Molenkamp}},\ }\bibfield  {title} {\bibinfo {title} {Induced
			superconductivity in the three-dimensional topological insulator hgte},\
	}\href {https://link.aps.org/doi/10.1103/PhysRevLett.109.186806} {\bibfield
		{journal} {\bibinfo  {journal} {Phys. Rev. Lett.}\ }\textbf {\bibinfo
			{volume} {109}},\ \bibinfo {pages} {186806} (\bibinfo {year}
		{2012})}\BibitemShut {NoStop}%
	\bibitem [{\citenamefont {Liao}\ \emph {et~al.}(2018)\citenamefont {Liao},
		\citenamefont {Zang}, \citenamefont {Guan}, \citenamefont {Li}, \citenamefont
		{Gong}, \citenamefont {Zhu}, \citenamefont {Hu}, \citenamefont {Zhang},
		\citenamefont {Xu}, \citenamefont {Wang}, \citenamefont {He}, \citenamefont
		{Ma}, \citenamefont {Zhang},\ and\ \citenamefont {Xue}}]{Liao_np_2018}%
	\BibitemOpen
	\bibfield  {author} {\bibinfo {author} {\bibfnamefont {M.}~\bibnamefont
			{Liao}}, \bibinfo {author} {\bibfnamefont {Y.}~\bibnamefont {Zang}}, \bibinfo
		{author} {\bibfnamefont {Z.}~\bibnamefont {Guan}}, \bibinfo {author}
		{\bibfnamefont {H.}~\bibnamefont {Li}}, \bibinfo {author} {\bibfnamefont
			{Y.}~\bibnamefont {Gong}}, \bibinfo {author} {\bibfnamefont {K.}~\bibnamefont
			{Zhu}}, \bibinfo {author} {\bibfnamefont {X.-P.}\ \bibnamefont {Hu}},
		\bibinfo {author} {\bibfnamefont {D.}~\bibnamefont {Zhang}}, \bibinfo
		{author} {\bibfnamefont {Y.}~\bibnamefont {Xu}}, \bibinfo {author}
		{\bibfnamefont {Y.-Y.}\ \bibnamefont {Wang}}, \bibinfo {author}
		{\bibfnamefont {K.}~\bibnamefont {He}}, \bibinfo {author} {\bibfnamefont
			{X.-C.}\ \bibnamefont {Ma}}, \bibinfo {author} {\bibfnamefont {S.-C.}\
			\bibnamefont {Zhang}},\ and\ \bibinfo {author} {\bibfnamefont {Q.-K.}\
			\bibnamefont {Xue}},\ }\bibfield  {title} {\bibinfo {title}
		{Superconductivity in few-layer stanene},\ }\href
	{https://doi.org/10.1038/s41567-017-0031-6} {\bibfield  {journal} {\bibinfo
			{journal} {Nature Physics}\ }\textbf {\bibinfo {volume} {14}},\ \bibinfo
		{pages} {344} (\bibinfo {year} {2018})}\BibitemShut {NoStop}%
	\bibitem [{\citenamefont {Falson}\ \emph {et~al.}(2020)\citenamefont {Falson},
		\citenamefont {Xu}, \citenamefont {Liao}, \citenamefont {Zang}, \citenamefont
		{Zhu}, \citenamefont {Wang}, \citenamefont {Zhang}, \citenamefont {Liu},
		\citenamefont {Duan}, \citenamefont {He} \emph {et~al.}}]{falson2020type}%
	\BibitemOpen
	\bibfield  {author} {\bibinfo {author} {\bibfnamefont {J.}~\bibnamefont
			{Falson}}, \bibinfo {author} {\bibfnamefont {Y.}~\bibnamefont {Xu}}, \bibinfo
		{author} {\bibfnamefont {M.}~\bibnamefont {Liao}}, \bibinfo {author}
		{\bibfnamefont {Y.}~\bibnamefont {Zang}}, \bibinfo {author} {\bibfnamefont
			{K.}~\bibnamefont {Zhu}}, \bibinfo {author} {\bibfnamefont {C.}~\bibnamefont
			{Wang}}, \bibinfo {author} {\bibfnamefont {Z.}~\bibnamefont {Zhang}},
		\bibinfo {author} {\bibfnamefont {H.}~\bibnamefont {Liu}}, \bibinfo {author}
		{\bibfnamefont {W.}~\bibnamefont {Duan}}, \bibinfo {author} {\bibfnamefont
			{K.}~\bibnamefont {He}}, \emph {et~al.},\ }\bibfield  {title} {\bibinfo
		{title} {Type-ii ising pairing in few-layer stanene},\ }\href
	{https://www.science.org/doi/full/10.1126/science.aax3873} {\bibfield
		{journal} {\bibinfo  {journal} {Science}\ }\textbf {\bibinfo {volume}
			{367}},\ \bibinfo {pages} {1454} (\bibinfo {year} {2020})}\BibitemShut
	{NoStop}%
	\bibitem [{\citenamefont {Goll}\ \emph {et~al.}(2008)\citenamefont {Goll},
		\citenamefont {Marz}, \citenamefont {Hamann}, \citenamefont {Tomanic},
		\citenamefont {Grube}, \citenamefont {Yoshino},\ and\ \citenamefont
		{Takabatake}}]{Goll_pbcm_2008}%
	\BibitemOpen
	\bibfield  {author} {\bibinfo {author} {\bibfnamefont {G.}~\bibnamefont
			{Goll}}, \bibinfo {author} {\bibfnamefont {M.}~\bibnamefont {Marz}}, \bibinfo
		{author} {\bibfnamefont {A.}~\bibnamefont {Hamann}}, \bibinfo {author}
		{\bibfnamefont {T.}~\bibnamefont {Tomanic}}, \bibinfo {author} {\bibfnamefont
			{K.}~\bibnamefont {Grube}}, \bibinfo {author} {\bibfnamefont
			{T.}~\bibnamefont {Yoshino}},\ and\ \bibinfo {author} {\bibfnamefont
			{T.}~\bibnamefont {Takabatake}},\ }\bibfield  {title} {\bibinfo {title}
		{Thermodynamic and transport properties of the non-centrosymmetric
			superconductor {LaBiPt}},\ }\href
	{https://doi.org/10.1016/j.physb.2007.10.089} {\bibfield  {journal} {\bibinfo
			{journal} {Physica B: Condensed Matter}\ }\textbf {\bibinfo {volume}
			{403}},\ \bibinfo {pages} {1065} (\bibinfo {year} {2008})}\BibitemShut
	{NoStop}%
	\bibitem [{\citenamefont {Butch}\ \emph {et~al.}(2011)\citenamefont {Butch},
		\citenamefont {Syers}, \citenamefont {Kirshenbaum}, \citenamefont {Hope},\
		and\ \citenamefont {Paglione}}]{butch_prb_2011}%
	\BibitemOpen
	\bibfield  {author} {\bibinfo {author} {\bibfnamefont {N.~P.}\ \bibnamefont
			{Butch}}, \bibinfo {author} {\bibfnamefont {P.}~\bibnamefont {Syers}},
		\bibinfo {author} {\bibfnamefont {K.}~\bibnamefont {Kirshenbaum}}, \bibinfo
		{author} {\bibfnamefont {A.~P.}\ \bibnamefont {Hope}},\ and\ \bibinfo
		{author} {\bibfnamefont {J.}~\bibnamefont {Paglione}},\ }\bibfield  {title}
	{\bibinfo {title} {Superconductivity in the topological semimetal yptbi},\
	}\href {https://link.aps.org/doi/10.1103/PhysRevB.84.220504} {\bibfield
		{journal} {\bibinfo  {journal} {Phys. Rev. B}\ }\textbf {\bibinfo {volume}
			{84}},\ \bibinfo {pages} {220504} (\bibinfo {year} {2011})}\BibitemShut
	{NoStop}%
	\bibitem [{\citenamefont {Nakajima}\ \emph {et~al.}(2015)\citenamefont
		{Nakajima}, \citenamefont {Hu}, \citenamefont {Kirshenbaum}, \citenamefont
		{Hughes}, \citenamefont {Syers}, \citenamefont {Wang}, \citenamefont {Wang},
		\citenamefont {Wang}, \citenamefont {Saha}, \citenamefont {Pratt} \emph
		{et~al.}}]{Nakajima_sa_2015}%
	\BibitemOpen
	\bibfield  {author} {\bibinfo {author} {\bibfnamefont {Y.}~\bibnamefont
			{Nakajima}}, \bibinfo {author} {\bibfnamefont {R.}~\bibnamefont {Hu}},
		\bibinfo {author} {\bibfnamefont {K.}~\bibnamefont {Kirshenbaum}}, \bibinfo
		{author} {\bibfnamefont {A.}~\bibnamefont {Hughes}}, \bibinfo {author}
		{\bibfnamefont {P.}~\bibnamefont {Syers}}, \bibinfo {author} {\bibfnamefont
			{X.}~\bibnamefont {Wang}}, \bibinfo {author} {\bibfnamefont {K.}~\bibnamefont
			{Wang}}, \bibinfo {author} {\bibfnamefont {R.}~\bibnamefont {Wang}}, \bibinfo
		{author} {\bibfnamefont {S.~R.}\ \bibnamefont {Saha}}, \bibinfo {author}
		{\bibfnamefont {D.}~\bibnamefont {Pratt}}, \emph {et~al.},\ }\bibfield
	{title} {\bibinfo {title} {Topological r pdbi half-heusler semimetals: A new
			family of noncentrosymmetric magnetic superconductors},\ }\href
	{https://doi.org/10.1126/sciadv.1500242} {\bibfield  {journal} {\bibinfo
			{journal} {Science advances}\ }\textbf {\bibinfo {volume} {1}},\ \bibinfo
		{pages} {e1500242} (\bibinfo {year} {2015})}\BibitemShut {NoStop}%
\end{thebibliography}

\begin{thebibliography}{6}%
	\makeatletter
	\providecommand \@ifxundefined [1]{%
		\@ifx{#1\undefined}
	}%
	\providecommand \@ifnum [1]{%
		\ifnum #1\expandafter \@firstoftwo
		\else \expandafter \@secondoftwo
		\fi
	}%
	\providecommand \@ifx [1]{%
		\ifx #1\expandafter \@firstoftwo
		\else \expandafter \@secondoftwo
		\fi
	}%
	\providecommand \natexlab [1]{#1}%
	\providecommand \enquote  [1]{``#1''}%
	\providecommand \bibnamefont  [1]{#1}%
	\providecommand \bibfnamefont [1]{#1}%
	\providecommand \citenamefont [1]{#1}%
	\providecommand \href@noop [0]{\@secondoftwo}%
	\providecommand \href [0]{\begingroup \@sanitize@url \@href}%
	\providecommand \@href[1]{\@@startlink{#1}\@@href}%
	\providecommand \@@href[1]{\endgroup#1\@@endlink}%
	\providecommand \@sanitize@url [0]{\catcode `\\12\catcode `\$12\catcode
		`\&12\catcode `\#12\catcode `\^12\catcode `\_12\catcode `\%12\relax}%
	\providecommand \@@startlink[1]{}%
	\providecommand \@@endlink[0]{}%
	\providecommand \url  [0]{\begingroup\@sanitize@url \@url }%
	\providecommand \@url [1]{\endgroup\@href {#1}{\urlprefix }}%
	\providecommand \urlprefix  [0]{URL }%
	\providecommand \Eprint [0]{\href }%
	\providecommand \doibase [0]{https://doi.org/}%
	\providecommand \selectlanguage [0]{\@gobble}%
	\providecommand \bibinfo  [0]{\@secondoftwo}%
	\providecommand \bibfield  [0]{\@secondoftwo}%
	\providecommand \translation [1]{[#1]}%
	\providecommand \BibitemOpen [0]{}%
	\providecommand \bibitemStop [0]{}%
	\providecommand \bibitemNoStop [0]{.\EOS\space}%
	\providecommand \EOS [0]{\spacefactor3000\relax}%
	\providecommand \BibitemShut  [1]{\csname bibitem#1\endcsname}%
	\let\auto@bib@innerbib\@empty
	\bibitem [{\citenamefont {Kitaev}(2001)}]{Kitaev_pu_2001}%
	\BibitemOpen
	\bibfield  {author} {\bibinfo {author} {\bibfnamefont {A.~Y.}\ \bibnamefont
			{Kitaev}},\ }\href {https://doi.org/10.1070/1063-7869/44/10s/s29} {\bibfield
		{journal} {\bibinfo  {journal} {Physics-Uspekhi}\ }\textbf {\bibinfo {volume}
			{44}},\ \bibinfo {pages} {131} (\bibinfo {year} {2001})}\BibitemShut
	{NoStop}%
	\bibitem [{\citenamefont {Budich}\ and\ \citenamefont
		{Ardonne}(2013)}]{budich_prb_2013}%
	\BibitemOpen
	\bibfield  {author} {\bibinfo {author} {\bibfnamefont {J.~C.}\ \bibnamefont
			{Budich}}\ and\ \bibinfo {author} {\bibfnamefont {E.}~\bibnamefont
			{Ardonne}},\ }\href {https://link.aps.org/doi/10.1103/PhysRevB.88.075419}
	{\bibfield  {journal} {\bibinfo  {journal} {Phys. Rev. B}\ }\textbf {\bibinfo
			{volume} {88}},\ \bibinfo {pages} {075419} (\bibinfo {year}
		{2013})}\BibitemShut {NoStop}%
	\bibitem [{\citenamefont {Alexandradinata}\ \emph {et~al.}(2014)\citenamefont
		{Alexandradinata}, \citenamefont {Dai},\ and\ \citenamefont
		{Bernevig}}]{aris2014wilson}%
	\BibitemOpen
	\bibfield  {author} {\bibinfo {author} {\bibfnamefont {A.}~\bibnamefont
			{Alexandradinata}}, \bibinfo {author} {\bibfnamefont {X.}~\bibnamefont
			{Dai}},\ and\ \bibinfo {author} {\bibfnamefont {B.~A.}\ \bibnamefont
			{Bernevig}},\ }\href {https://doi.org/10.1103/PhysRevB.89.155114} {\bibfield
		{journal} {\bibinfo  {journal} {Phys. Rev. B}\ }\textbf {\bibinfo {volume}
			{89}},\ \bibinfo {pages} {155114} (\bibinfo {year} {2014})}\BibitemShut
	{NoStop}%
	\bibitem [{\citenamefont {Yang}\ \emph {et~al.}(2015)\citenamefont {Yang},
		\citenamefont {Morimoto},\ and\ \citenamefont {Furusaki}}]{yang2015topo}%
	\BibitemOpen
	\bibfield  {author} {\bibinfo {author} {\bibfnamefont {B.-J.}\ \bibnamefont
			{Yang}}, \bibinfo {author} {\bibfnamefont {T.}~\bibnamefont {Morimoto}},\
		and\ \bibinfo {author} {\bibfnamefont {A.}~\bibnamefont {Furusaki}},\ }\href
	{https://doi.org/10.1103/PhysRevB.92.165120} {\bibfield  {journal} {\bibinfo
			{journal} {Phys. Rev. B}\ }\textbf {\bibinfo {volume} {92}},\ \bibinfo
		{pages} {165120} (\bibinfo {year} {2015})}\BibitemShut {NoStop}%
	\bibitem [{\citenamefont {Zhang}\ \emph {et~al.}(2020)\citenamefont {Zhang},
		\citenamefont {Hsu},\ and\ \citenamefont {Das~Sarma}}]{zhang2020higher}%
	\BibitemOpen
	\bibfield  {author} {\bibinfo {author} {\bibfnamefont {R.-X.}\ \bibnamefont
			{Zhang}}, \bibinfo {author} {\bibfnamefont {Y.-T.}\ \bibnamefont {Hsu}},\
		and\ \bibinfo {author} {\bibfnamefont {S.}~\bibnamefont {Das~Sarma}},\ }\href
	{https://doi.org/10.1103/PhysRevB.102.094503} {\bibfield  {journal} {\bibinfo
			{journal} {Phys. Rev. B}\ }\textbf {\bibinfo {volume} {102}},\ \bibinfo
		{pages} {094503} (\bibinfo {year} {2020})}\BibitemShut {NoStop}%
	\bibitem [{\citenamefont {Sancho}\ \emph {et~al.}(1985)\citenamefont {Sancho},
		\citenamefont {Sancho}, \citenamefont {Sancho},\ and\ \citenamefont
		{Rubio}}]{sancho_JPF_1985}%
	\BibitemOpen
	\bibfield  {author} {\bibinfo {author} {\bibfnamefont {M.~L.}\ \bibnamefont
			{Sancho}}, \bibinfo {author} {\bibfnamefont {J.~L.}\ \bibnamefont {Sancho}},
		\bibinfo {author} {\bibfnamefont {J.~L.}\ \bibnamefont {Sancho}},\ and\
		\bibinfo {author} {\bibfnamefont {J.}~\bibnamefont {Rubio}},\ }\href
	{https://dx.doi.org/10.1088/0305-4608/15/4/009} {\bibfield  {journal}
		{\bibinfo  {journal} {Journal of Physics F: Metal Physics}\ }\textbf
		{\bibinfo {volume} {15}},\ \bibinfo {pages} {851} (\bibinfo {year}
		{1985})}\BibitemShut {NoStop}%
\end{thebibliography}

%

\clearpage
\onecolumngrid

\begin{center}
	\bf\large	Supplemental Material for ``Topological Superconducting Vortex From Trivial Electronic Bands"
\end{center}


\section{Supplementary Note 1: Topological Invariants of Vortex Lines}

In this part, we provide mathematical expressions of the $\mathbb{Z}_2$ and $\mathbb{Z}$-type topological invariants defined for the quasi one-dimensional (1D) vortex line Hamiltonian in the main text.

\subsection{1.1 $\mathbb{Z}_2$ Topological Invariant}

For $C_n$-symmetric vortex lines,  $\mathbb{Z}_2$ topological invariant $\nu_{J_z}$ characterizes the gapped vortex-line topology of Caroli-de Gennes-Matricon (CdGM) states that belong to a particle-hole symmetry (PHS) invariant angular momentum sector, i.e., $J_z=0$ or $J_z=n/2$ for spin-singlet s-wave pairing. Therefore, $\nu_{J_z}$ is exactly the $\mathbb{Z}_2$ topological invariant for 1D class D systems but with an additional $J_z$ index. In this case, the quasi-1D system is always fully gapped without a topological phase transition. Following Ref.~\cite{Kitaev_pu_2001}, under the Majorana basis, the vortex-line Hamiltonian matrix ${\cal H_M}^{(J_z)}(k_z)$ for CdGM states in the $J_z$ sector is antisymmetric and that is why its Pfaffian is well-defined. The $\mathbb{Z}_2$ topological invariant is thus defined as 
\begin{equation}
	\nu_{J_z} = \text{sgn}\{\text{Pf}[{\cal H_M}^{(J_z)}(k_z=0)] \} \text{sgn}\{\text{Pf}[{\cal H_M}^{(J_z)}(k_z=\pi)] \} \in \mathbb{Z}_2.
	\label{eq-sm-pfaffian}
\end{equation}
The equivalence between the Pfaffian invariant in the Majorana representation and the quantized Berry phase of the occupied BdG bands in the Nambu basis has been established~\cite{budich_prb_2013}. As a result, $\nu_{J_z}$ can be further expressed as 
\begin{equation}
	\nu_{J_z} = \frac{1}{2\pi} \text{tr} \int_{-\pi}^{\pi} {\cal A}^{(J_z)} (k_z) dk_z,
\end{equation}
where the non-Abelian Berry connection ${\cal A}^{(J_z)}_{nm} (k_z) = i\langle u_n^{(J_z)}| \partial_{k_z} | u_m^{(J_z)} \rangle$ is defined for all occupied CdGM bands carrying $J_z$. 
The above Berry phase formula can be further simplified if the 1D CdGM system features additional out-of-plane mirror symmetry ${\cal M}_z$. Notice that the superconducting vortex line is aligned along z-axis. To unambiguously extract the value of $\nu_{J_z}$, we just need the knowledge of the pattern of symmetry eigenvalues for the BdG occupied bands at high-symmetry momenta $k_z=0,\pi$~\cite{aris2014wilson}. In particular, let us define $m_{J_z,-}(0)$ and $m_{J_z,-}(\pi)$ as the number of occupied BdG bands at $k_z=0$ and $k_z=\pi$, respectively, with a ${\cal M}_z=-1$ label, then we have 
\begin{equation}
	\nu_{J_z} = m_{J_z,-}(0) - m_{J_z,-} (\pi),\ \ \ \text{mod }2.
\end{equation}
This symmetry-based expression of $\nu_{J_z}$ aligns with the spirit of symmetry indicator theory.

\subsection{1.2 $C_n$ Topological Charge}

In addition to the above $\mathbb{Z}_2$ topological invariant, we also define the $C_n$ topological charge ${\cal Q}_{J_z}\in \mathbb{Z}$ that will indicate the number of symmetry-protected Dirac nodal crossings in the quasi-1D vortex-line spectrum. Our definition is similar to the topological charges defined for 3D Dirac semimetals~\cite{yang2015topo} and for 3D Dirac superconductors~\cite{zhang2020higher}. 

\begin{figure*}[!htbp]
	\centering
	\includegraphics[width=0.8\linewidth]{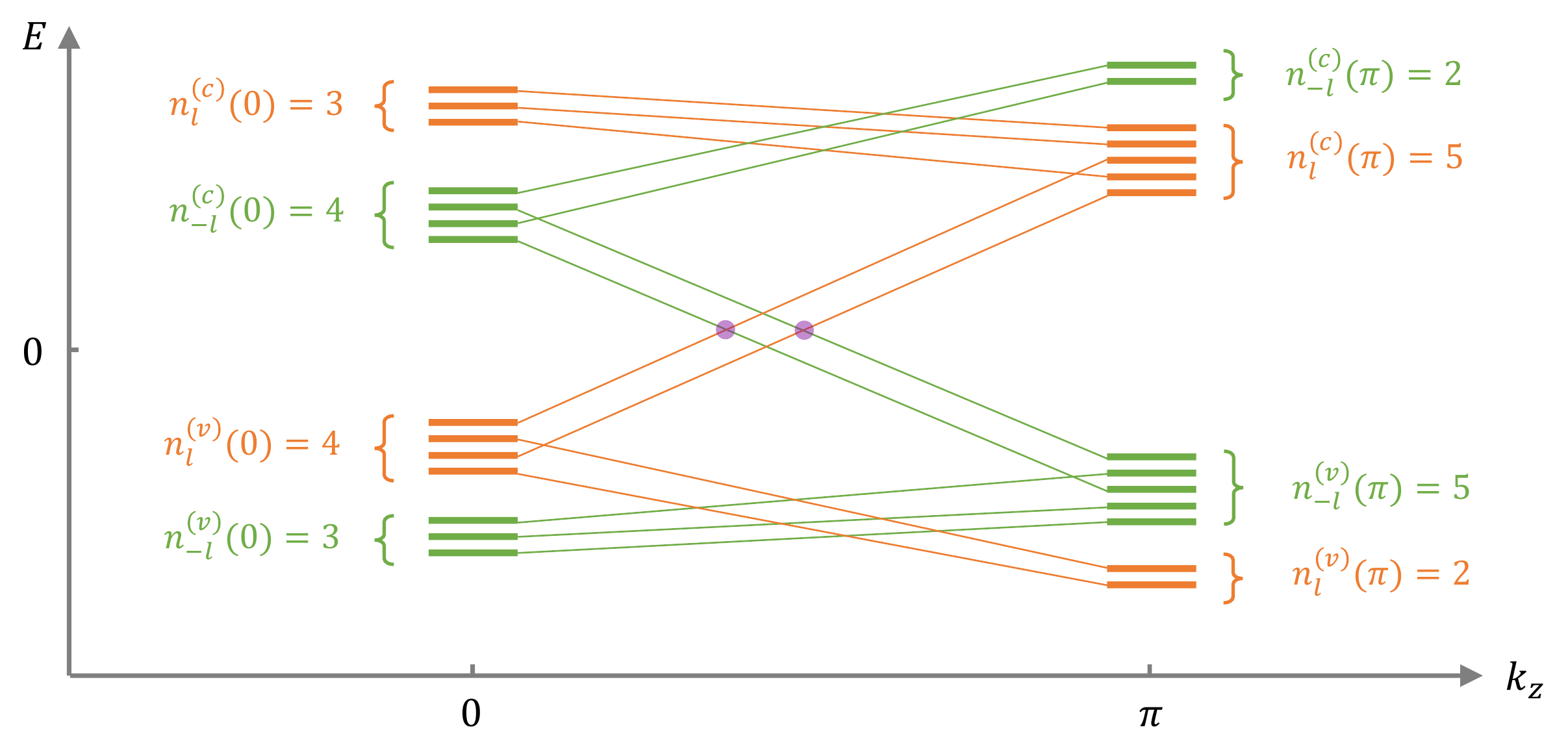}
	\caption{Symmetry data counting and $C_n$ topological charge ${\cal Q}_{J_z}$. The pattern of symmetry eigenvalues must satisfy both the particle-hole symmetry and the band connectivity relation. In the band configuration shown here, ${\cal Q}_l = n_{l}^{(v)} (0) - n_{l}^{(v)} (\pi) = 4 - 2 = 2$, indicating the existence of two 1D Dirac nodes denoted by the purple dots.}
	\label{sm-fig1}
\end{figure*}

Here, ${\cal Q}_{J_z}$ are defined for $J_z$ sectors that are not PHS-invariant. PHS generally flips $J_z$ to $-J_z$, forming a pair of PHS-related $J_z$ sectors. Namely, if there exists a $J_z$-labeled CdGM state at $k_z$ with an energy $E$, PHS mandates the existence of another partner state at $-k_z$ and energy $-E$, which is $-J_z$ labeled. The effective vortex Hamiltonian $h_\text{vort} (k_z)$ is generally gapped at $k_z=0, \pi$ and we will focus on the occupied CdGM states with $E<0$ at these high-symmetry momenta. Now we define $n_{J_z}^{(\alpha)}(k_i)$ as the number of occupied (unoccupied) $J_z$-labeled CdGM states at high-symmetry momentum $k_z=k_i$ (e.g.,~$k_i=0,\pi$) with $\alpha=v$ ($\alpha=c$). The $C_n$ symmetry charge is defined as
\begin{eqnarray}
	{\cal Q}_{J_z} \equiv n_{J_z}^{(v)} (0) - n_{J_z}^{(v)} (\pi) ,
	\label{eq_Cn_charge}
\end{eqnarray}
for $J_z=1$ for $C_{3,4}$ and $J_z=1,2$ for $C_6$.  Because of the connectivity of energy bands, 
\begin{equation}
	n_{J_z}^{(c)} (0) + n_{J_z}^{(v)} (0) = n_{J_z}^{(c)} (\pi) + n_{J_z}^{(v)} (\pi).
	\label{eq-band-connectivity}
\end{equation}
Meanwhile, PHS requires that $n_{J_z}^{(c)} (k_i) = n_{-J_z}^{(v)} (k_i)$ and $n_{J_z}^{(v)} (k_i) = n_{-J_z}^{(c)} (k_i)$. It is then easy to show that equivalently,
\begin{equation}
	{\cal Q}_{J_z} \equiv  n_{J_z}^{(c)} (\pi) - n_{J_z}^{(c)} (0) = n_{-J_z}^{(c)} (0) - n_{-J_z}^{(c)} (\pi) = n_{-J_z}^{(v)} (\pi) - n_{-J_z}^{(v)} (0).
\end{equation}
$|{\cal Q}_{J_z}|$ determines the number of $C_n$-protected 1D Dirac nodes from $k_z=0$ to $k_z=\pi$, which is also the number of pairs of 1D Dirac nodes in the CdGM spectrum. 
These nodes can not be removed without (i) breaking $C_n$ symmetry; and/or (ii) closing the energy gap at $k_z=0,\pi$. As schematically shown in Fig.~\ref{sm-fig1}, the physical meaning of Eq.~\eqref{eq_Cn_charge} can be understood as follows:
\begin{enumerate}
	\item[(i)] Consider a PHS-related sector $(l,-l)$ and assume $n_{l}^{(v)} (0) = n, n_{l}^{(v)} (\pi) = m$, and $n_{-l}^{(v)} (0) = n'$. PHS requires $n_{-l}^{(c)} (0) = n, n_{l}^{(c)} (0) = n', n_{-l}^{(c)} (\pi) = m$.
	\item[(ii)] Eq.~\eqref{eq-band-connectivity} requires $n_{l}^{(c)}(\pi) = n_{-l}^{(v)}(\pi) = n+n'-m$. 
	\item[(iii)] To ensure the connectivity of the bands, there must be $N_l$ number of $l$-indexed bands starting from the occupied bands at $k_z=0$, crossing the zero energy, and ending at the conduction bands at $k_z=\pi$, where
	\begin{equation}
		N_l \equiv {\cal Q}_l = n_l^{(v)}(0) - n_l^{(v)}(\pi) = n-m.
	\end{equation}
	Note that $N_l$ is exactly our choice of $C_n$ topological charge. Similarly, one can find $N_{-l} = n_{-l}^{(v)}(0) - n_{-l}^{(v)}(\pi) = m-n = -N_l$. If either $N_l$ or $N_{-l}$ is negative, this indicates the existence of left-moving modes along $\Gamma-Z$, instead of right-moving ones.
	\item[(iv)] As a result, there are $|{\cal Q}_l|$ pairs of left-movers and right-movers along $\Gamma-Z$. They together form $|{\cal Q}_l|$ 1D Dirac nodes that are $C_n$-protected. 
\end{enumerate}
For the example shown in Fig.~\ref{sm-fig1}, we have $n=4$ and $m=2$, and this is why there are $|n-m|=2$ Dirac nodes.

We note that when a pair of CdGM bands get inverted around a generic momentum $k_z\neq 0, \pi$, they can contribute to an additional pair of Dirac nodes that are not captured by ${\cal Q}_{J_z}$. These Dirac nodes, however, can be eliminated without closing the energy gap at $k_z=0, \pi$. As a result, gapping these Dirac nodes will not lead to any topologically gapped state like a Kitaev vortex state. Therefore, we do not term vortices carrying these Dirac nodes as ``topological" nodal vortex and we leave a discussion of these states to future works. 

\section{Supplementary Note 2: Vortex in Luttinger Semimetals: Numerical Results}
\label{sec-LSM}

In this section, we introduce the model Hamiltonian of a generalized Luttinger semimetal and study its vortex topological phase diagram.

\subsection{2.1 Luttinger Semimetal from 6-band Kane Model}

We first introduce the spin-$\frac{3}{2}$ matrices
\begin{equation}\label{sm-eq-jxjyjz}
	J_x=\begin{pmatrix}
		0 & \frac{\sqrt{3}}{2} & 0 & 0 \\
		\frac{\sqrt{3}}{2} & 0 & 1 & 0 \\
		0 & 1 & 0 & \frac{\sqrt{3}}{2} \\
		0 & 0 & \frac{\sqrt{3}}{2} & 0 \\
	\end{pmatrix},\ \ 
	J_y=\begin{pmatrix}
		0 & -\frac{\sqrt{3}i}{2} & 0 & 0 \\
		\frac{\sqrt{3}i}{2} & 0 & -i & 0 \\
		0 & i & 0 & -\frac{\sqrt{3}i}{2} \\
		0 & 0 & \frac{\sqrt{3}i}{2} & 0 \\
	\end{pmatrix},\ \ 
	J_z=\begin{pmatrix}
		\frac{3}{2} & 0 & 0 & 0 \\
		0 & \frac{1}{2} & 0 & 0 \\
		0 & 0 & -\frac{1}{2} & 0 \\
		0 & 0 & 0 & -\frac{3}{2} \\
	\end{pmatrix}.
\end{equation}
It is easy to check that $[J_i, J_j]=i\epsilon_{ijk}J_k$. Then the isotropic Luttinger Hamiltonian formed by the $\Gamma_8$ bands is
\begin{eqnarray}\label{sm-eq-lsm-ham0}
	h_8 ({\bf k}) &=& (\lambda_1 + \frac{5}{2} \lambda_2) {\bf k}^2 - 2 \lambda_2 ({\bf k \cdot J})^2 \nonumber \\
	&=& -\lambda_1 {\bf k}^2 +
	\begin{pmatrix}
		\lambda_2(k_x^2+k_y^2-2k_z^2) & -2\sqrt{3}\lambda_2 k_z k_- & -\sqrt{3}\lambda_2 k_-^2 & 0 \\
		-2\sqrt{3}\lambda_2 k_z k_+	 & -\lambda_2(k_x^2+k_y^2-2k_z^2) & 0 & -\sqrt{3}\lambda_2 k_-^2 \\
		-\sqrt{3}\lambda_2 k_+^2 & 0 & -\lambda_2(k_x^2+k_y^2-2k_z^2) & 2\sqrt{3}\lambda_2 k_z k_- \\
		0 & -\sqrt{3}\lambda_2 k_+^2 & 2\sqrt{3}\lambda_2 k_z k_+ & \lambda_2(k_x^2+k_y^2-2k_z^2) \\
	\end{pmatrix}.
\end{eqnarray}
The Hamiltonian form in terms of gamma matrices are defined in the main text. To fully incorporate the band topology of relevant quantum materials, we need to generalize the Luttinger model into a 6-band Kane model by including the $\Gamma_6$ bands. Therefore, we have
\begin{equation}
	H_{\text Kane} = \begin{pmatrix}
		h_6 ({\bf k}) & T({\bf k}) \\
		T^{\dagger}({\bf k}) & h_8 ({\bf k}) \\
	\end{pmatrix},
\end{equation}
where $h_6 = (E_c + \lambda_3 k^2) \sigma_0$ and 
\begin{equation}
	T({\bf k})=v\begin{pmatrix}
		-\frac{1}{\sqrt{2}} k_+ & \sqrt{\frac{2}{3}} k_z & \frac{1}{\sqrt{6}} k_- & 0 \\
		0 & -\frac{1}{\sqrt{6}} k_+ & \sqrt{\frac{2}{3}} k_z & \frac{1}{\sqrt{2}} k_- \\
	\end{pmatrix}
\end{equation}
Note that $H_\text{Kane}$ is essentially the same as $H_\text{Kane}(1, {\bf k})$, but in a slightly different form. To identify the conditions for Luttinger semimetallic phase in the Kane model, we now project everything onto the $\Gamma_8$ bases, following
\begin{equation}
	H_\text{eff} ({\bf k}) = h_8 - T^{\dagger} h_6^{-1} T.
\end{equation}
As required by the $O(3)$ symmetry, the effective Hamiltonian must take the same form as $h_8$ but the band parameters will get renormalized accordingly, where
\begin{equation}
	\lambda_1 \rightarrow \lambda_1' = \lambda_1 + \frac{v^2}{3 E_c},\ \ \  \lambda_2 \rightarrow \lambda_2' = \lambda_2 - \frac{v^2}{6 E_c}.
\end{equation}
In this case, the energy spectrum for $H_\text{eff}$ is given by
\begin{align}\label{sm-eq-proj-lsm-band}
	\begin{split}
		E_{\frac{1}{2}} &= (-\lambda_1' + 2\lambda_2' ) k^2  = (-\lambda_1 + 2\lambda_2 - \frac{2v^2}{3E_c}) k^2 , \\
		E_{\frac{3}{2}} &= (-\lambda_1' - 2\lambda_2') k^2 =(-\lambda_1 - 2\lambda_2) k^2.
	\end{split}
\end{align}

For HgTe-class materials, $\Gamma_6$ and $\Gamma_8$ bands are electron-like and hole-like, respectively, leading to $\lambda_1>0, \lambda_2<0, \lambda_3>0$ and $\lambda_1>-2\lambda_2$. Meanwhile, the $\Gamma_6$-$\Gamma_8$ inversion requires $E_c<0$. Therefore, to achieve a semimetallic phase, $E_{\frac{1}{2}}$ must play the role of electron bands and $E_{\frac{3}{2}}$ will be the hole bands. As a result, the LSM condition is 	
\begin{eqnarray}
	&& \lambda_1>0,\  \lambda_2<0,\  \lambda_3>0,\ E_c<0 \nonumber \\
	&& -\lambda_1-2\lambda_2<0,\ -\lambda_1 + 2\lambda_2 - \frac{2v^2}{3E_c}>0.
\end{eqnarray}

Next, let us estimate the above projection parameters for the six-band Kane model, whose parameters are given by
\begin{align}
	\lambda_1 = \frac{4.1 P^2}{18.8}, \lambda_2 = -\frac{0.5 P^2}{18.8}, E_c = -0.303, \lambda_3 = \frac{P^2}{18.8}, v=P.
\end{align}
Note these parameters are all in unit of energy [eV]. Here $ a_0=6.46$ \AA \, is the in-plane lattice constant and $P = 8.46/a_0$. Now we use Eq.~\eqref{sm-eq-proj-lsm-band}, the parameters for the projected LSM around the $\Gamma$ point are given by
\begin{align}
	\lambda_1 ' = -1.51271, \lambda_2' = 0.897757.
\end{align}
Thus, the diagonal term for the $J_z=\pm 3/2$ bands reads 
\begin{align}
	-(\lambda_1+\frac{v^2}{3E_c} )(k_x^2+k_y^2+k_z^2 )+(\lambda_2-\frac{v^2}{6E_c} )(k_x^2+k_y^2-2k_z^2 ) \approx 2.41(k_x^2+k_y^2 )-3.3k_z^2.
\end{align}
We note that the sign is opposite with those for the LSM model used in the main text and Supplementary Note 2, where $-(k_x^2+k_y^2)+2k_z^2$ is used for the numerical simulation. 
This sign difference could directly give rise to the opposite vortex band dispersion between Fig. 1 and Fig. 3 in the main text. In summary, 
\begin{align}
	\begin{cases}
		\text{Fig.~1: LSM model,} -(k_x^2+k_y^2)+2k_z^2,  \text{the vortex-band with } J_z=+1 \text{ is a hole-like band}, \\
		\text{Fig.~3: Kane model, } 2.41(k_x^2+k_y^2 )-3.3k_z^2, \text{the vortex-band with } J_z=+1 \text{ is a electron-like band}.
	\end{cases}
\end{align}
The opposite effective mass due to the sign switching will be explicitly shown after deriving the low-energy vortex Hamiltonian from the perturbation theory in the Supplementary Note 3.
However, the physics we want to address in the main text would not be affected (see Eq. 4). No matter $m_1$ is positive or negative, the vortex phase of a LSM is a Kitaev$\oplus$Nodal phase, thus, the vortex phase of a Kane model is nodal because of Eq. 4. Therefore, the choice of parameters of LSM completely does not affect our conclusion.


\subsection{2.2 Bogoliubov-de Gennes Hamiltonian}
We now discuss in details the Bogoliubov-de Gennes Hamiltonian of LSM and the numerical mapping of its vortex topological phase diagram. As shown in the main text, Hamiltonian for a general anisotropic LSM consists of four $\Gamma_8$ bands,
\begin{align}
	\mathcal{H}_{\text{LSM}} = \lambda_1 k^2\gamma_0 + M({\bf k})\gamma_5 + v_z k_z (k_x \gamma_{45}+ k_y \gamma_{35}) - \sqrt{3}\lambda_2 ((k_x^2-k_y^2)\gamma_{25} + 2k_xk_y\gamma_{15}).
\end{align} 
Here, $M({\bf k})=m_1(k_x^2+k_y^2)+m_2k_z^2$ and the $4\times4$ $\gamma$-matrices are defined as 
\begin{align}
	\gamma_1 = \sigma_x \otimes s_z,\ \gamma_2 = \sigma_y \otimes s_z,\ \gamma_3 = \sigma_0 \otimes s_x,\ \gamma_4 = \sigma_0 \otimes s_y,\ \gamma_5 = \sigma_z \otimes s_z 
\end{align}
with $\gamma_{mn}=-i\gamma_m\gamma_n$ and $\gamma_0=\sigma_0\otimes s_0$ the identity matrix. ${\cal H}_\text{LSM}$ satisfies time-reversal symmetry $\Theta = i\gamma_{13} \mathcal{K}$ with $\mathcal{K}$ being the complex conjugate, as well as an out-of-plane mirror symmetry $\mathcal{M}_z = i\gamma_5$.
Without loss of generality, we choose $\lambda_1=0$ for simplicity. Similar to the Kane model discussed in the main text, we further include the effect of lattice strain described by
\begin{align}
	\mathcal{H}_{\text{str}} \triangleq \Sigma_{\text{str}}\gamma_{5} = \Sigma_{\text{str}} \begin{pmatrix}
		1 & & & \\
		& -1 & & \\
		& & -1 & \\
		& & & 1 \\
	\end{pmatrix}.
\end{align} 
$\Sigma_{\text{str}}$ will become an important tuning parameter in our vortex topological phase diagram, as will be shown soon.

We now turn on an isotropic $s$-wave spin-singlet pairing potential and consider a vortex-line configuration along the $z$-axis. The corresponding Bogoliubov de-Gennes Hamiltonian (i.e., Eq.~(1) in the main text) is given by
\begin{equation}\label{sm-eq-bdg-ham-lsm}
	\mathcal{H}_\text{BdG} = \begin{pmatrix}
		\mathcal{H}_{\text{LSM}}(\mathbf{k})-\mu  & \mathcal{H}_\Delta \\
		\mathcal{H}_\Delta^\dagger & \mu-\mathcal{H}_{\text{LSM}}^\ast(-\mathbf{k})
	\end{pmatrix}, 
\end{equation}
where $\mu$ is the chemical potential. The pairing function is captured by $\mathcal{H}_\Delta= i\Delta(\mathbf{r})\gamma_{13}$. Apparently, ${\cal H}_\text{BdG}$ carries a trivial BdG bulk topology because of the s-wave pairing. The Nambu basis for ${\cal H}_\text{BdG}$ is 
\begin{eqnarray}\label{sm-eq-basis-1-bdg}
	|\Psi_{\text{BdG}} \rangle = \left\{ |\frac{3}{2} \uparrow\rangle_e, |\frac{1}{2} \downarrow\rangle_e, |-\frac{1}{2} \uparrow\rangle_e, |-\frac{3}{2} \downarrow\rangle_e, |\frac{3}{2} \uparrow\rangle_h, |\frac{1}{2} \downarrow\rangle_h, |-\frac{1}{2} \uparrow\rangle_h, |-\frac{3}{2} \downarrow\rangle_h \right\}^T, 
\end{eqnarray}
where the atomic basis with a subscript e or h carries a crystal momentum ${\bf k}$ or $-{\bf k}$, respectively. In particular, the particle-hole symmetry 
\begin{equation}
	\Xi |J_z, s\rangle_e \rightarrow |-J_z,-s\rangle_h,
\end{equation}
and a constant pairing term between $|J_z, s\rangle_e$ and $|J_z, s\rangle_h$ describes a spin-singlet s-wave Cooper pairing in our notation.
Under this basis, we have
\begin{equation}\label{sm-eq-bdg-ham-lsm}
	\mathcal{H}_\text{BdG} = \begin{pmatrix}
		F_1 + F_2 & v k_z k_- & \tilde{v} k_-^2 & 0 & \Delta e^{i\theta} & 0 & 0 & 0 \\
		v k_z k_+ & F_1 - F_2  & 0 & \tilde{v} k_-^2 & 0 & -\Delta e^{i\theta} & 0 & 0 \\
		\tilde{v} k_+^2 & 0 & F_1 - F_2  & -v k_z k_- & 0 & 0 & \Delta e^{i\theta} & 0 \\
		0 & \tilde{v} k_+^2 & -v k_z k_+ & F_1 + F_2  & 0 & 0 & 0 & -\Delta e^{i\theta} \\
		\Delta e^{-i\theta} & 0 & 0 & 0 & -F_1 -F_2  & v k_z k_- & -\tilde{v} k_-^2 & 0 \\
		0 & -\Delta e^{-i\theta} & 0 & 0 & v k_z k_+ & -F_1 + F_2  & 0 & -\tilde{v} k_-^2 \\
		0 & 0 & \Delta e^{-i\theta} & 0 & -\tilde{v} k_+^2 & 0 & -F_1 + F_2  & -v k_z k_- \\
		0 & 0 & 0 & -\Delta e^{-i\theta} & 0  & -\tilde{v}k_+^2 & -vk_z k_+ & -F_1 - F_2 \\
	\end{pmatrix},
\end{equation}
where $\tilde{v}=\sqrt{3}\lambda_2$ is used for short.
Here $(r,\theta)$ describe the in-plane polar coordinates and $k_z$ remains a good quantum number. The vortex line centering at $r=0$ is described by $\Delta(\mathbf{r})=\Delta_0\tanh(r/\xi_0) e^{i\theta}$, where $\xi_0$ is the SC coherence length.
Here $F_1=\lambda_1 {\bf k}^2 - \mu, F_2 = \Sigma_{\text{str}}+ \lambda_2 (k_x^2+k_y^2 - 2k_z^2)$. We define
\begin{align}
	\Xi = \tau_x \sigma_x s_x \mathcal{K} \triangleq \left(\begin{array}{cccc|cccc}
		0 & 0 & 0 & 0 & 0 & 0 & 0 & 1 \\
		0 & 0 & 0 & 0 & 0 & 0 & 1 & 0 \\
		0 & 0 & 0 & 0 & 0 & 1 & 0 & 0 \\
		0 & 0 & 0 & 0 & 1 & 0 & 0 & 0 \\ \hline
		0 & 0 & 0 & 1 & 0 & 0 & 0 & 0 \\
		0 & 0 & 1 & 0 & 0 & 0 & 0 & 0 \\
		0 & 1 & 0 & 0 & 0 & 0 & 0 & 0 \\
		1 & 0 & 0 & 0 & 0 & 0 & 0 & 0 \\
	\end{array} \right)  \mathcal{K},\ \ \ \ 
	U_{\text{BdG}} =	\left( \begin{array}{cccccccc}
		1 & 0 & 0 & 0 & 0 & 0 & 0 & 0 \\
		0 & 0 & 0 & 0 & 0 & 0 & 1 & 0 \\
		0 & 0 & 0 & 0 & 1 & 0 & 0 & 0 \\
		0 & 0 & 1 & 0 & 0 & 0 & 0 & 0 \\
		0 & 1 & 0 & 0 & 0 & 0 & 0 & 0 \\
		0 & 0 & 0 & 0 & 0 & 0 & 0 & 1 \\
		0 & 0 & 0 & 0 & 0 & 1 & 0 & 0 \\
		0 & 0 & 0 & 1 & 0 & 0 & 0 & 0 \\
	\end{array}	\right),
\end{align}
where $\Xi$ is the operation of particle-hole symmetry (PHS). $U_\text{BdG}$ is a unitary transformation transforming the original Nambu basis in Eq.~\eqref{sm-eq-basis-1-bdg} into a new basis $|\Psi_{\text{BdG}} \rangle' = U_{\text{BdG}} |\Psi_{\text{BdG}} \rangle $, with 
\begin{align}\label{sm-eq-new-basis-bdg-lsm}
	|\Psi_{\text{BdG}} \rangle' = \left\{ |\frac{3}{2} \uparrow\rangle_e,  |-\frac{1}{2} \uparrow\rangle_h,  |\frac{3}{2} \uparrow\rangle_h,   |-\frac{1}{2} \uparrow\rangle_e,  |\frac{1}{2} \downarrow\rangle_e,   |-\frac{3}{2} \downarrow\rangle_h,  |\frac{1}{2} \downarrow\rangle_h,  |-\frac{3}{2} \downarrow\rangle_e    \right\}.
\end{align}
Under this transformation, the new Hamiltonian becomes
\begin{align}\label{sm-eq-bdg-new-basis}
	\mathcal{H}_{\text{BdG}} = H_0({\bf k}_\parallel) + H_1({\bf k}_\parallel, k_z), 
\end{align}
with
\begin{align}\label{sm-eq-ham0-new-basis}
	H_0({\bf k}_\parallel) = \left(
	\begin{array}{cc|cc|cc|cc}
		0 & 0 & \Delta  e^{i \theta } &  \tilde{v} k_-^2  & 0 & 0 & 0 & 0 \\
		0 & 0 & -\tilde{v} k_+^2  & \Delta  e^{-i \theta } & 0 & 0 & 0 & 0 \\ \hline
		\Delta  e^{-i \theta } & -\tilde{v} k_-^2 & 0 & 0 & 0 & 0 & 0 & 0 \\
		\tilde{v} k_+^2  & \Delta  e^{i \theta } & 0 & 0 & 0 & 0 & 0 & 0 \\ \hline
		0 & 0 & 0 & 0 & 0 & 0 & - \Delta e^{i \theta } &  \tilde{v}k_-^2  \\
		0 & 0 & 0 & 0 & 0 & 0 & -\tilde{v} k_+^2  & -\Delta e^{-i \theta } \\ \hline
		0 & 0 & 0 & 0 & -\Delta e^{-i \theta } & -\tilde{v} k_-^2 & 0 & 0 \\
		0 & 0 & 0 & 0 & \tilde{v} k_+^2 & -\Delta e^{i \theta } & 0 & 0 \\
	\end{array}	\right) \triangleq \left(  \begin{array}{cc}
		h_{\Delta}({\bf k}_{\parallel},\theta) & 0 \\
		0 & h_{-\Delta}({\bf k}_{\parallel},\theta) \\
	\end{array}\right).
\end{align}
and
\begin{align} \label{sm-eq-ham-bdg-1}
	H_1({\bf k}_\parallel, k_z) = \left(
	\begin{array}{cc|cc|cc|cc}
		F_1+F_2  & 0 & 0 & 0 &  v k_z k_- & 0 & 0 & 0 \\
		0 & -F_1+F_2  & 0 & 0 & 0 & -v k_z k_- & 0 & 0 \\ \hline
		0 & 0 & -F_1-F_2  & 0 & 0 & 0 &  v k_z k_- & 0 \\
		0 & 0 & 0 & F_1-F_2 & 0 & 0 & 0 & - v k_z k_- \\ \hline
		v k_z k_+ & 0 & 0 & 0 & F_1-F_2  & 0 & 0 & 0 \\
		0 & -v k_z k_+ & 0 & 0 & 0 & -F_1-F_2  & 0 & 0 \\ \hline
		0 & 0 & v k_z k_+ & 0 & 0 & 0 & -F_1+F_2  & 0 \\
		0 & 0 & 0 & -v k_z k_+ & 0 & 0 & 0 & F_1+F_2  \\
	\end{array} \right).
\end{align}

\subsection{2.3 Surface Local Density of States}

The continuum LSM Hamiltonian can be regularized into a tight-binding (TB) model by replacing $k_i, k_i^2$ with $\sin(k_i), 2(1-\cos(k_i))$, respectively. Compared to the continuum Hamiltonian, the advantages of a TB model are
\begin{itemize}
	\item[(1)] it facilitates the studies of rotational symmetry breaking effects on the vortex-line topology.
	\item[(2)] it allows for an iterative Green's function method to calculate the surface local density of states (LDOS) for a semi-infinite slab geometry.
\end{itemize}

In Fig.~2 of the main text, we have discussed two vortex topological phase diagrams for the isotropic Luttinger semimetal. Below, we briefly review the standard recursive Green's function method~\cite{sancho_JPF_1985} to calculate the surface LDOS $D(\vec{r}_\parallel,\omega)$. The same methodology has been applied to generate similar surface LDOS plots for HgTe-class systems in Fig.~4 of the main text.
\begin{align}
	\begin{split}
		&\text{Step 1: initialize the in-plane vortex Hamiltonian $\mathcal{H}_{\parallel}$ and the z-direction hoping Hamiltonian ${\cal H}_{z}$}, \\
		&\text{Step 2: initialize the first iteration, } T_{1,i} = {\cal H}_z, T_{2,i}={\cal H}_z^\dagger, H_{1,i} = \mathcal{H}_{\parallel}, H_{2,i}= \mathcal{H}_{\parallel}, \\
		&\text{Step 3: intermediate matrices, } A = [\omega+i\eta - H_{1,i}]^{-1}, M_1 = A\cdot T_{1,i}, M_2 = A\cdot T_{2,i}, \\
		&\text{Step 4: the $(i+1)^{th}$ matrices, } T_{1,i+1} = T_{1,i} \cdot M_1, T_{2,i+1} = T_{2,i} \cdot M_2, \\
		& \qquad \qquad \qquad \qquad \quad \; \qquad \qquad H_{1,i+1} = H_{1,i} + T_{1,i} \cdot M_2 + T_{2,i} \cdot M_1,
		H_{2,i+1} =  H_{2,i} + T_{1,i} \cdot M_2, \\
		&\text{Step 5: after convergence (typically iteration number $N_{itr} \sim 13$), } G_n = [\omega+i\eta - H_{2,N_{itr}+1}]^{-1}, \\
		&\text{Step 6: the surface Green's function } G_{\text{surf}}(\vec{r}_\parallel,\omega)  = [\omega+i\eta - {\cal H}_z \cdot G_n \cdot {\cal H}_z^\dagger]^{-1}.
	\end{split}
\end{align}
Then the spin-resolved surface LDOS are defined as
\begin{align}
	& D_{\text{tot}} (\vec{r}_\parallel,\omega) = D_\uparrow (\vec{r}_\parallel,\omega) + D_\downarrow (\vec{r}_\parallel,\omega), \\
	& D_\uparrow(\vec{r}_\parallel,\omega) = -\frac{1}{\pi} \text{Im}\left(\text{Tr}[ M_\uparrow \cdot G_{\text{surf}}(\vec{r}_\parallel,\omega) ] \right) , \\
	& D_\downarrow(\vec{r}_\parallel,\omega) = -\frac{1}{\pi} \text{Im}\left( \text{Tr} [ M_\downarrow \cdot G_{\text{surf}}(\vec{r}_\parallel,\omega) ] \right),
\end{align}
where $M_{\uparrow}$ and $M_{\downarrow}$ are the projection operator onto spin-up and spin-down subspace, respective. For the Luttinger semimetal model, they are
\begin{align}
	\begin{split}
		M_\uparrow &= \frac{\tau_0+\tau_z}{2} \otimes \sigma_0 \otimes \frac{s_0+s_z}{2}, \\
		M_\downarrow &= \frac{\tau_0+\tau_z}{2} \otimes \sigma_0 \otimes \frac{s_0-s_z}{2}.
	\end{split}
\end{align}
Similarly, the spin projection operators for the six-band Kane model are
\begin{align}
	\begin{split}
		M_\uparrow &= \frac{\tau_0+\tau_z}{2} \otimes \text{Diag}[0,1,0,1,0,1], \\
		M_\downarrow &= \frac{\tau_0+\tau_z}{2} \otimes \text{Diag}[1,0,1,0,1,0].
	\end{split}
\end{align}

\begin{figure*}[tbp]
	\centering
	\includegraphics[width=0.9\linewidth]{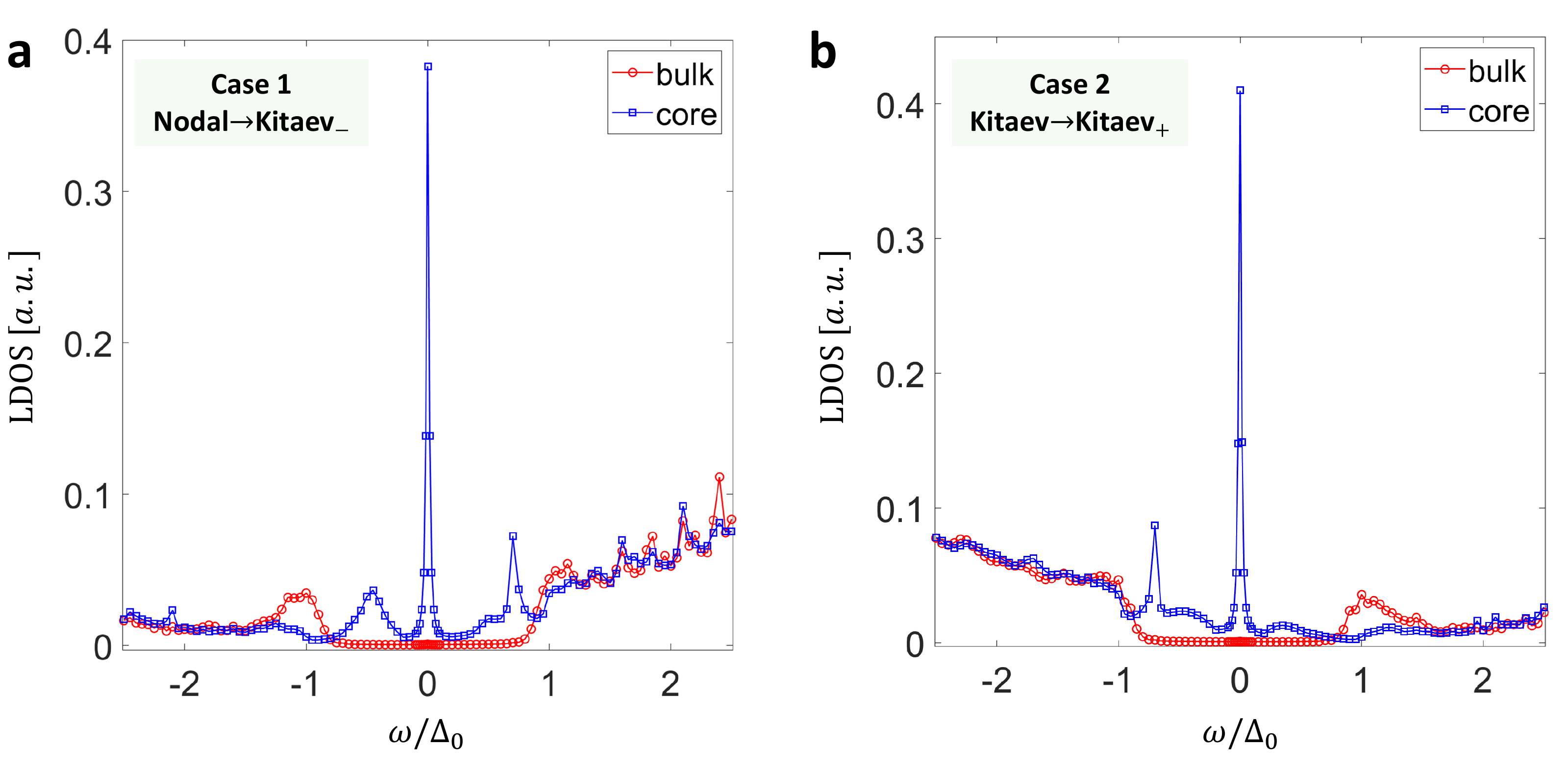}
	\caption{Surface LDOS for Kitaev$_\pm$ vortex phases are shown in (a) and (b). The blue curve represents the LDOS $D_{\text{tot}} (\vec{r}_c,\omega)$ at the vortex core center $\vec{r}_c=(20,20)$ and the red curve is for the LDOS $D_{\text{tot}} (\vec{r}_b,\omega)$ at a position $\vec{r}_b=(30,30)$ far away from the vortex core. Here a.u. stands for arbitrary units.
	}
	\label{sm-fig3} 
\end{figure*}

Vortex MZMs of both Kitaev$_-$ and Kitaev$_+$ vortex phases will induce pronounced zero-bias peaks (ZBPs) in both the total and spin-resolved LDOS at the vortex core. The simulation is performed on a $39\times 39$ lattice with the following parameter set,
\begin{align}
	m_1=-1, m_2=2, \lambda_1=0, \lambda_2=-1, v_z=-2\sqrt{3}\lambda_2, \Delta_0=0.4, \Sigma_{\text{str}}= 0.3, \Sigma_{\text{sb}} = 0.2.
\end{align}
The Kitaev$_\pm$ vortex phases are achieved when $\mu=\mp 0.5$. We set the energy resolution $\eta=\Delta_0/80$ and show the numerical results in Fig.~\ref{sm-fig3}, where the blue curve represents the LDOS $D_{\text{tot}} (\vec{r}_c,\omega)$ at the vortex core center $\vec{r}_c=(20,20)$ and the red curve is for the LDOS $D_{\text{tot}} (\vec{r}_b,\omega)$ at a position $\vec{r}_b=(30,30)$ far away from the vortex core. As shown in Fig.~\ref{sm-fig3}, both Kitaev$_\pm$ phases show the significant zero-bias peak signature at vortex core center. This clearly demonstrates that vortex Majorana bound states can be indeed generated by doping a topologically trivial band insulator.

\begin{figure*}[tbp]
	\centering
	\includegraphics[width=\linewidth]{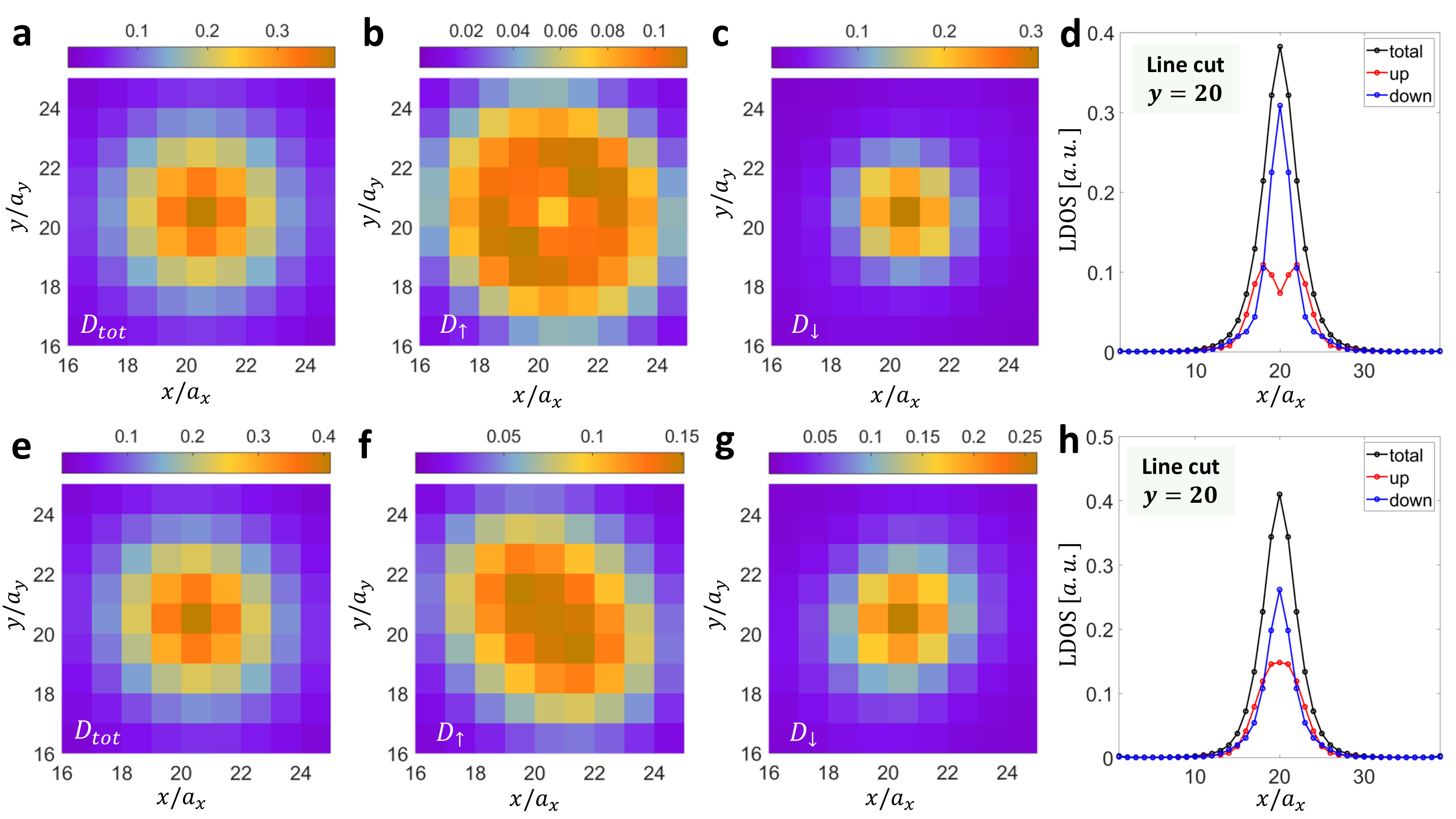}
	\caption{The color map plots of the spin-resolved surface LDOS for Kitaev$_-$ and Kitaev$_+$ phases are shown in (a-d) and (e-h) near the vortex core center, respectively. The effect of rotation symmetry breaking is clearly visible. In particular, (d) and (h) shows the LDOS for Kitaev$_-$ and Kitaev$_+$ phases along the line cut with $y=20$, both of which shows that the spin-down sector has a more pronounced zero-bias peak.
	}
	\label{sm-fig4} 
\end{figure*}

The spin-resolved LDOSs $D_\uparrow(\vec{r}_\parallel,\omega=0)$ and $D_\downarrow(\vec{r}_\parallel,\omega=0)$ at a zero bias voltage for both Kitaev$_-$ and Kitaev$_+$ vortex phases are shown in Fig.~\ref{sm-fig4}, where (a-d) is for the Kitaev$_-$ vortex and (e-h) is for the Kitaev$_+$ vortex. The $C_2$-symmetric vortex profile images in Fig.~\ref{sm-fig4} (a) and (e) confirm the breaking of rotational symmetry by $\Sigma_{\text{sb}}$, which should be experimentally detectable. In addition, we also notice that both cases show that $D_\uparrow(\vec{r}_c,\omega=0) < D_\downarrow(\vec{r}_c,\omega=0)$ at the vortex core center $\vec{r}_c=(20,20)$, which is consistent with the LDOSs of the Kitaev$_-$ vortex phase of the six-band Kane model [see Fig.~4 in the main text]. This is reasonable since the Kitaev$_-$ vortex in the Kane model originates from its LSM physics.

\section{Supplementary Note 3: Vortex Topology in Luttinger Semimetal: Analytical Theory}

In this part, we will combine both analytical and numerical expertise to show that how $H_1({\bf k})$ in Eq.~\eqref{sm-eq-ham-bdg-1} will make the zero modes disperse $k_z$ and further develop vortex-line topology in the superconducting Luttinger semimetal. 

The chiral winding number calculation described in Methods is powerful to analytically prove the existence of four zero modes. For our purpose, we will need additional information of the zero-mode wavefunction, which turns out to be analytically challenging to obtain. Instead, we choose to extract the general form of the zero-mode wavefunctions through a large-scale numerical calculation. In particular, we find that two of the zero modes carry $J_z=0$ and the other two belong to the $J_z=\pm 1$ sector:
\begin{align}
	\begin{split}
		&\vert \Phi(0,r,\theta)\rangle_{1}  =   \Big{\{} u_1(-1,r) e^{-i\theta}, u_2(0,r),  0, 0, 0,  0,  0, 0 \Big{\}}^T, \\
		&\vert \Phi(0,r,\theta)\rangle_{2}  =   \Big{\{}  0, 0, 0,  0, u_2(0,r), u_1(-1,r) e^{i\theta},  0, 0 \Big{\}}^T, \\
		&\vert \Phi(1,r,\theta)\rangle_{1}  =   \Big{\{} u_1(0,r), u_2(1,r) e^{i\theta},  0, 0, 0,  0,  0, 0 \Big{\}}^T, \\
		&\vert \Phi(-1,r,\theta)\rangle_{2}  =   \Big{\{}  0, 0, 0,  0, u_2(1,r) e^{-i\theta}, u_1(0,r),  0, 0 \Big{\}}^T.
		\label{eq-zero-mode}
	\end{split}
\end{align} 
Here $u_1(-1,r), u_2(0,r)$ and $u_1(0,r), u_2(1,r)$ follow the definition in Methods in the main text and their coefficients can be determined numerically. As will be shown next, these coefficient details do not matter in terms of the vortex topological conclusion. 

The four zero modes in Eq.~\eqref{eq-zero-mode} span the following low-energy basis function that is crucial for understanding vortex topology,
\begin{align}
	\vert \Psi_{\text{vortex}} \rangle = \left\{\vert \Phi(0,r,\theta)\rangle_{1} ,   \vert \Phi(0,r,\theta)\rangle_{2}, \vert \Phi(1,r,\theta)\rangle_{1},  \vert \Phi(-1,r,\theta)\rangle_{2} \right\}.
\end{align}
Projecting $H_1({\bf k})$ onto the zero-mode manifold, we arrive at a 4 by 4 vortex Hamiltonian consisting of two 2 by 2 decoupled blocks,
\begin{align}\label{sm-eq-vortex-ham}
	\begin{split}
		h_{\text{vortex}} &= \langle \Psi_{\text{vortex}} \vert H_1(\bm{k}_\parallel, k_z) \vert \Psi_{\text{vortex}} \rangle
		= \left(\begin{array}{c|c}
			h_{\text{Kitaev}} & 0 \\ \hline
			0 & h_{\text{Nodal}}
		\end{array}\right).
	\end{split}	                  
\end{align}
In particular, we find that
\begin{align}
	h_{\text{Kitaev}} &= 
	\left(\begin{array}{cc}
		K_0 + K_z k_z^2 & K_vk_z  \\
		K_vk_z & -K_0 - K_zk_z^2
	\end{array}\right), \\
	h_{\text{Nodal}} &= 
	\left(\begin{array}{cc}
		N_0 + N_z k_z^2 & 0 \\
		0 &  -N_0 - N_z k_z^2
	\end{array}\right),
\end{align}
with 
\begin{equation}
	K_0 = K_{\mu} +  K_{\parallel} + K_{\Sigma},\ \ \ N_0 = N_{\mu} +  N_{\parallel} + N_{\Sigma}.
\end{equation}
The explicit form of each projection coefficient is given by
\begin{align} 
	K_v &= v \times \int_{0}^{2\pi}\, \frac{d\theta}{2\pi} \, \int_{0}^{R_{\text{disk}}} \, r dr \, \left\lbrack (u_1^\ast(-1,r) e^{i\theta}) [k_-] (u_2(0,r)) + (u_2^\ast(0,r)) [k_-] (u_1(1,r)e^{i\theta})   \right\rbrack, \\
	K_{\mu} &= \mu \times \int_{0}^{2\pi}\, \frac{d\theta}{2\pi} \, \int_{0}^{R_{\text{disk}}} \, r dr \, \left\lbrack  \vert u_1(-1,r)\vert^2 - \vert u_2(0,r)\vert ^2  \right\rbrack, \\
	K_{\parallel} &= m_1 \times \int_{0}^{2\pi}\, \frac{d\theta}{2\pi} \,   \int_{0}^{R_{\text{disk}}} \, r dr \, \left\lbrack  (u_1^\ast(-1,r) e^{i\theta}) [k_+k_-] (u_1(-1,r) e^{-i\theta}) + (u_2^\ast(0,r)) [k_+k_-] (u_2(0,r))  \right\rbrack, \\
	K_{z} & = m_2 \times \int_{0}^{2\pi}\, \frac{d\theta}{2\pi} \,  \int_{0}^{R_{\text{disk}}} \, r dr \, \left\lbrack  \vert u_1(-1,r)\vert^2 + \vert u_2(0,r)\vert ^2  \right\rbrack, \\
	K_{\Sigma} & = \Sigma_{\text{str}} \times \int_{0}^{2\pi}\, \frac{d\theta}{2\pi} \,  \int_{0}^{R_{\text{disk}}} \, r dr \, \left\lbrack  \vert u_1(-1,r)\vert^2 + \vert u_2(0,r)\vert ^2  \right\rbrack, \\
	N_{\mu} &= \mu \times \int_{0}^{2\pi}\, \frac{d\theta}{2\pi} \,  \int_{0}^{R_{\text{disk}}} \, r dr\, \left\lbrack  \vert u_1(0,r)\vert^2 - \vert u_2(1,r)\vert ^2  \right\rbrack, \\
	N_{\parallel} &= m_1 \times  \int_{0}^{2\pi}\, \frac{d\theta}{2\pi} \,  \int_{0}^{R_{\text{disk}}} \, r dr \, \left\lbrack  (u_1^\ast(0,r)) [k_+k_-] (u_1(0,r)) + (u_2^\ast(1,r)e^{-i\theta}) [k_+k_-] (u_2(1,r)e^{i\theta})  \right\rbrack, \\
	N_{z} & = m_2 \times \int_{0}^{2\pi}\, \frac{d\theta}{2\pi} \,  \int_{0}^{R_{\text{disk}}} \, r dr \, \left\lbrack  \vert u_1(0,r)\vert^2 + \vert u_2(1,r)\vert ^2  \right\rbrack, \\
	N_{\Sigma} & = \Sigma_{\text{str}}\times \int_{0}^{2\pi}\, \frac{d\theta}{2\pi} \,  \int_{0}^{R_{\text{disk}}} \, r dr \, \left\lbrack  \vert u_1(0,r)\vert^2 + \vert u_2(1,r)\vert ^2  \right\rbrack.
\end{align}
where we set $F_1=0$ and $F_2=\Sigma_{\text{str}}+m_1(k_x^2+k_y^2)+m_2k_z^2$ for simplicity and the relation $u(-1,r)=-u(1,r)$ has been applied because of $J_n(r) = (-1)^n J_{-n}(r)$. 

To prove that $h_{\text{Kitaev}}$ and $h_{\text{Nodal}}$ describe a Kitaev vortex and a nodal vortex, respectively, we first note that
\begin{equation}
	\text{sgn}[K_{\Sigma}] = \text{sgn}[N_{\Sigma}] = \text{sgn}[\Sigma_{\text{str}}]
\end{equation} 
We now prove the following relations: 
\begin{align}\label{sm-eq-m1mz-relation}
	\text{sgn}[K_{\parallel}] = \text{sgn}[N_{\parallel}] = \text{sgn}[m_1],\ \ \ \text{sgn}[K_{z}] = \text{sgn}[N_{z}] = \text{sgn}[m_2].	
\end{align}
Take $K_{\parallel}$ as an example and we need to evaluate the integral into two parts:
\begin{itemize}
	\item  $u_2(0,r)$:  According to the Bessel function expansion in Methods, we have
	\begin{align}
		u_2(0,r) = \sum_{j=1}^{N} c_{j,0} \phi(0,r,\alpha_j),
	\end{align}
	which gives rise to
	\begin{align}
		\begin{split}
			& \int_{0}^{2\pi}\, \frac{d\theta}{2\pi} \, \int_{0}^{R_{\text{disk}}} \, r dr \, \left\lbrack (u_2^\ast(0,r)) [k_+k_-] (u_2(0,r))  \right\rbrack \\
			=& \int_{0}^{2\pi}\, \frac{d\theta}{2\pi} \,   \int_{0}^{R_{\text{disk}}} \, r dr \, \left\lbrack (\sum_{j=1}^{N} c_{j,0}^\ast \phi(0,r,\alpha_j)) [k_+k_-] (\sum_{l=1}^{N} c_{l,0} \phi(0,r,\alpha_l))  \right\rbrack \\ 	
			=& \sum_{j=1}^{N}\sum_{l=1}^{N} c_{j,0}^\ast \times c_{l,0} \times \frac{\alpha_l^2}{R_{\text{disk}}^2} \times \left\lbrack \int_{0}^{R_{\text{disk}}} \, r dr \, \phi(0,r,\alpha_j) \phi(0,r,\alpha_l) \right\rbrack \\
			=& \sum_{j=1}^{N}\sum_{l=1}^{N} c_{j,0}^\ast \times c_{l,0} \times \frac{\alpha_l^2}{R_{\text{disk}}^2} \times \delta_{j,l} \\
			=& \sum_{j=1}^{N} \vert c_{j,0}\vert^2  \times \frac{\alpha_j^2}{R_{\text{disk}}^2} \ge 0.
		\end{split}
	\end{align}
	Here, $\alpha_j$ is the zero of $\mathcal{J}_0(r)$ and we have used the fact that $[k_+k_-] \phi(0,r,\alpha_l) = \frac{\alpha_l^2}{R_{\text{disk}}^2}$.
	\item $u_1(-1,r)$: Similarly, we can easily prove that
	\begin{align}
		\begin{split}
			\int_{0}^{2\pi}\, \frac{d\theta}{2\pi} \, \int_{0}^{R_{\text{disk}}} \, r dr \, \left\lbrack (u_1^\ast(-1,r) e^{i\theta}) [k_+k_-] (u_1(-1,r) e^{-i\theta})  \right\rbrack = \sum_{j=1}^{N} \vert c_{j,-1}\vert^2  \times \frac{\alpha_j^2}{R_{\text{disk}}^2} \ge 0,
		\end{split}
	\end{align}
	where $\alpha_j$ is the zero of $\mathcal{J}_1(r)$.
\end{itemize}
Combining the above two contributions together, we have proved that 
\begin{equation}
	K_{\parallel} = m_1 [\sum_{j=1}^{N} (\vert c_{j,0}\vert^2 + \vert c_{j,-1}\vert^2) \times \frac{\alpha_j^2}{R_{\text{disk}}^2}],
\end{equation}
and clearly $\text{sgn}(K_{\parallel}) = \text{sgn}(m_1)$. Similarly, we can prove the other relations in Eq.~\eqref{sm-eq-m1mz-relation}. This complete our proof of the nontrivial topological properties of vortex Hamiltonian in Eq.~\eqref{sm-eq-vortex-ham}. For LSM, $m_1=\lambda_2<0$ and $m_2=-2\lambda_2>0$. With $\mu=\Sigma_{\text{str}}=0$, we have 
\begin{align}
	K_{\parallel}  K_{z}<0,\ \ N_{\parallel} N_{z} <0.
\end{align}
This immediately indicates the inverted band structure for both $h_\text{Kitaev}$ and $h_\text{Nodal}$, leading to 
\begin{equation}
	\nu_0 = 1, {\cal Q}_1 =1 .
\end{equation}
The above topological invariants explain the coexistence of Kitaev vortex and nodal vortex for LSM. The mapping between bulk and vortex coefficients further allow us to qualiatively understand the strain-induced vortex topological phase diagram. For example, a negative $\Sigma_{\text{str}}<0$ enhances the vortex-mode band inversion and further stabilizes the Kitaev$\oplus$nodal vortex phase. A positive $\Sigma_{\text{str}}$, however, weakens the vortex topology and make other phases in the VTPD (e.g. Kitaev, nodal, trivial vortex phases) to emerge.

\begin{figure*}[!htbp]
	\centering
	\includegraphics[width=\linewidth]{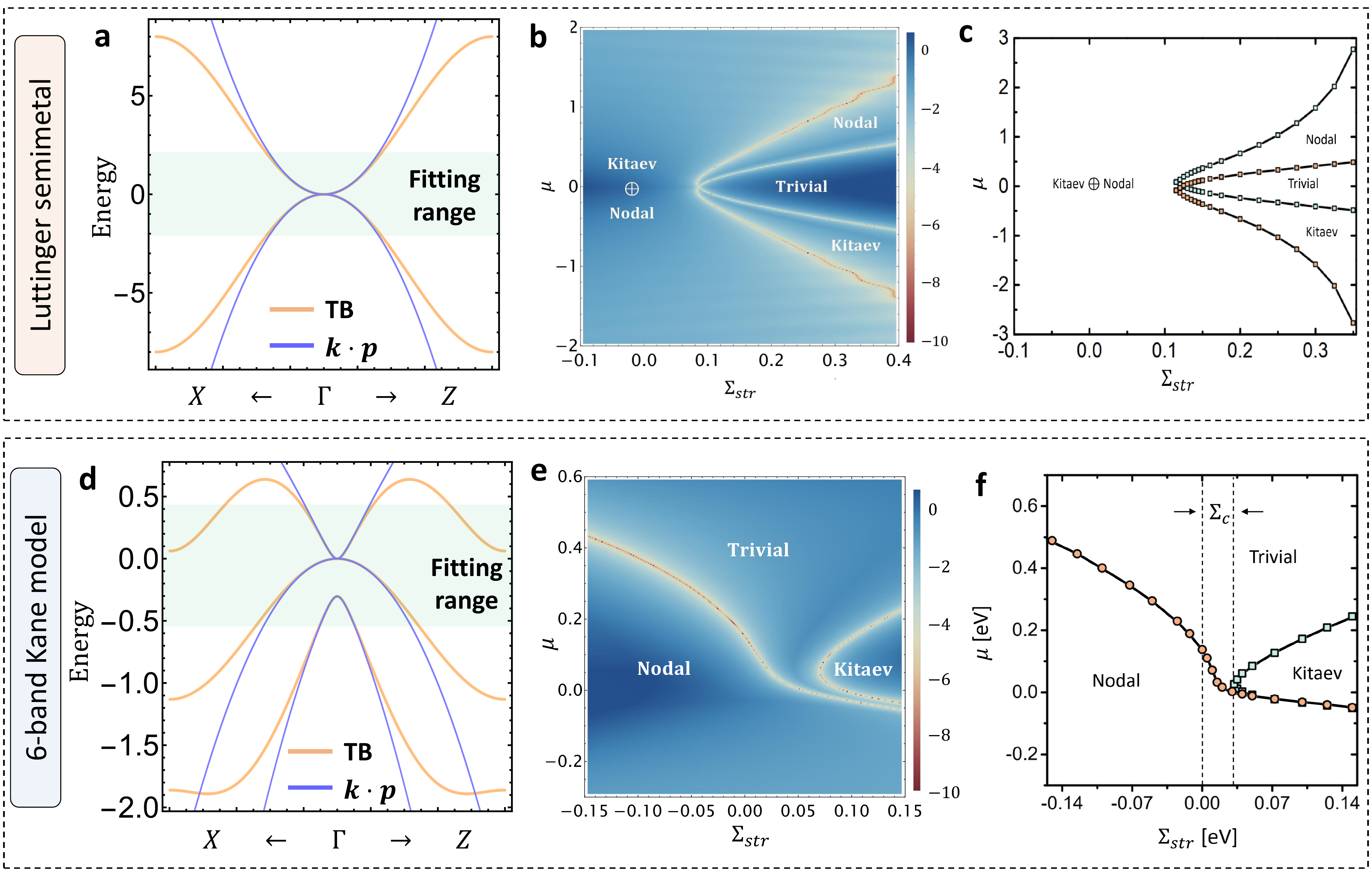}
	\caption{Continuum models v.s. lattice models.
		(a) and (d) are the bulk band structures, where the green shaded region denotes the energy range where continuum and lattice models fit well with each other. 
		Clearly, within these energy windows, the vortex topological phase diagrams (VTPDs) shown in (b) and (c), as well as in (e) and (f), also agree well. Here (b) and (e) are obtained from the tight-binding models, while (c) and (f) are based on the continuum models.
		For the results based on tight-binding model, the VTPDs are achieved by mapping out the vortex energy gap at $k_z=0$, whose logarithmic value is shown by the colors in (b) and (e). 
	}
	\label{sm-fig5} 
\end{figure*}

\section{Supplementary Note 4: Continuum Model v.s. Lattice Model}

In Fig.~\ref{sm-fig5}, we provide a comprehensive comparison between the lattice model and the continuum ${\bf  k}\cdot{\bf p}$ model for both LSM and HgTe, including bulk band structures and vortex topological phase digrams. The lattice models are regularized on a in-plane square lattice, with $k_z$ still being a good quantum number. Clearly, the results from both lattice and continuum models agrees well in a quantitative manner, for both LSM model and Kane model. Therefore, the main conclusions of our work are robust and do not depend on the explicit types of models that are adopted in the numerical simulations.

\end{document}